\documentclass[usenatbib]{mn2e}
\usepackage{graphicx,ifthen,url}
\def\ltsima{$\; \buildrel < \over \sim \;$}
\def\simlt{\lower.5ex\hbox{\ltsima}}
\def\gtsima{$\; \buildrel > \over \sim \;$}
\def\simgt{\lower.5ex\hbox{\gtsima}}
\newcommand {\uJy}{$\mu$Jy}
\newcommand {\um}{$\mu$m}
\newcommand {\co}{\rm CO}
\newcommand {\Mpy}{M$_\odot$\,yr$^{-1}$}
\newcommand {\aj}{AJ}
\newcommand {\aap}{A\&A}
\newcommand {\apj}{ApJ}
\newcommand {\apjs}{ApJS}
\newcommand {\apjl}{ApJL}
\newcommand {\mnras}{MNRAS}

\newcommand {\pasj}{PASJ}
\newcommand {\physrep}{Phys. Rep.}
\newcommand {\nat}{Nature}
\newcommand {\araa}{ARA\&A}
\newcommand {\qjras}{QJRAS}
\def\um     {$\mu$m}
\def\ts     {\thinspace}
\def\kms    {\ifmmode{{\rm \ts km\ts s}^{-1}}\else{\ts km\ts s$^{-1}$}\fi}
\def\msol   {\ifmmode{{\rm M}_{\odot}}\else{M$_{\odot}$}\fi}
\def\lsol   {\ifmmode{{\rm L}_{\odot}}\else{L$_{\odot}$}\fi}
\def\zsol   {\ifmmode{{\rm Z}_{\odot}}\else{Z$_{\odot}$}\fi}
\def\degree {$^{o}$}
\def\etal   {{\rm et\ts al.}}
\def\aco    {\ifmmode{^{12}{\rm CO}(J\!=\!1\! \to \!0)}\else{$^{12}${\rm CO}($J$=1$\to$0)}\fi}
\def\bco    {\ifmmode{^{12}{\rm CO}(J\!=\!2\! \to \!1)}\else{$^{12}${\rm CO}($J$=2$\to$1)}\fi}
\def\cco    {\ifmmode{^{12}{\rm CO}(J\!=\!3\! \to \!2)}\else{$^{12}${\rm CO}($J$=3$\to$2)}\fi}
\def\dco    {\ifmmode{^{12}{\rm CO}(J\!=\!4\! \to \!3)}\else{$^{12}${\rm CO}($J$=4$\to$3)}\fi}
\def\gco    {\ifmmode{^{12}{\rm CO}(J\!=\!7\! \to \!6)}\else{$^{12}${\rm CO}($J$=7$\to$6)}\fi}
\def\rabf     {\rm RGJ123644}
\def\rbbf     {\rm RGJ123711}
\def\rcbf     {\rm RGJ123707}
\def\rdbf     {\rm RGJ123653}
\def\rebf     {\rm RGJ131208}
\def\rfbf     {\rm RGJ105209}
\def\rgbf     {\rm RGJ123645}
\def\rhbf     {\rm RGJ123642}
\def\ribf     {\rm RGJ123718}
\def\chap     {\rm RGJ163655} 
\def\chan     {\rm RGJ131236} 
\def\dada     {\rm RGJ123626} 
\def\dadb     {\rm RGJ123710} 
\def\tett     {\rm RGJ021827} 
\def\tptt     {\rm RGJ105239} 
\def\tott     {\rm RGJ131207} 
\def\rabflong     {\rm RGJ123644.13+621450.7}
\def\rbbflong     {\rm RGJ123711.34+621331.0}
\def\rcbflong     {\rm RGJ123707.82+621057.6}
\def\rdbflong     {\rm RGJ123653.37+621139.6}
\def\rebflong     {\rm RGJ131208.34+424144.4}
\def\rfbflong     {\rm RGJ105209.31+572202.8}
\def\rgbflong     {\rm RGJ123645.88+620754.2}
\def\rhbflong     {\rm RGJ123642.96+620958.1}
\def\ribflong     {\rm RGJ123718.58+621315.0}
\def\chaplong     {\rm RGJ163655.04+410432.0} 
\def\chanlong     {\rm RGJ131236.01+424044.1} 
\def\dadalong     {\rm RGJ123626.53+620835.3} 
\def\dadblong     {\rm RGJ123710.60+622234.6} 
\def\tettlong     {\rm RGJ021827.35--050055.9} 
\def\tpttlong     {\rm RGJ105239.84+572509.1} 
\def\tottlong     {\rm RGJ131207.74+423945.0} 
\def\ci     {\ifmmode{{\rm C}{\rm \small I}}\else{C\ts {\scriptsize I}}\fi}
\def\hi     {\ifmmode{{\rm H}{\rm \small I}}\else{H\ts {\scriptsize I}}\fi}
\def\hh     {\ifmmode{{\rm H}_2}\else{H$_2$}\fi}
\def\cone {\ifmmode{{\rm C}{\rm \small I}(^3\!P_1\!\to^3\!P_0)}
     \else{C\ts {\scriptsize I}{\small$(^3\!P_1\!\to\,^3\!P_0)$}}\fi}
\def\ctwo {\ifmmode{{\rm C}{\rm \small I}(^3\!P_2\!\to\,^3\!P_1)}
     \else{C\ts {\scriptsize I}{\small$(^3\!P_2\!\to\,^3\!P_1)$}}\fi}
\def\cij {\ifmmode{{\rm C}{\rm \small I}\,(^3P_i\to^3P_j)}\else{C\ts {\scriptsize I}\,{\small$(^3P_i\to^3P_j)$}}\fi}
\def\cii    {\ifmmode{{\rm C}{\rm \small II}}\else{C\ts {\scriptsize II}}\fi}
\def\tex {\ifmmode{{T}_{\rm ex}}\else{$T_{\rm ex}$}\fi}
\def\tmb {\ifmmode{{T}_{\rm mb}}\else{$T_{\rm mb}$}\fi}
\def\tkin {\ifmmode{{T}_{\rm kin}}\else{$T_{\rm kin}$}\fi}
\def\microns {\ifmmode{\mu{\rm m}}\else{$\mu$m}\fi}
\def\nhh   {\ifmmode{n({\rm H}_2)}\else{$n$(H$_2$)}\fi}

\newcommand{\kpc}{{\rm\,kpc}}

\newcommand{\msun}{{\rm\,M$_\odot$}}
\newcommand{\sfr}{{\rm\,M$_\odot$\,yr$^{-1}$}}
\newcommand{\lsun}{{\rm\,L$_\odot$}}

\loadboldmathitalic 
\title [CO in z$>$1 submm-faint ULIRGs]
{Molecular Gas in Submillimetre-Faint, Star-Forming Ultraluminous Galaxies at z$>$1}
\author[C.~M. Casey et al.]
{
C.~M. Casey$^{1,2}$\thanks{Hubble Fellow; cmcasey@ifa.hawaii.edu}, S.~C. Chapman$^1$, R. Neri$^3$, F. Bertoldi$^4$, I. Smail$^5$, K. Coppin$^{5,6}$,
\newauthor T.~R. Greve$^7$, M.~S. Bothwell$^1$, R.~J. Beswick$^8$, A.~W. Blain$^9$, P. Cox$^3$, R. Genzel$^{10}$, 
\newauthor T.~W.~B. Muxlow$^7$, A. Omont$^{11}$, A.~M. Swinbank$^{5}$\\
$^1$ Institute of Astronomy, Madingley Road, Cambridge, CB3 0HA \\
$^2$ Institute for Astronomy, University of Hawai'i, 2680 Woodlawn Dr, Honolulu, HI 96822, U.S.A.\\
$^3$ Institut de Radio Astronomie Millimetrique (IRAM), St. Martin d'Heres, France \\
$^4$ Argenlander Institute for Astronomy, University of Bonn, Auf dem H\"{u}gel 71, 53121 Bonn, Germany \\
$^5$ Institute for Computational Cosmology, Durham University, South Road, Durham DH1 3LE \\
$^6$ McGill University, 3600 rue University, Montreal, QC, H3A 2T8, Canada \\
$^7$ Dark Cosmology Centre, Niels Bohr Institute, University of Copenhagen, Juliane Maries Vej 30, DK-2100 Copenhagen \O, Denmark \\
$^8$ Jodrell Bank Observatory, University of Manchester, Macclesfield, SK11 9DL \\
$^9$ Department of Physics \&\ Astronomy, University of Leicester, University Road, Leicester, LE1 7RH \\
$^{10}$ MPE, Giessenbachstrasse 1, D-85741 Garching, Germany \\
$^{11}$ Institut d'Astrophysique de Paris, CNRS and Universit{\'e} Pierre et Marie Curie, 
         98 Bis Boulevard Arago, 75014 Paris, France \\
}
\date{Accepted 10 April 2011.}
\pagerange{\pageref{firstpage}--\pageref{lastpage}} \pubyear{2009}

\begin{document} 
\maketitle 
\label{firstpage}

\begin{abstract}
  We present interferometric CO observations of twelve z$\sim$2
  submillimetre-faint, star-forming radio galaxies (SFRGs) which are
  thought to be ultraluminous infrared galaxies (ULIRGs) possibly
  dominated by warmer dust (T$_{\rm dust}\,\simgt\,$40\,K) than
  submillimetre galaxies (SMGs) of similar luminosities.  Four other
  CO-observed SFRGs are included from the literature, and all
  observations are taken at the Plateau de Bure Interferometer (PdBI)
  in the compact configuration.  Ten of the sixteen SFRGs observed in
  CO (63\%) are detected at $>$4$\sigma$ with a mean inferred
  molecular gas mass of $\sim$2$\times$10$^{10}$\,\msun.
SFRGs trend slightly above the local ULIRG L$_{\rm
  FIR}$-L$^\prime_{\rm CO}$ relation.  Since SFRGs are about two times
fainter in radio luminosity but exhibit similar CO luminosities to
SMGs, this suggests SFRGs are slightly more efficient star formers
than SMGs at the same redshifts.  SFRGs also have a narrow
mean CO line width, 320$\pm$80\,\kms.  Many SMGs have
similarly narrow CO line widths, but very broad features
($\sim$900\,\kms) are present in a few SMGs and are absent from SFRGs.
  SFRGs bridge the gap between properties of very luminous
  $>$5$\times$10$^{12}$\,\lsun SMGs and those of local ULIRGs and are
  consistent with intermediate stage major mergers.  We suspect that
  more moderate-luminosity SMGs, not yet surveyed in CO, would show
  similar molecular gas properties to SFRGs.  The AGN fraction of
  SFRGs is consistent with SMGs and is estimated to be 0.3$\pm$0.1,
  suggesting that SFRGs are observed near the peak phase of star
  formation activity and not in a later, post-SMG enhanced AGN phase.
  Excitation analysis of one SFRG is consistent with CO excitation
  observed in SMGs (turning over beyond \dco).  This CO survey of
  SFRGs serves as a pilot project for the much more extensive survey
  of {\it Herschel} and {\sc SCUBA-2} selected sources which only
  partially overlap with SMGs.  Better constraints on the CO
  properties of a diverse high-$z$ ULIRG population are needed from
  ALMA to determine the evolutionary origin of extreme starbursts, and
  what role ULIRGs serve in catalyzing the formation of massive
  stellar systems in the early Universe.
\end{abstract}

\section{Introduction}\label{s:introduction}

Ultraluminous infrared galaxies (ULIRGs) exhibit some of the most
extreme star formation rates in the Universe.  The volume density of
higher redshift ULIRGs peaks at z$\sim$2-3 \citep{chapman05a}$-$this
is also the peak epoch in the cosmic star formation rate density and
volume density of active galactic nuclei \citep[AGN;
e.g.][]{fan01a,richards06a}.  Not only does this indicate a possible
link between supermassive black hole growth and rapid star formation,
but it also signals the most active phase in galaxy evolution and
formation.  ULIRGs exhibit very intense (SFR\simgt200\,\Mpy),
short-lived bursts ($\tau\,\sim\,$100\,Myr) of star formation.  The
possible life cycle of a ULIRG, from star-formation dominated,
dust-enshrouded galaxy, to obscured AGN and then luminous quasar
\citep[e.g.][]{sanders88b,veilleux09a}, provides a testable
evolutionary sequence.

The best studied ULIRGs at high redshift are submillimetre galaxies
\citep[SMGs;][]{blain02a} which are characterised by their detection
at 850\um\, with S$_{850}\simgt5$\,mJy.  While SMGs put powerful
constraints on galaxy evolution theories and the environments of
extreme star formation
\citep{greve05a,tacconi06a,tacconi08a,chapman05a,pope06a}, their
selection is susceptible to strong temperature biasing
\citep{eales00a,blain04a}.  At the mean redshift of radio SMGs,
z$\sim$2.2, observations at 850\um\ sample the Rayleigh-Jeans tail of
blackbody emission where the observed flux density may be approximated
by S$_{850}\propto$L$_{\rm FIR}\,T_{dust}^{-3.5}$.  Due to the strong
dependence on dust temperature, the 850\um\ flux density of warm-dust
ULIRGs (T$_{d}\simgt$40\,K) might be much lower than cooler dust
specimens (T$_{d}=$20-40\,K) thus causing the warmer-dust galaxies to
evade submm detection.  This selection bias suggests that a large
fraction of the z$\sim2$ ULIRG population has not been accounted for
in current work on high-z star formation.

\citet{chapman04a} describe the first observational effort to identify
warm-dust ULIRGs as a population, via the selection of submm-faint
radio galaxies (SFRGs) with starburst-consistent rest-UV spectra.
While they were thought to be ULIRGs by their similarities to SMGs
(similar radio luminosities, optical spectra, stellar masses), without
detection in the far-infrared (FIR), there was no direct evidence that
their luminosities were in excess of 10$^{12}$\,\lsun.  Ideally,
detection at shorter wavelengths in the infrared, at
$\lambda\,\le\,$500\,\um\, must be used to confirm a ULIRG's
luminosity in the absence of submm detection; \citet{casey09b} used
70\,\um\ detection to confirm that a subset of the SFRG population
contains a dominating warmer-dust component with $<T_{d}>$\,=\,52\,K.
While a population of warm-dust ULIRGs has been shown to exist, some
fundamental questions still remain unanswered: are warm-dust ULIRGs in
a post-SMG AGN heated phase?  Could they be triggered by different
mechanisms than the major mergers said to give rise to cold-dust SMGs?

Investigating the molecular gas content is fundamental to the
characterisation of star formation properties and gas dynamics of a
galaxy population.  Molecular line transitions from carbon monoxide
(CO) are a direct probe of the vast gas reservoirs that are needed to
fuel high star formation rates
\citep{frayer99a,greve05a,tacconi06a,tacconi08a,chapman08a}.  The gas
dynamics which are derived from these observations shed light on
galaxies' evolutionary sequences by measuring how disturbed their gas
reservoirs are and how long they can maintain their star formation
rates with the observed fuel supply.  Recent simulations work hint
that a ULIRG phase may be triggered by either major merger
interactions \citep[e.g.][]{narayanan09a} or from steady bombardment
from low mass fragments \citep[e.g.][]{dave10a}; linking observations
with these different evolutionary scenarios is an essential step in
understanding galaxy evolution in the early Universe.

In this paper, we present CO molecular gas observations, taken with
the IRAM Plateau de Bure Interferometer (PdBI), of twelve SFRGs (and
four additional SFRGs from the literature) to compare the population
with SMGs and other high redshift star forming galaxies.  Section
\ref{s:observations} describes the sample selection, molecular gas
observations and ancillary data, while section \ref{s:results}
presents our results, in the form of derived gas and star formation
quantities of the SFRG sample.  Section \ref{s:discussion} discusses
the gas properties of the sample, compares the population to other
high redshift galaxies, and hypothesizes on the role of SFRGs in a
broader galaxy evolution context relative to local ULIRGs and SMGs
while section \ref{s:conclusions} concludes.  Throughout, we use a
$\Lambda$ CDM cosmology with $H_{\rm 0} = 71$\kms~Mpc$^{-1}$,
$\Omega_{\rm \Lambda}=0.73$ and $\Omega_{\rm m}=0.27$
\citep{hinshaw09a}.

\section{Observations \& Reduction}\label{s:observations}

Our sample is drawn from a set of \uJy\ radio galaxies in the GOODS-N,
Lockman Hole, Elais-N2, SSA13 and the UDS fields using the
\citet{chapman04a} selection of submm-faint star forming radio
galaxies (SFRGs).  They were detected in ultra-deep VLA radio maps
\citep{biggs06a,ivison02a,ivison07a,fomalont06a} with S$_{\rm
  1.4\,GHz}\simgt$15\uJy\ at $>$3\,$\sigma$ with an approximate upper
limit of S$_{\rm 1.4\,GHz}\simlt$1\,mJy since strong AGN and
radio-bright local galaxies were removed from the sample.  The
\uJy\ radio galaxy population was identified in an effort to isolate
bright star formers at high redshift, so only the sources with non-AGN
photometric redshifts of $z\simgt$1 were included
\citep[e.g.][]{chapman03b}.  Spectroscopic follow up with Keck LRIS
revealed starburst spectral features \citep{chapman04a,reddy06a},
mostly at redshifts z\simgt1.  The SFRGs with the most reliable
spectroscopic redshifts (often due to a strong Ly-$\alpha$ emission
peak, $\sigma_{z}\simlt$0.005) were chosen for CO observations at the
IRAM Plateau de Bure Interferometer.

We note that all galaxies in our sample satisfy the $BzK$ 'active'
galaxy selection criterion \citep{daddi04a}, and 10/14 satisfy the
Dust Obscured Galaxy (DOG) selection \citep{dey08a}.  All SFRGs have
poor rest-UV photometry, so their selection with respect to BX/BM
\citep{steidel04a} is not constrained.

\begin{figure}
  \centering
  \includegraphics[width=0.90\columnwidth]{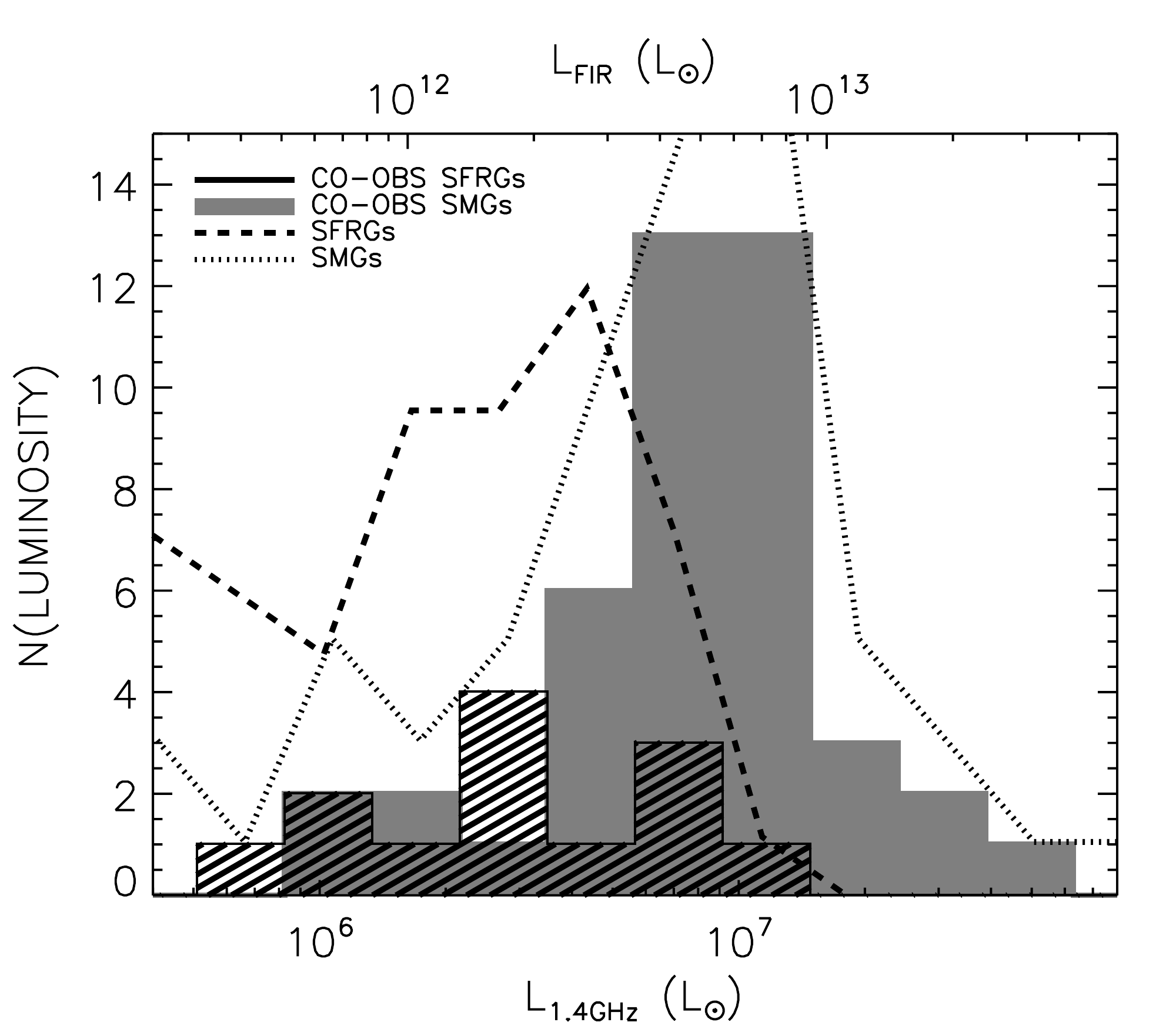}
  \caption{The distribution in radio luminosity (and inferred FIR
    luminosity) of CO observed SFRGs (black line-filled) and SMGs
    (gray filled) relative to the distribution in radio luminosities
    of their parent populations: spectroscopically confirmed SFRGs
    (dashed line) and SMGs (dotted line) without CO observations.  }
  \label{fig:radiodist}
\end{figure}

Figure~\ref{fig:radiodist} shows the distribution in radio luminosity
of the CO-observed SFRG sample relative to the distributions of parent
SFRGs, CO-observed SMGs, and parent SMGs.  We note that the CO
observed sample in this paper is about two times less radio luminous
than the CO-observed SMGs which were analyzed in \citet{neri03a},
\citet{greve05a}, and \citet{tacconi06a}$-$an aspect of their
selection which traces back to the removal of more luminous
spectroscopic AGN from the SFRG sample.  The equivalent class of
spectroscopic AGN are not removed from the SMG sample since their
detection in the FIR provides sufficient evidence that the SMGs are
star-formation dominated.  This is likely represents the most dominant
selection bias between the populations, however we address more in the
discussion in section \ref{ss:volumedensity}.  We also discuss the AGN
fraction of SFRGs at length in section \ref{ss:agn}.  While lower
luminosity SMGs have been observed with PdBI (representing the low
luminosity tail on the SMGs in Fig~\ref{fig:radiodist}; Bothwell \etal,
in preparation), only the published CO observed SMGs are included in
this paper for comparison.

\subsection{PdBI Observations}\label{ss:pdbiobs}

Nine SFRGs were observed from June 2008 to October 2008, while three
additional sources were observed in August through October of 2009.
The mean redshift of the sample is $z=1.9\pm0.6$.

\begin{table*}
\begin{center}
\caption{PdBI Observation Properties and FIR, Radio Data for SFRGs}
\label{tab:observations}
\begin{tabular}{c@{ }c@{ }c@{ }c@{ }c@{ }c@{ }c@{ }c@{ }c@{ }c@{ }c@{ }c@{ }c@{ }c@{ }c@{ }c}
\hline\hline
NAME & z$_{\rm opt}$ & Obs. $^{12}$CO & $\nu_{\rm obs}$     & {\footnotesize BW$^{a}$} & {\footnotesize RMS$_{ct}$} & {\footnotesize RMS$_{ch}$} & Bin & Beamsize & $i$ & S$_{24}$ & S$_{70}$ & S$_{350}$ & S$_{850}$ & S$_{1200}$ & S$_{1.4\,GHz}$ \\
        &                   &   Transition  &   (GHz)      & {\footnotesize (GHz)}  & {\scriptsize (mJy)} & {\scriptsize (mJy)} & (MHz) & (\arcsec\,$\times$\,\arcsec, \degree) & (mag) & (mJy)    & (mJy)   & (mJy)     & (mJy)     & (mJy)      & (\uJy)     \\
\hline
\tett    & 1.361   &  2$\to$1  &  97.810 & 0.9 & 0.11 & 0.71 & 20 & 7.1$\times$4.4, 150\degree & 24.5    & ...        & ...          & ...        & $<$3.5 & ...    & 83.7$\pm$7.0   \\ 
\rfbf    & 2.112   &  3$\to$2  & 111.117 & 1.8 & 0.08 & 0.56 & 40 & 5.1$\times$3.9, 133\degree & $>$25.1 & 167$\pm$47 & $<$1.8       & $<$33      & $<$4.2 & $<$1.8 & 34.5$\pm$5.5   \\
\tptt    & 1.819   &  2$\to$1  &  81.780 & 0.9 & 0.10 & 0.65 & 20 & 6.1$\times$5.1, 98\degree & 22.8    & 150$\pm$30 & $<$6.2       & ...         & $<$1.8 & $<$1.4 & 25.6$\pm$6.2   \\
\rhbf    & 3.661   &  4$\to$3  &  98.915 & 1.8 & 0.08 & 0.56 & 40 & 7.6$\times$3.9, 49\degree & 25.9    & $<$15      & $<$1.9       & ...         & $<$3.4 & $<$0.9 & 20.1$\pm$8.2   \\
\rabf    & 2.095   &  3$\to$2  & 111.727 & 1.6 & 0.10 & 0.66 & 40 & 6.6$\times$3.6, 46\degree & 24.2    & 123$\pm$29 & $<$1.8       & ...         & $<$3.6 & $<$2.8 & 39.6$\pm$8.7   \\
\rgbf    & 1.433   &  2$\to$1  &  94.755 & 1.8 & 0.06 & 0.39 & 40 & 5.0$\times$4.8, 86\degree & 23.5    & 172$\pm$34 & 4.8$\pm$0.4  & ...         & $<$10.8 & $<$1.6 & 83.4$\pm$9.8   \\
\rdbf    & 1.275   &  2$\to$1  & 101.335 & 1.8 & 0.08 & 0.56 & 40 & 4.9$\times$4.0, 75\degree & 22.7    & 164$\pm$33 & 6.6$\pm$0.4  & ...         & $<$1.2 & $<$0.7 & 86.7$\pm$8.3   \\
\rcbf    & 1.489   &  2$\to$1  &  92.623 & 1.8 & 0.08 & 0.56 & 40 & 6.1$\times$4.4, 65\degree & 22.9    & 588$\pm$63 & $<$1.7       & ...         & $<$3.2 & $<$1.7 & 24.1$\pm$8.6   \\
\rbbf    & 1.996   &  4$\to$3  & 153.885 & 1.8 & 0.09 & 0.60 & 40 & 4.2$\times$2.7, 53\degree & 24.2    & 473$\pm$57 & 1.4$\pm$0.4  & $<$24       & $<$2.4 & $<$4.4 & 126.3$\pm$8.6  \\
\ribf    & 1.512   &  2$\to$1  &  91.775 & 1.8 & 0.06 & 0.39 & 40 & 5.5$\times$4.4, 80\degree & 23.1    & 73$\pm$23  & $<$1.7       & ...         & $<$3.8 & $<$0.6 & 15.2$\pm$6.8   \\
\tott    & 1.532   &  2$\to$1  &  91.050 & 0.9 & 0.12 & 0.82 & 20 & 5.8$\times$4.3, 127\degree & 24.8    & 105$\pm$15$^\ddag$ & ...  & ...        & $<$3.8 & ...    & 44.9$\pm$2.4   \\
\rebf    & 2.237   &  3$\to$2  & 106.826 & 1.8 & 0.07 & 0.44 & 40 & 4.5$\times$4.0, 82\degree & 25.2    & 279$\pm$14$^\ddag$ & ...  & $<$82       & $<$3.0 & ...    & 37.6$\pm$4.0   \\
{\bf Lit SFRGs:} & & & & & & & & & & &   &  \\
RG\,J123626$^{a}$   & 1.465  & 2$\to$1 & 93.525  & 1.8 & 0.09 & 0.78 & 12 & & 24.3    & 94$\pm$26  & $<$1.7       & ...          & $<$7.0 & $<$1.1 & 37.9$\pm$9.3   \\
RG\,J123710$^{a}$   & 1.522  & 2$\to$1 & 91.411  & 1.8 & 0.10 & 0.87 & 12  & & 24.2   & 227$\pm$39 & 3.9$\pm$0.5  & ...          & $<$1.8 & $<$1.2 & 38.3$\pm$10.1  \\
{\it ( RG\,J123711$^{b}$} & {\it 1.996 }  & {\it 3$\to$2 } & {\it 115.410} & 0.9 & 0.19 & 1.34 & 18 ) & &        &  &  &  & & &  \\
RG\,J131236$^{b}$   & 2.224  & 3$\to$2 & 106.727 & 0.9 & 0.10 & 0.69 & 18 & & 24.1    & ...        & ...          & ...          & $<$2.2 & ...    & 43.9$\pm$7.1   \\
RG\,J163655$^{b}$   & 2.186  & 3$\to$2 & 108.536 & 1.2 & 0.07 & 0.49 & 20 & & 23.5    & $<$150     & ...          & 13.7$\pm$6.9 & $<$2.2 & ...    & 48.7$\pm$4.3   \\
\hline\hline
\end{tabular}
\end{center}
{\small {\bf Table Notes.}  The top 12 sources were observed in our
  program and the bottom five CO-observed SFRGs are taken from the
  literature: $^a$ from \citet{daddi08a} and $^b$ from
  \citet{chapman08a}.  BW$^{a}$ denotes the bandwidth of observations.
  Non-detections are 2$\sigma$ upper limits, magnitudes are in AB, and
  ellipses denote that no data have been taken of that galaxy at the
  given wavelength (24\um, 70\um, 350\um, and 1200\um). \rbbf\ is
  listed twice, once for its observations taken under our program, and
  once for the observations discussed in \citet{chapman08a}.  The
  24\um\ flux densities of the SSA13 field sources are marked by
  $^\ddag$: they are derived from IRS spectral observations since no
  MIPS imaging exists for this field (Casey et al., in preparation).
  The RMS$_{ct}$ is the noise of the CO observations averaged over the
  bandwidth of observations (also over one beamsize at phase centre),
  while RMS$_{ch}$ is the noise per frequency channel which has width
  with size Bin, in MHz.  The frequency bins are chosen for optimum
  presentation of the spectra, as shown in Figure~\ref{fig:cospec} and
  are based on bandwidth. The uncertainties on the radio flux
  densities are errors on flux integrated measurements and not
  statistical errors.  \tett\ was targeted at a redshift of 1.357,
  which is the redshift of a nearby source; the correct UV
  spectroscopic redshift is 1.361, which is within the bandwidth of
  our CO observations. \chap\ has an H$\alpha$ redshift of 2.192 and a
  UV redshift of 2.186, both within the bandwidth of the literature
  observations.  The redshift for \chan\ has been corrected from
  earlier measurement of 2.240.  Its CO observations were taken at
  2.240, so unfortunately the CO[3-2] line at 2.224 falls at the edge
  of the PdBI bandwidth, which makes it difficult to put constraining
  limits on its CO emission.  }
\end{table*}

CO Observations were carried out with PdBI in the 5 dish
D-configuration (i.e. compact).  We used the 2\,mm and 3\,mm receivers
tuned to the appropriate frequencies of redshifted CO transitions, as
detailed in Table~\ref{tab:observations}.  Pointing centres were at
the VLA positions of the galaxies, and example phase calibrators which
we used were 0221+067, 1044+719, 1418+546, 1308+326, and 0954+658 with
flux calibrators like MWC349, 3C84, and 3C345.  The
synthesised beam size for the configuration varies from
$\approx\,$3-6\arcsec\ FWHM.  Receiver Noise Temperature calibration
was obtained every 12\,min using the standard hot/cold--load absorber
measurements.  The antenna gains were found to be consistent with a
standard range of values from 24-31\,Jy\,K$^{-1}$.  We estimate the
flux density scales to be accurate to about $\pm15$\%.

Data were recorded using both polarisations, offset in frequency,
covering a 1.8\,GHz\ bandwidth, except in the case of \tett, \tott,
and \tptt\ which had 0.9\,GHz bandwidth coverage (overlapping
polarisations), and \rabf\ which had 1.6\,GHz coverage (semi
overlapping polarisations).  \rabf\ had an uncertain spectroscopic
redshift and upon initial observations with a 0.9\,GHz bandwidth, a
feature was identified at the edge of the bandwidth and subsequent
observations were shifted in redshift, thus totaling 1.6\,GHz
coverage.  The total on-source integration time varied between $\sim$4-12
hours.  The data were processed using the {\sc GILDAS} packages {\sc
  CLIC} and {\sc MAPPING} and analyzed with our own IDL-based
routines.  The RMS noise of each object's map is also given in
Table~\ref{tab:observations}.  For clarity of presentation, we have
re-gridded the data to a spectral resolution of 40\,MHz for data with
$>$1.5GHz bandwidth and to 20\,MHz for data with $<$1.5GHz bandwidth.
The bandwidths, binsizes and noise properties of the maps are given in
Table~\ref{tab:observations}.

The search for CO line detection was performed by integrating over all
possible combination of channels at every location in the observed
maps using our own iterative {\sc GILDAS} script.  From the result we
measure the data's signal to noise ratio (S/N) at all points in the
map, and subsequently investigate signal peaks greater than 4$\sigma$.
If a detection with signal strength $>$4$\sigma$ exists within
5\arcsec\ of the target position, the observed galaxy is classified as
detected.

\subsubsection{Offset CO Detections}

Searching for detections within a 5\arcsec\ radius imply that
significant positional offsets are acceptable.  In our galaxies, only
three of our sources, \tett, \rfbf, and \tptt), have offsets
\simgt2\arcsec.  Nominally, an offset $>$2\arcsec\ would be too large
to attribute to the targeted source, however we note a few factors
which can increase the positional uncertainty to $\sim$5\arcsec.  PdBI
D-configuration positional uncertainty is governed by the source S/N
(which is proportional to beamsize/2(S/N), $\simlt$1\arcsec, but might
diverge at small S/N), the baseline model uncertainty
($\sim$0.25\arcsec\ for D-config), and the quality of phase
calibration (which can translate to 1.5-3\arcsec\ seeing).  Since the
S/N of these galaxies is small, then the nominal uncertainty likely
increases from $\sim$2\arcsec to $\sim$3\arcsec.  It is also possible
for the bulk of the gas reservoir in a system to be offset from the
primary source of starlight, especially in the case of major mergers.
In addition, we have found large positional offsets
$\sim$5\arcsec\ with strong $>$5$\sigma$ CO detections in the larger
samples of SMGs (i.e. CO positional offset from radio centroid, Smail
\&\ Bothwell, private communication).  Since these detections are
centred on the correct redshifts and there does not seem to be other
radio/UV/IR sources corresponding with their positions, we treat
\tett, \rfbf, and \tptt\ as detected; however, we label their spectra
and CO maps as `OFFSET' and list them separately in
Table~\ref{tab:co}, both as tentatively detected and as undetected,
giving the upper limits for phase center observations.  We plot them
in subsequent figures as detected, although mark them with different
symbols to indicate exclusion from key calculations in our analysis.

\subsubsection{Remarks on Individual Sources}

\chap, \chan, and \cco\ \rbbf\ are taken from \citet{chapman08a}, the
pilot study for the sample observed in this paper.  These data, when
presented in this paper, are different from the spectra and maps
presented in Chapman \etal\ due to a different reduction.  We decided
to re-reduce their data to be consistent with our reduction, making
significant improvements in both phase and amplitude calibration.  The
differences do not change the results of Chapman \etal, but a few
changes in minor conclusions are noted later on in the results and
discussion.  The Chapman \etal\ CO observations of \chan\ were carried
out assuming $z=$2.240 however more recent rest-UV spectroscopic
observations indicate a different rest-UV redshift, $z=$2.224.  The
final channel in the data cube might suggest a flux excess at the
correct redshift, although we cannot distinguish between a bright line
cutoff at the edge, a broad faint line, continuum, or noise spike.  We
treat this source as undetected with the associated noise properties
of our observations.  We also note that \rbbf\ was observed as part of
\citet{chapman08a} in \cco, but newer, higher S/N observations in
\dco\ were taken as part of this program. For comparison, we include
the Chapman \etal\ results in Tables~\ref{tab:observations} and
\ref{tab:co}.  In the re-reduction of \chap\ observations, we find a
possible companion CO source at the same redshift offset by
$\sim$8\arcsec\ to the northwest which is not seen in the Chapman
\etal\ results, however significant improvements on phase and
amplitude calibration have been made since.  This source is discussed
briefly in section \ref{ss:derivedprops}.  The new data taken for
\rbbf\ reveals a marginal \dco\ detection for the SMG
SMM\,J123711.98+621325.7 (also called HDF\,255); its integrated CO
flux (detected at 4.5$\sigma$) is
$I_{\dco}\,=\,$0.54$\pm$0.12\,Jy\,\kms; its spectrum and properties
will be discussed more at length in a paper summarising CO properties
of observed SMGs to date (Bothwell \etal\ in preparation).

\subsection{Archival Observations}\label{ss:archivalobs}

This paper also includes the CO-observations of two additional SFRGs
from the literature.  \citet{daddi08a} analyzed \dadb\ and \dada\ as
being high redshift normal spiral galaxies which have very low star
formation efficiencies \citep[a study expanded upon in][]{daddi10a}.
The active $BzK$ galaxies which Daddi \etal\ survey have significant
overlap with the SFRG population, since their CO sources require radio
detection, thus they have ULIRG luminosities implied from the radio.
Both \dadb\ and \dada\ are selected as SFRGs via the
\citet{chapman04a} method and are likely very-luminous star formers,
with higher SFRs than most active $BzK$s.  \dadb\ has also been
detected at 70\um\ \citep{casey09b}, directly confirming that it is a
ULIRG with a warm dust temperature.  Interpreting the $BzK$ active
galaxy population as ULIRGs rather than ``high-$z$ normal spirals''
is perhaps sensible for these reasons.

\begin{table*}
\begin{center}
\caption{Derived Gas Properties of submm-faint ULIRGs}
\label{tab:co}
\begin{tabular}{c@{ }c@{ }c@{ }c@{ }c@{ }c@{ }c@{ }c@{ }c@{ }c}
\hline\hline
NAME & z$_{\rm optical}$ & z$_{CO}$ & Obs. $^{12}CO$      & S/N & I$_{CO}$            & L$^\prime_{CO}$           & I$_{CO(1-0)}$  & L$^\prime_{CO(1-0)}$            & $\Delta V_{CO}$ \\
     &                   &          & {\small Transition} &    & (Jy\,km\,s$^{-1}$)  & (K\,km\,s$^{-1}$\,pc$^2$)  & (Jy\,km\,s$^{-1}$) & (K\,km\,s$^{-1}$\,pc$^2$)  & (km\,s$^{-1}$) \\
\hline
{\bf CO-Detected SFRGs:} &  &              &      &    &  &   &  \\
\dadalong    & 1.465             & 1.465$\pm$0.002 & \bco & 6.8 & 0.60$\pm$0.09   & (1.7$\pm$0.3)$\times$10$^{10}$  & 0.20$\pm$0.10 & (2.3$\pm$0.3)$\times$10$^{10}$ & $\sim$350      \\
\rabflong    & 2.095             & 2.090$\pm$0.001 & \cco & 4.3 & 0.79$\pm$0.22   & (1.9$\pm$0.5)$\times$10$^{10}$  & 0.12$\pm$0.03 & (2.5$\pm$0.7)$\times$10$^{10}$ & 214$\pm$132    \\
\rgbflong    & 1.433             & 1.434$\pm$0.001 & \bco & 4.1 & 0.64$\pm$0.17   & (1.8$\pm$0.5)$\times$10$^{10}$  & 0.22$\pm$0.06 & (2.4$\pm$0.6)$\times$10$^{10}$ & 320$\pm$144    \\
\dadblong    & 1.522             & 1.522$\pm$0.002 & \bco & 8.9 & 0.85$\pm$0.10   & (2.5$\pm$0.3)$\times$10$^{10}$  & 0.28$\pm$0.08 & (3.3$\pm$0.3)$\times$10$^{10}$ & $\sim$250      \\
\rbbflong    & 1.996             & 1.988$\pm$0.002 & \dco & 6.5 & 1.02$\pm$0.16   & (1.3$\pm$0.2)$\times$10$^{10}$  & 0.10$\pm$0.02 & (2.0$\pm$0.4)$\times$10$^{10}$ & 558$\pm$121    \\
             &                   & 1.996$\pm$0.002 & \dco & 7.3 & 0.61$\pm$0.08   & (7.8$\pm$1.1)$\times$10$^{9}$   & 0.06$\pm$0.01 & (1.2$\pm$0.1)$\times$10$^{10}$ & 318$\pm$86     \\
             &                   & {\it COMBINED:} & \dco &10.4 & 1.87$\pm$0.18   & (2.4$\pm$0.2)$\times$10$^{10}$  & 0.19$\pm$0.02 & (3.7$\pm$0.4)$\times$10$^{10}$ & (1400$\pm$100) \\
{\it (RGJ123711.34+621331.0} & {\it 1.996} & {\it 1.995} & \cco &{\it 3.2} & {\it 0.70$\pm$0.22} &{\it (1.5$\pm$0.5)$\times$10$^{10}$} & 0.10$\pm$0.03 & {\it (2.0$\pm$0.6)$\times$10$^{10}$)} & ... \\
\rebflong    & 2.237             & 2.237$\pm$0.001 &  \cco &5.6 & 0.88$\pm$0.15   & (2.4$\pm$0.5)$\times$10$^{10}$  & 0.13$\pm$0.02 & (3.3$\pm$0.6)$\times$10$^{10}$ & 439$\pm$84     \\
\chaplong    & 2.186             & 2.187$\pm$0.002  &  \cco &4.9 & 0.29$\pm$0.06     & (7.8$\pm$1.6)$\times$10$^{9}$ & 0.04$\pm$0.01 & (1.0$\pm$0.2)$\times$10$^{10}$  & 252$\pm$40  \\
{\bf Offset-CO SFRGs:} &  &              &      &   &   &   &   \\
{\emph (as detections)} &  &              &      &   &   &   &   \\
\tettlong    & 1.361             & 1.362$\pm$0.002 &  \bco &4.2 & 0.44$\pm$0.10   & (1.1$\pm$0.2)$\times$10$^{10}$  & 0.15$\pm$0.04 & (1.5$\pm$0.3)$\times$10$^{10}$ & 557$\pm$107    \\
\rfbflong    & 2.112             & 2.113$\pm$0.001 &  \cco &4.6 & 1.12$\pm$0.25   & (2.8$\pm$0.6)$\times$10$^{10}$  & 0.17$\pm$0.04 & (3.7$\pm$0.7)$\times$10$^{10}$ & 446$\pm$85     \\
\tpttlong    & 1.819             & 1.820$\pm$0.001 &  \bco &4.7 & 0.57$\pm$0.12   & (2.5$\pm$0.5)$\times$10$^{10}$  & 0.19$\pm$0.04 & (3.3$\pm$0.7)$\times$10$^{10}$ & 498$\pm$157    \\
{\emph (as non-detections)} &  &              &      &   &   &   &   \\
\tettlong    & 1.361             & ...             &  \bco & ... & $<$0.20        & $<$4.9$\times$10$^{9}$          & $<$0.07 & $<$6.5$\times$10$^{9}$        &  ... \\
\rfbflong    & 2.112             & ...             &  \cco & ... & $<$0.21        & $<$5.1$\times$10$^{9}$          & $<$0.03 & $<$6.9$\times$10$^{9}$        &  ... \\
\tpttlong    & 1.819             & ...             &  \bco & ... & $<$0.20        & $<$8.5$\times$10$^{9}$          & $<$0.07 & $<$1.1$\times$10$^{10}$       &  ... \\
{\bf CO-Undetected SFRGs:} &  &              &      &   &   &   &   \\
\rhbflong    & 3.661             & ...             &  \dco &... &  $<$0.22        & $<$7.8$\times$10$^{9}$          & $<$0.02 & $<$1.3$\times$10$^{10}$       & ... \\
\rdbflong    & 1.275             & ...             &  \bco &... &  $<$0.22        & $<$4.7$\times$10$^{9}$          & $<$0.07 & $<$6.2$\times$10$^{9}$        & ... \\
\rcbflong    & 1.489             & ...             &  \bco &... &  $<$0.23        & $<$6.7$\times$10$^{9}$          & $<$0.08 & $<$8.9$\times$10$^{9}$        & ... \\
\ribflong    & 1.512             & ...             &  \bco &... &  $<$0.15        & $<$4.7$\times$10$^{9}$          & $<$0.05 & $<$6.3$\times$10$^{9}$        & ... \\
\tottlong    & 1.532             & ...             &  \bco &... &  $<$0.24        & $<$7.3$\times$10$^{9}$          & $<$0.08 & $<$9.8$\times$10$^{9}$        & ... \\
\chanlong    & 2.224             & ...             &  \cco &... &  $<$0.18        & $<$4.7$\times$10$^{9}$          & $<$0.03 & $<$6.3$\times$10$^{9}$        & ... \\

\hline\hline
\end{tabular}
\end{center}
{\small {\bf Table Notes.}  The observed CO properties of the sample.
  The top seven sources have clean $>$4$\sigma$ detections of CO at
  phase center, the next three sources (labeled 'Offset') have CO
  detections at the correct redshift but offset $\sim$3-5\arcsec\ from
  phase center, and the remaining six sources are undetected in CO.
  $L_{CO}^{\prime}$ and I$_{CO}$ are given for the observed CO
  transition, which does not depend on source excitation.  Then
  assuming the \citet{weiss07a} SMG excitation ladder (see text for
  description), we convert to $I_{CO[1-0]}$ and $L_{CO[1-0]}^\prime$,
  which is given in the subsequent columns.  $\Delta V_{\rm \co}$ is
  the FWHM of the fitted feature shown in Fig.~\ref{fig:comaps}.  The
  SFRG with a double peaked feature (\rbbf) has the details of each
  feature listed separately.  I$_{\rm \co}$ 2-$\sigma$ limits for
  undetected SFRGs are calculated assuming a $\Delta V_{\rm
    \co}$\,=\,320\kms\ (the mean FWHM of the detected-SFRG sample).  }
\end{table*}

\subsection{Multiwavelength Data}\label{ss:otherdata}

Radio fluxes from VLA B-array maps (5\arcsec\ resolution) are taken
from \citet{richards00a} and \citet{morrison08a}, where the
uncertainty represents the error in the extracted flux measurement
rather than statistical error.  High-resolution observations from the
Multi-Element Radio Linked Interferometer Network
\citep[MERLIN;][]{thomasson86a} were obtained for the sources in
GOODS-N as described in \citet{muxlow05a}, and a combined MERLIN+VLA
map was constructed with an RMS noise of 4.0\,\uJy\,beam$^{-1}$.  A
similar map was constructed in Lockman Hole with an RMS noise of
4.5\,\uJy\,beam$^{-1}$ \citep*{biggs08a}. The combined MERLIN+VLA maps
have positional accuracies of tens of mas and restoring circular beam
sizes of 0.4\arcsec\ (GOODS-N) and 0.5\arcsec (Lockman Hole).

\rfbf, \rbbf\ and \rebf\ were all observed at 350\um\ with the {\sc
  SHARC2} camera \citep{dowell03a} on the Caltech Submillimeter
Observatory and reduced with {\sc CRUSH} software \citep{kovacs06a}.
With on-source integration times of 950s, 1080s and 200s, \rfbf,
\rbbf\ and \rebf\ were have flux densities of 19$\pm$7\,mJy,
6$\pm$9\,mJy, and 48$\pm$17\,mJy respectively.  A fourth SFRG, \chap,
was observed at 350\um\ with the CSO previously \citep{chapman08a}.
None are detected at $>$3$\sigma$.  While the acquisition of
ground-based 350\,\um\ observations is laborious and dependent on the
best weather conditions ($\tau_{\rm cso}\,<\,$0.06, measured at
225\,GHz), the relative depth of our observations is comparable to
expected confusion limits for the {\it Herschel Space Observatory} at
350\,\um.  While {\it Herschel} will detect $>$40\,mJy
350\,\um\ sources at high-$z$ regularly, sources with flux density
$\simlt$20\,mJy will need follow up from ground-based facilities like
the CSO for more precise flux density estimates and better SED
constraints.

The 1200\um\ flux limits for GOODS-N and Lockman Hole come from the
Max-Planck Millimeter Bolometer \citep[MAMBO, with a mean RMS of
  $\sim$0.8\,mJy;][]{greve08a} and 850\um\ flux limits from the Submm
Common User Bolometric Array \citep[SCUBA, with a mean RMS of
  $\sim$1.6\,mJy;][]{borys03a,coppin06a}.  All fields are covered with
$Spitzer$ IRAC (3.6, 4.5, 5.8, and 8.0\um) and MIPS (24 and 70\um),
however at greatest depths in all bands in GOODS-N. Optical photometry
in GOODS-N is from the $HST$ ACS\footnote{Based on observations made
  with the NASA/ESA Hubble Space Telescope, and obtained from the
  Hubble Legacy Archive, which is a collaboration between the Space
  Telescope Science Institute (STScI/NASA), the Space Telescope
  European Coordinating Facility (STECF/ESA) and the Canadian
  Astronomy Data Centre (CADC/NRC/CSA).}  using the F435W, F606W,
F814W, and F850LP filters (B, V, i and z bands). The Lockman Hole has
$HST$ ACS F814W (PI: Chapman HST 7057) in addition to extensive
optical photometry from Subaru/Suprime-Cam \citep{miyazaki02a}.  X-ray
fluxes are measured from the Chandra/XMM maps of GOODS-N
\citep{alexander03a}, Lockman Hole \citep{brunner08a}, SSA13
\citep{mushotzky00a}, and SXDF \citep{ueda08a}.

\subsection{\rfbf}

Recent work on mid-infrared spectra of SFRGs from the {\it Spitzer}
InfraRed Spectrograph (IRS) (Casey et al., in preparation) has
revealed that the 24\um\ source that lies at the position of \rfbf\
(at the radio position) has a dominant PAH redshift of 2.37, in
contrast to the CO and UV spectroscopic redshift of 2.112.  We suspect
that two different systems overlap to create these discrepant
redshifts, but note that only one source is visible in the
rest-UV. The high-resolution MERLIN+VLA radio imaging also shows only one
point source slightly offset 1.2\arcsec\ from the UV source.  Due to
positional uncertainties, it is difficult to pinpoint which source is
generating the radio emission.  For that reason, we assume in this
paper that the radio is associated with the 2.112 CO source.  We note
that similar positional overlap phenomena have occurred with star
forming galaxies before, for example there are several examples of
sources with two distinct redshifts for one continuum source in the
surveys of \citet{steidel04a} and \citet{reddy06a,reddy08a}, and the
density of bright star forming z$\sim$2 sources is high enough that
overlap will occur with non-negligible probability.

\section{Analysis and Results}\label{s:results}

\subsection{Derived Properties from CO Data}\label{ss:derivedprops}

\begin{figure*}
  \centering
  \includegraphics[width=0.67\columnwidth]{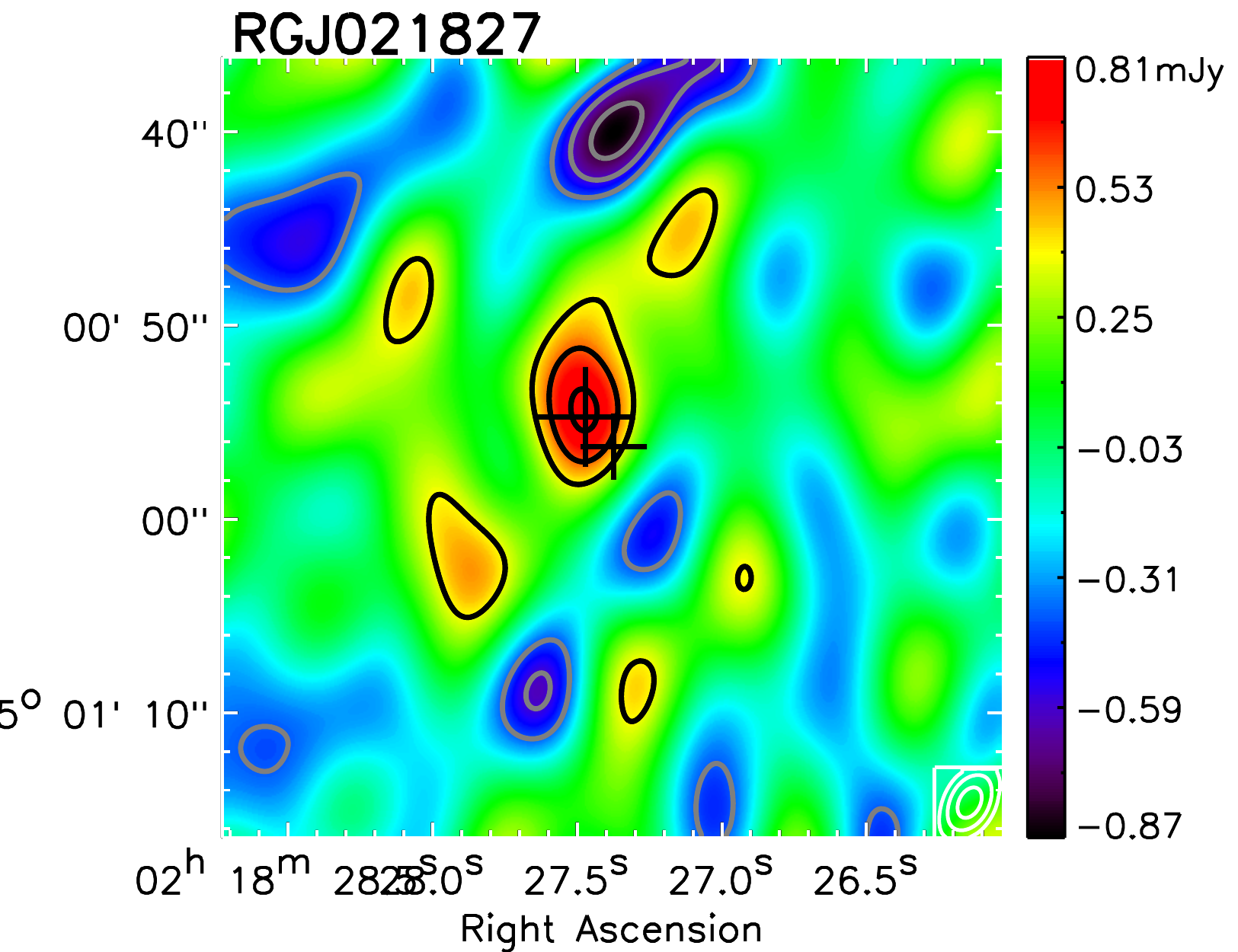}\hspace{0.1in}
  \includegraphics[width=0.67\columnwidth]{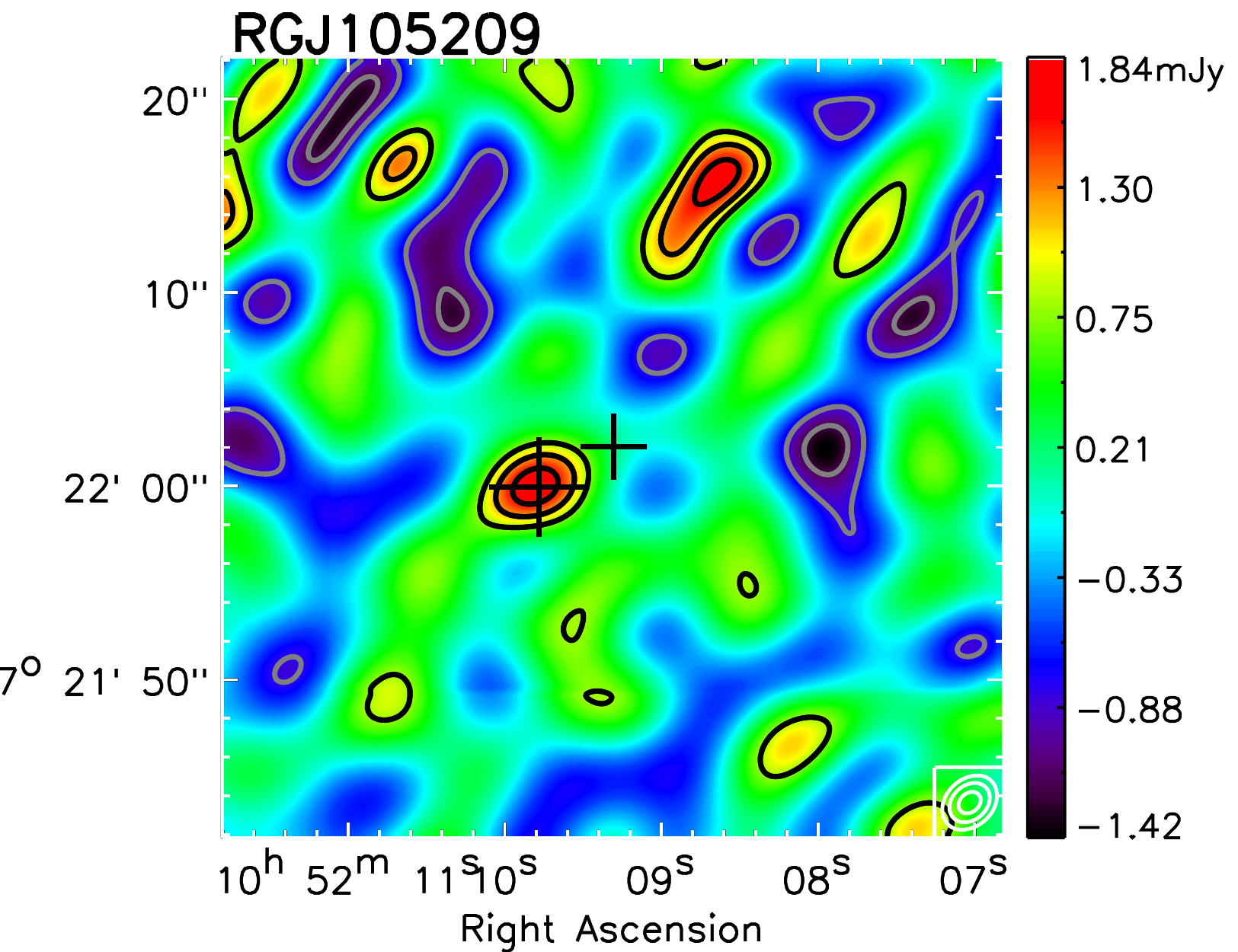}\hspace{0.1in}
  \includegraphics[width=0.67\columnwidth]{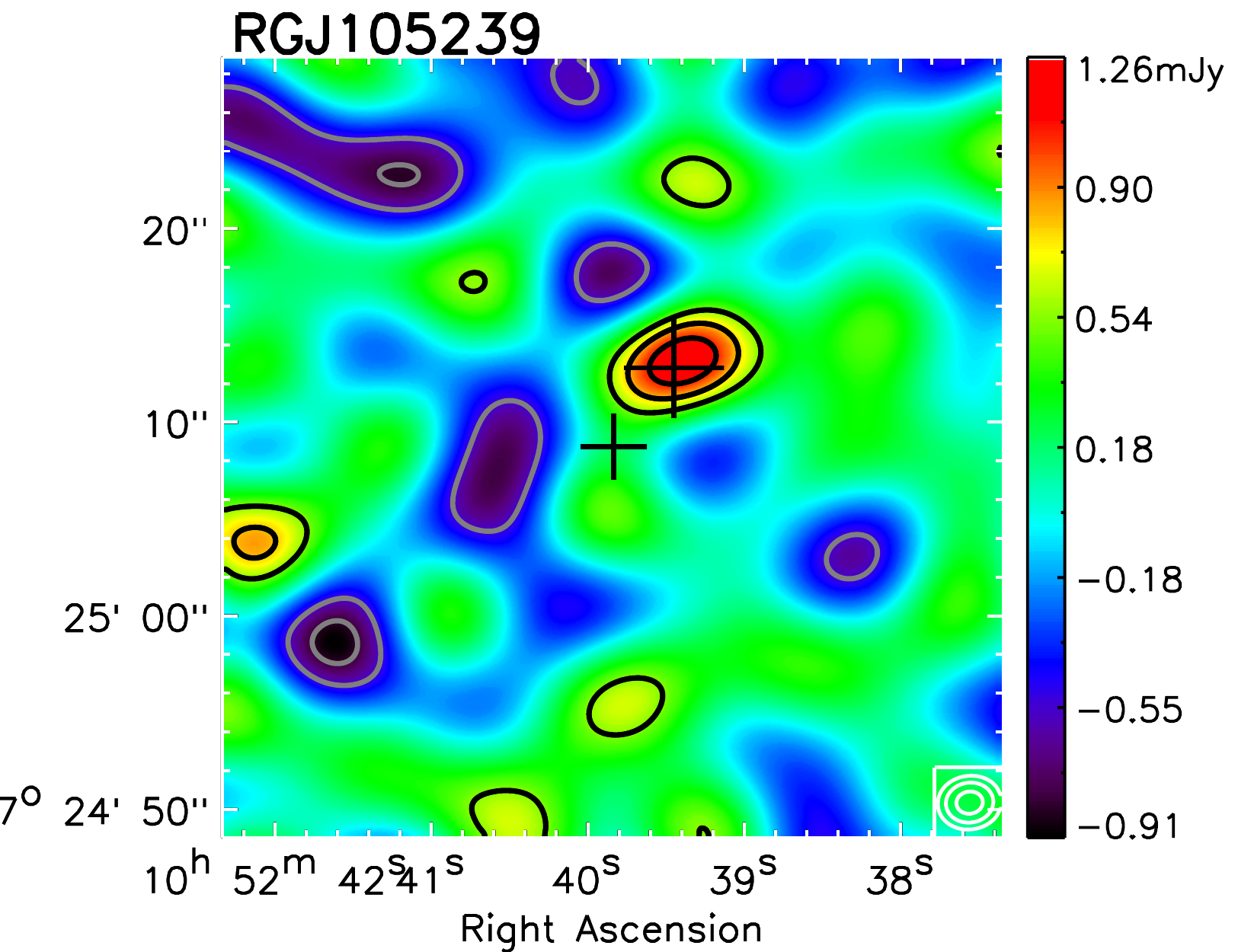}\\
  \includegraphics[width=0.67\columnwidth]{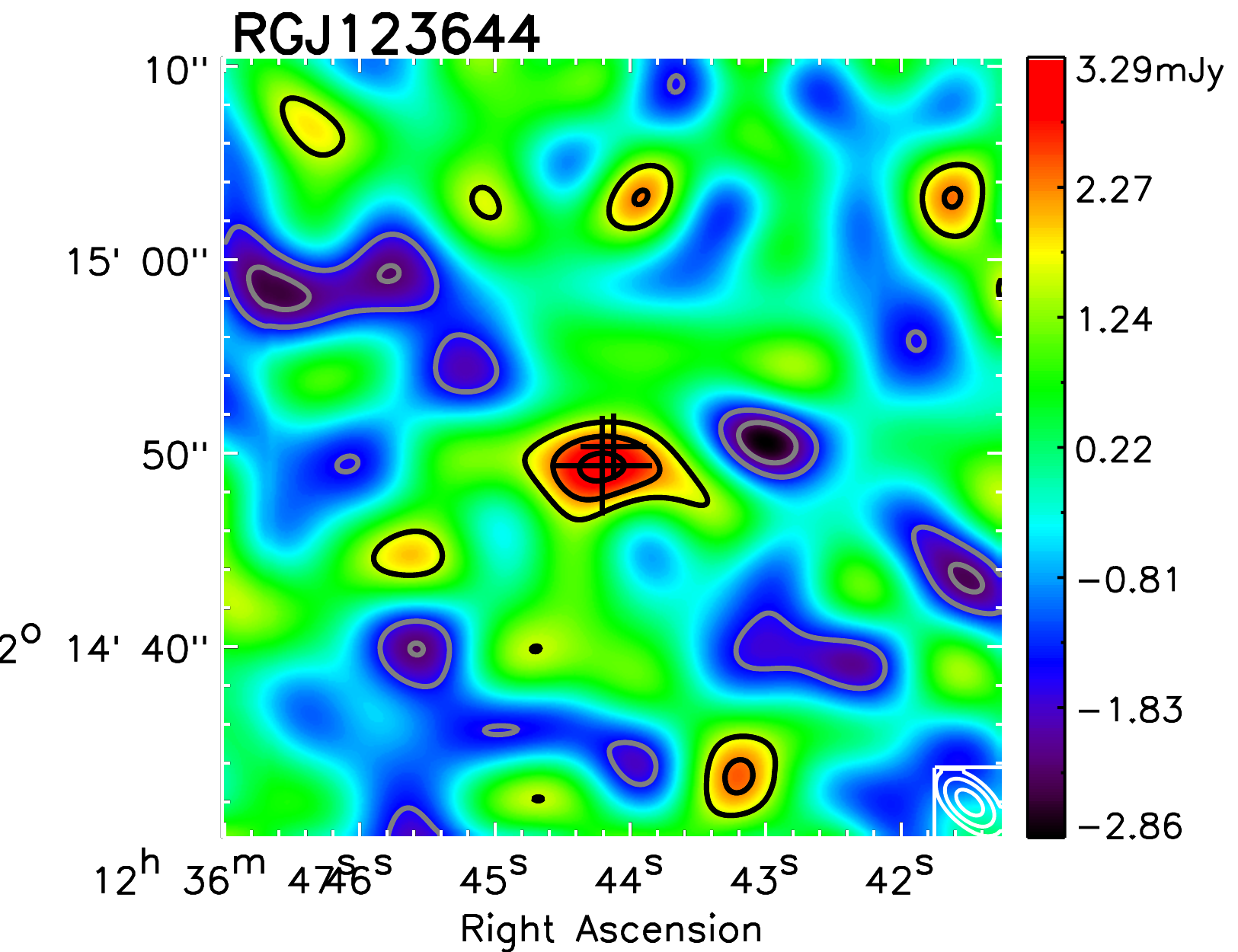}\hspace{0.1in}
  \includegraphics[width=0.67\columnwidth]{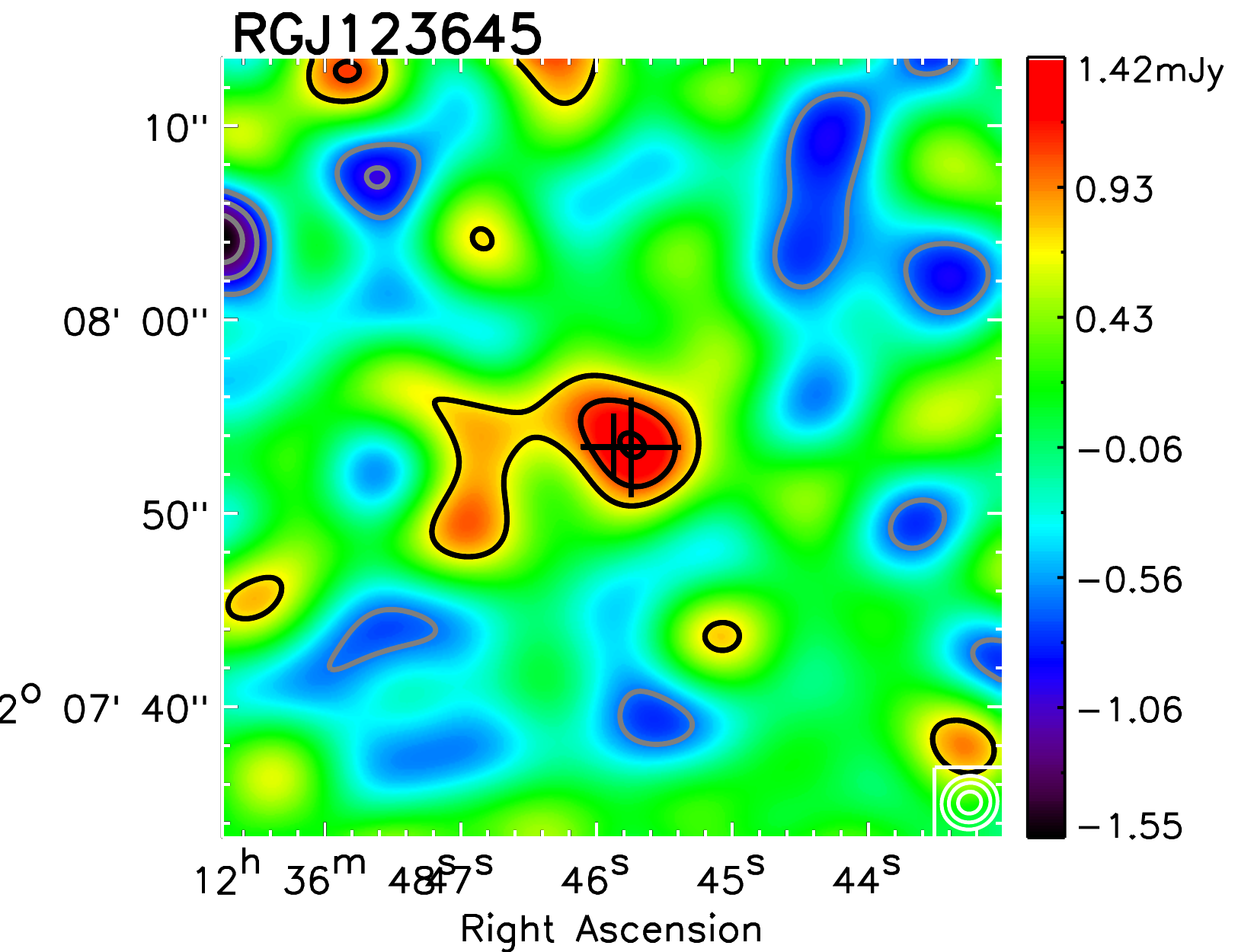}\hspace{0.1in}
  \includegraphics[width=0.67\columnwidth]{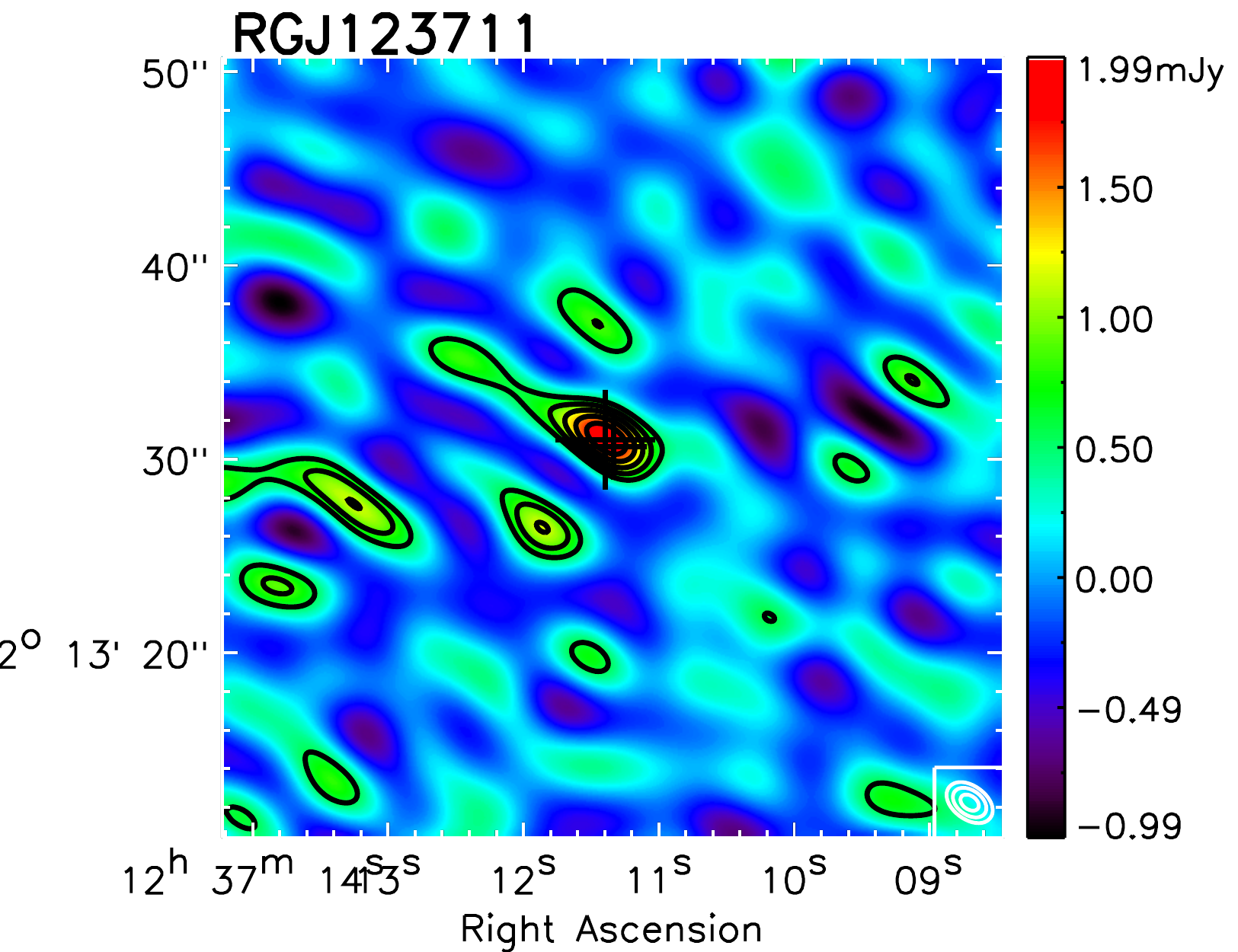}\\
  \includegraphics[width=0.67\columnwidth]{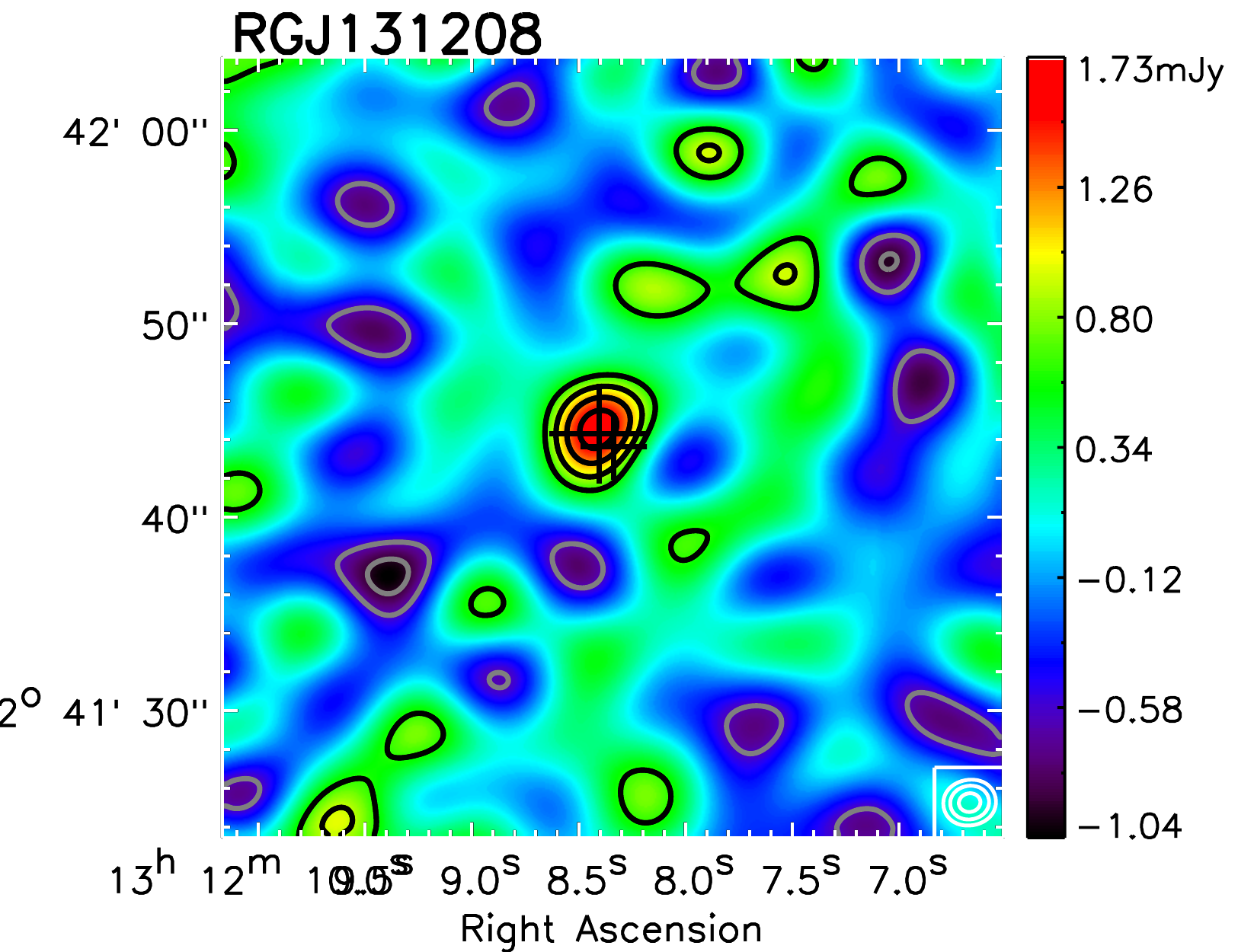}\hspace{0.1in}
  \includegraphics[width=0.67\columnwidth]{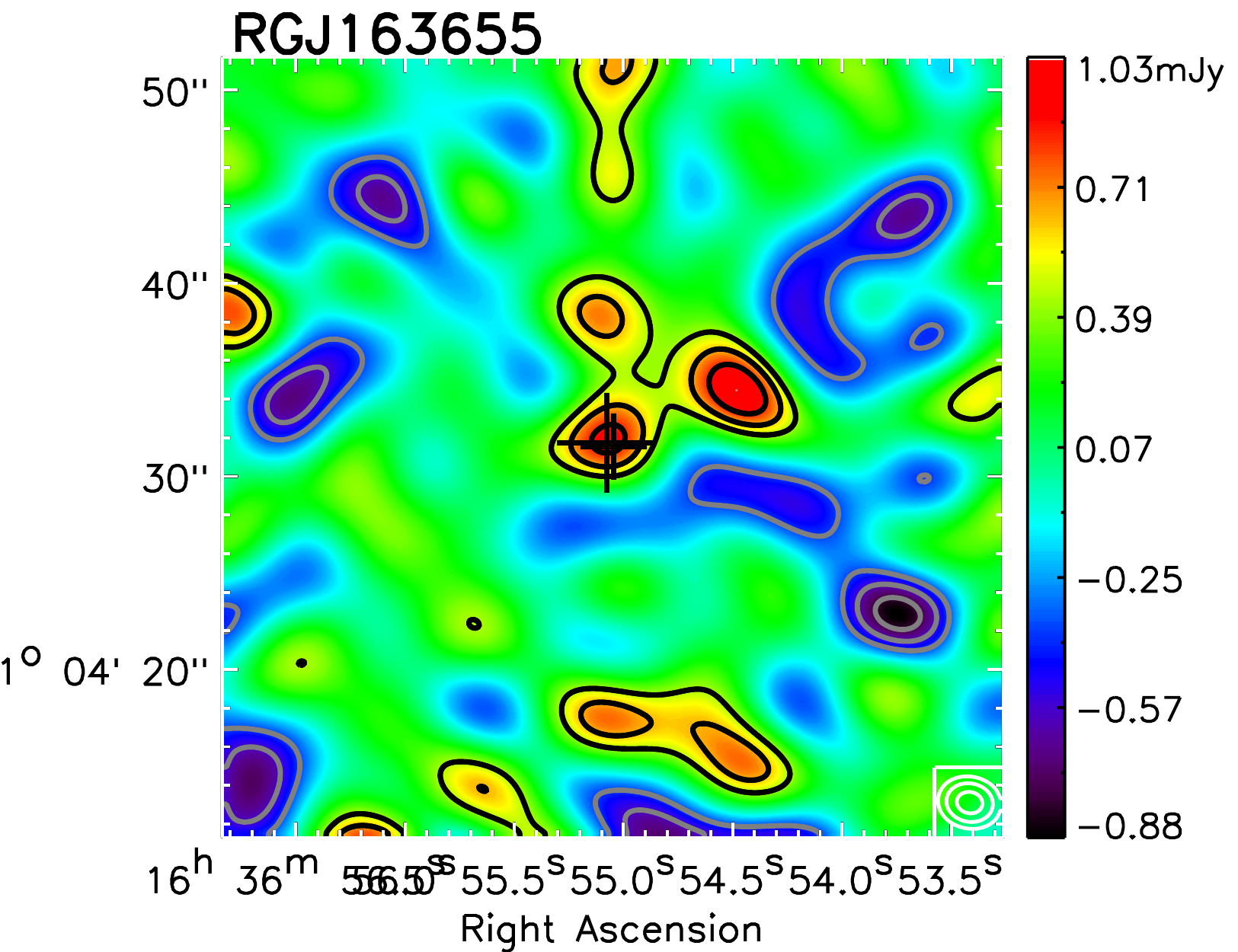}\\
  \includegraphics[width=0.67\columnwidth]{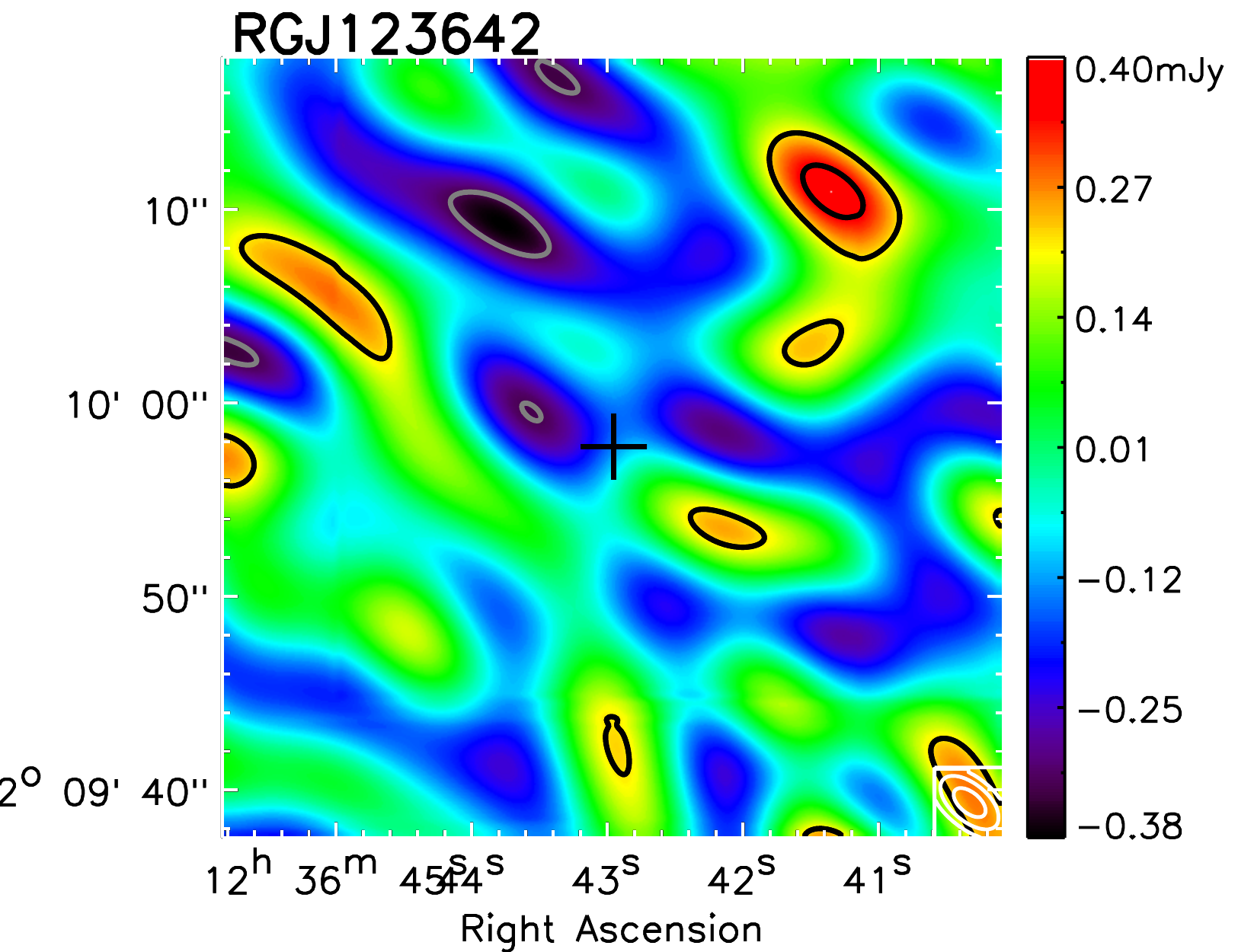}\hspace{0.1in}
  \includegraphics[width=0.67\columnwidth]{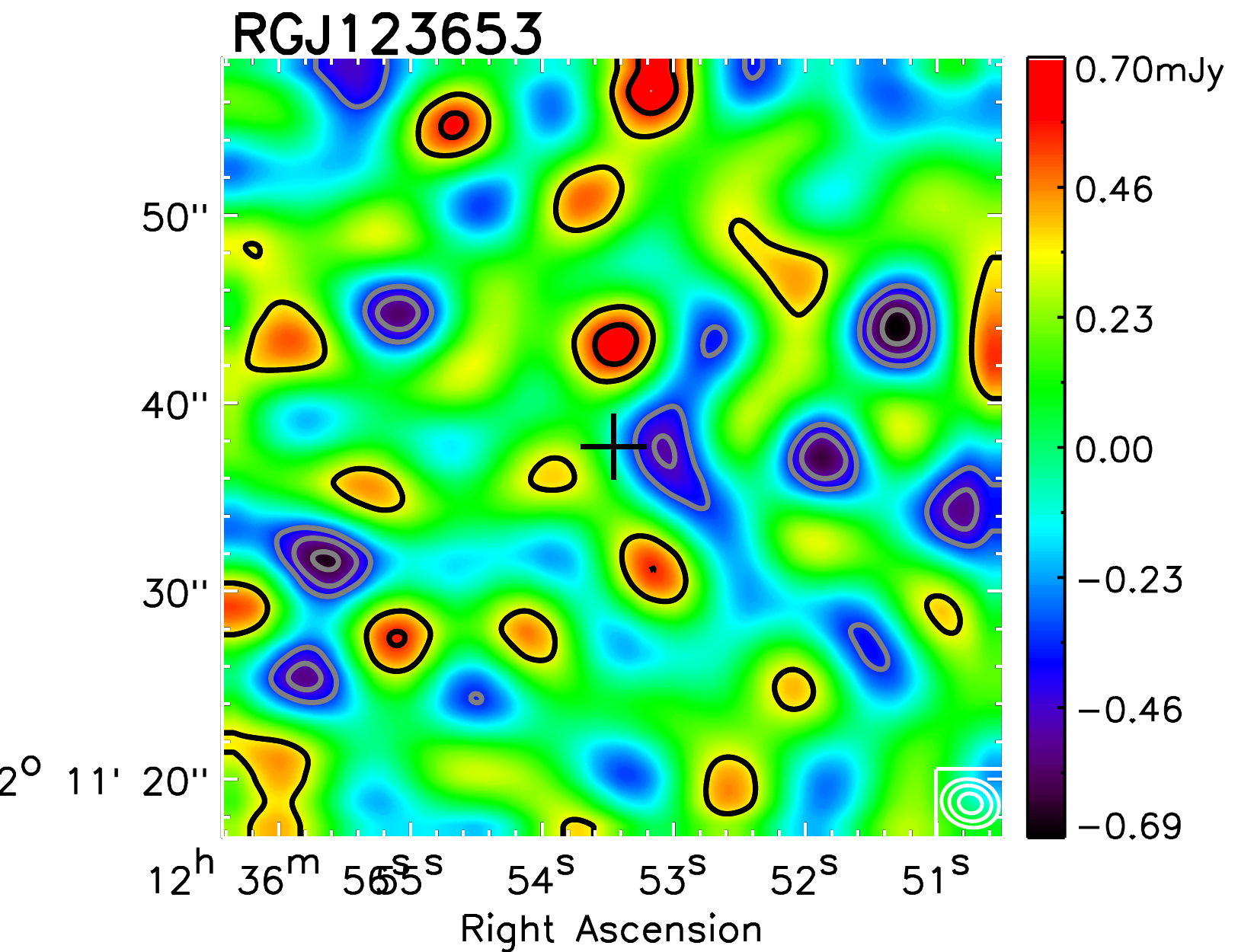}\hspace{0.1in}
  \includegraphics[width=0.67\columnwidth]{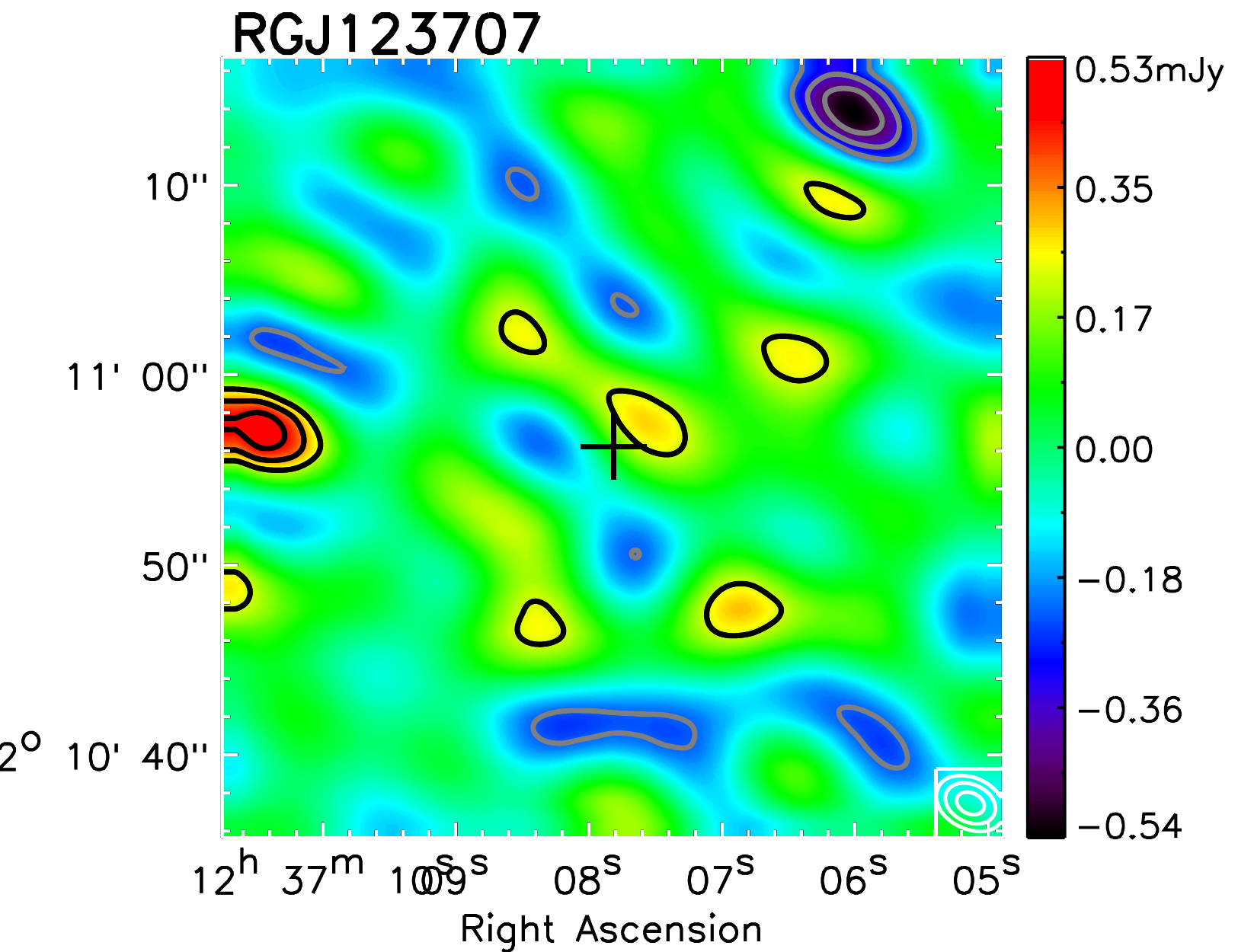}\\
  \includegraphics[width=0.67\columnwidth]{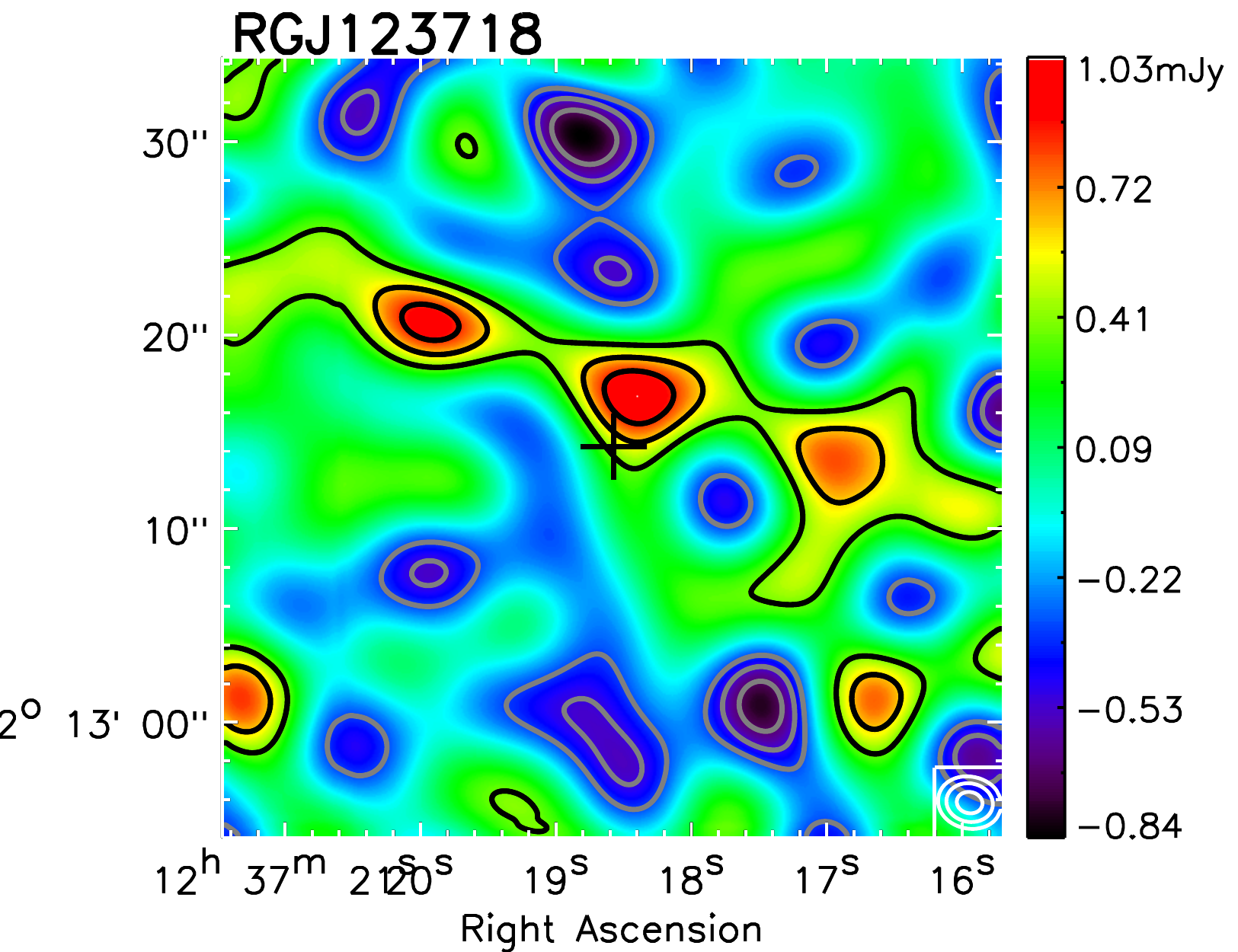}\hspace{0.1in}
  \includegraphics[width=0.67\columnwidth]{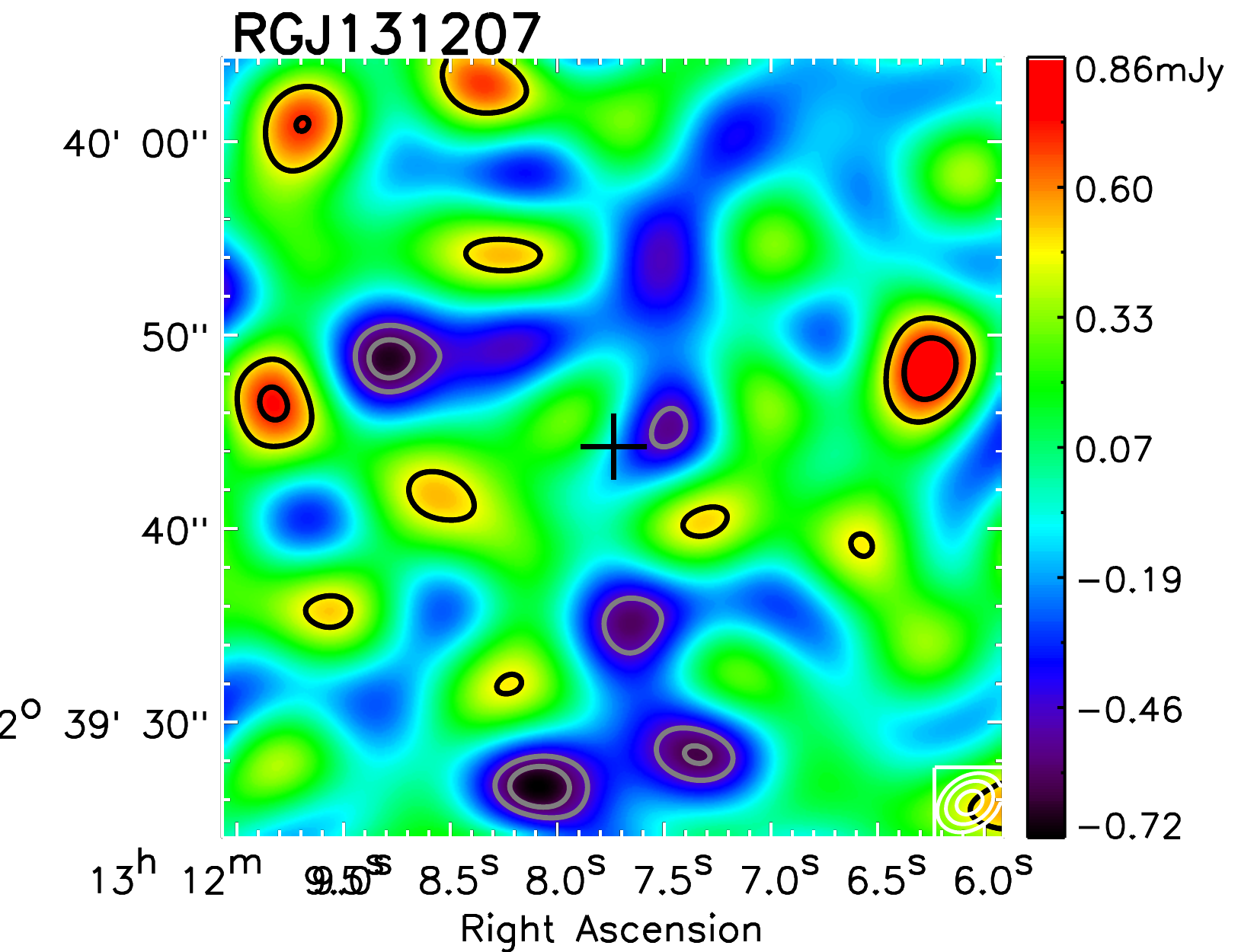}\hspace{0.1in}
  \includegraphics[width=0.67\columnwidth]{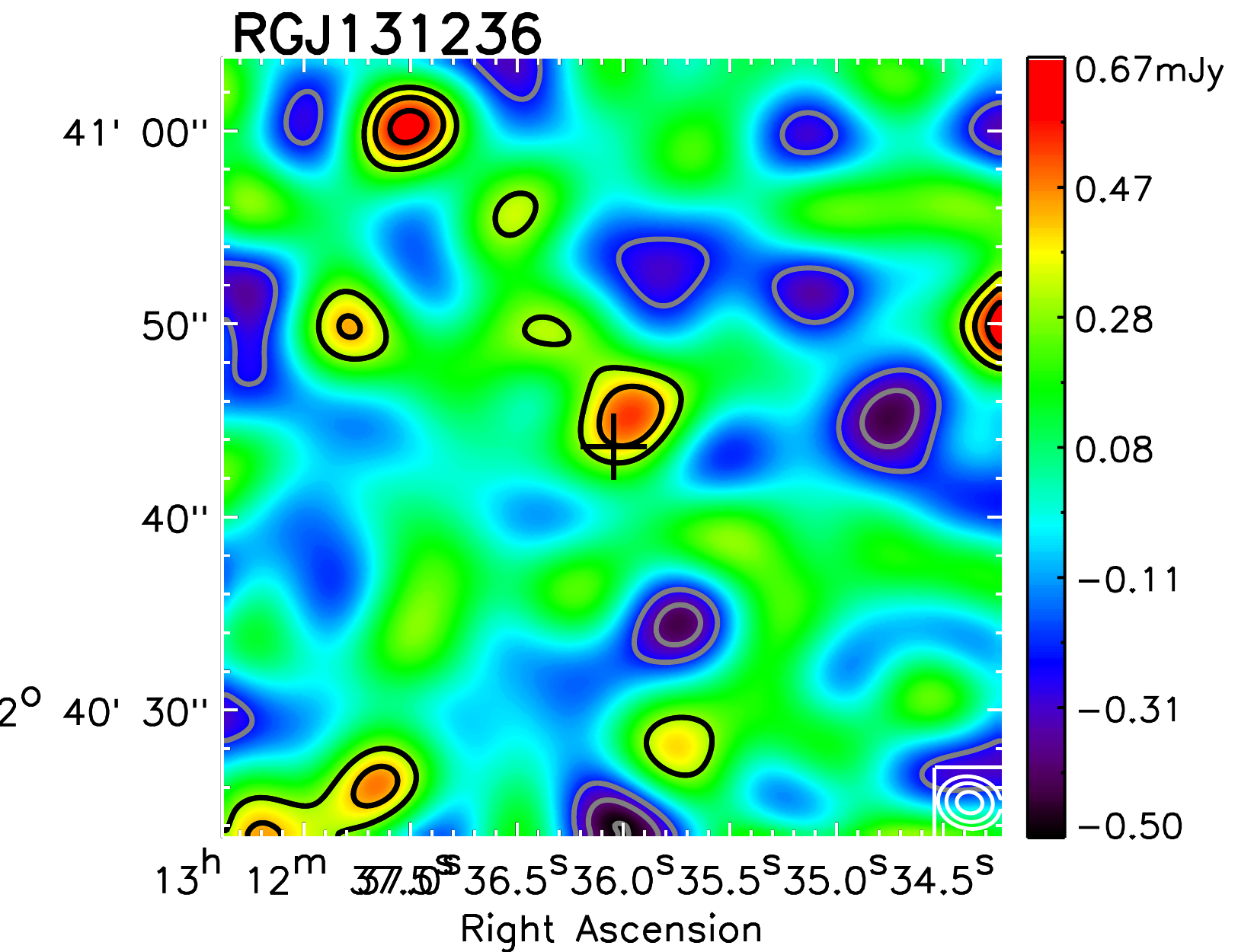}
\caption{CO maps of SFRG sample in
  40\arcsec\,$\times$\,40\arcsec\ cutouts, sorted first by detected CO
  (the first eight are detected), then by RA.  For those galaxies
  detected in CO, the maps are integrated over the optimal velocity
  channels corresponding to the detected CO line.  Black contours are
  integer multiples of the RMS starting at 2$\sigma$ and thin gray
  contours are negative integer multiples of RMS starting at
  -2$\sigma$.  The relative flux scales on the maps are indicated by
  the color-bars to the right.  The spectra in Figure \ref{fig:cospec}
  are extracted within one beamsize (white contours in the lower
  right, outer contour is FWHM) centred on the highest S/N point in
  the map (large cross).  The three sources at top are labelled
  ``Offset'' since their CO peak is $\sim$3-5\arcsec\ offset from
  phase center, a greater offset than would be caused by positional
  inaccuracies in the instrument.  For galaxies not detected in CO,
  the maps are integrated over all velocity channels and spectra are
  extracted from the map center (small cross).  }
  \label{fig:comaps}
\end{figure*}

\begin{figure*}
\centering
  \includegraphics[width=0.68\columnwidth]{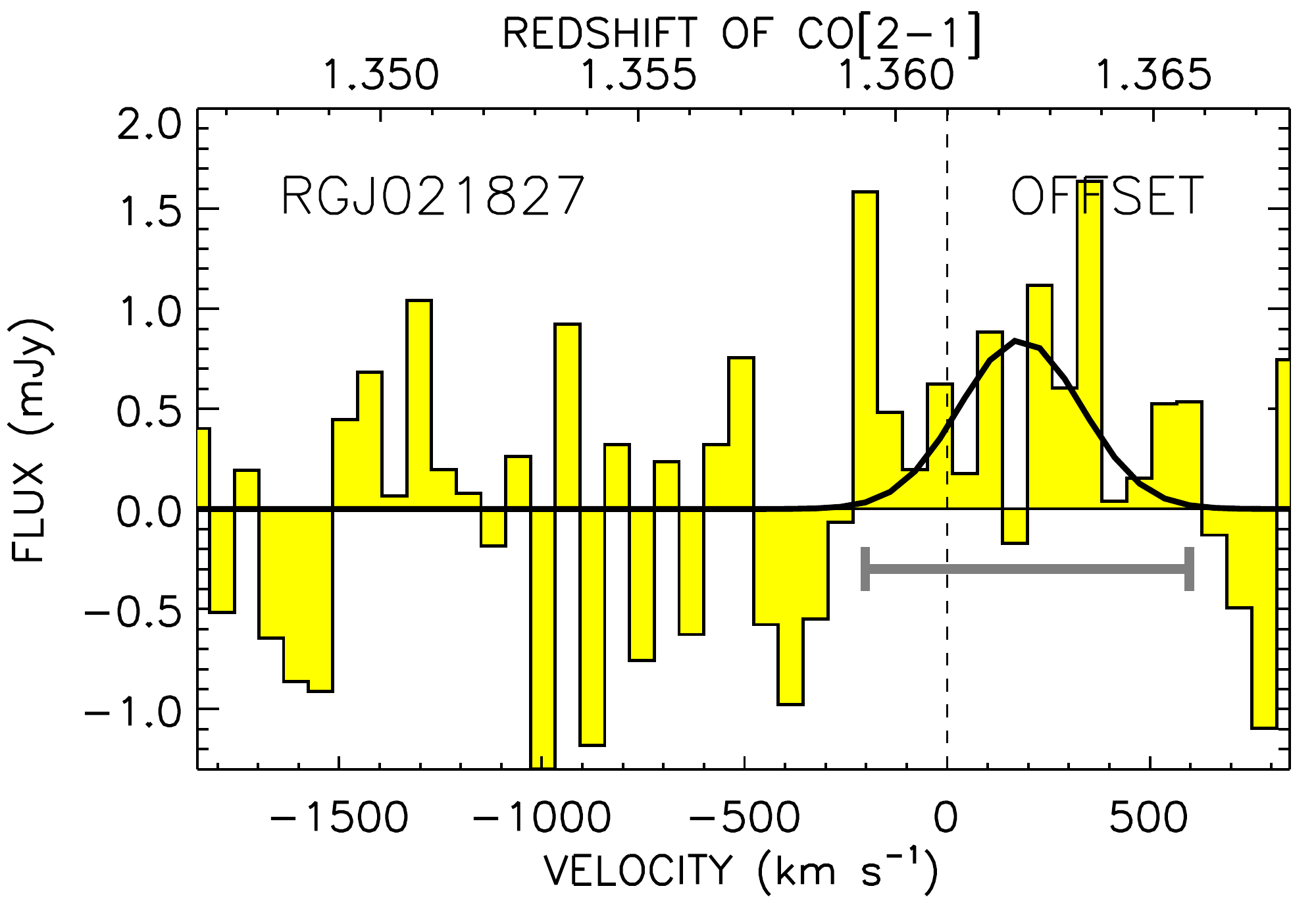}
  \includegraphics[width=0.68\columnwidth]{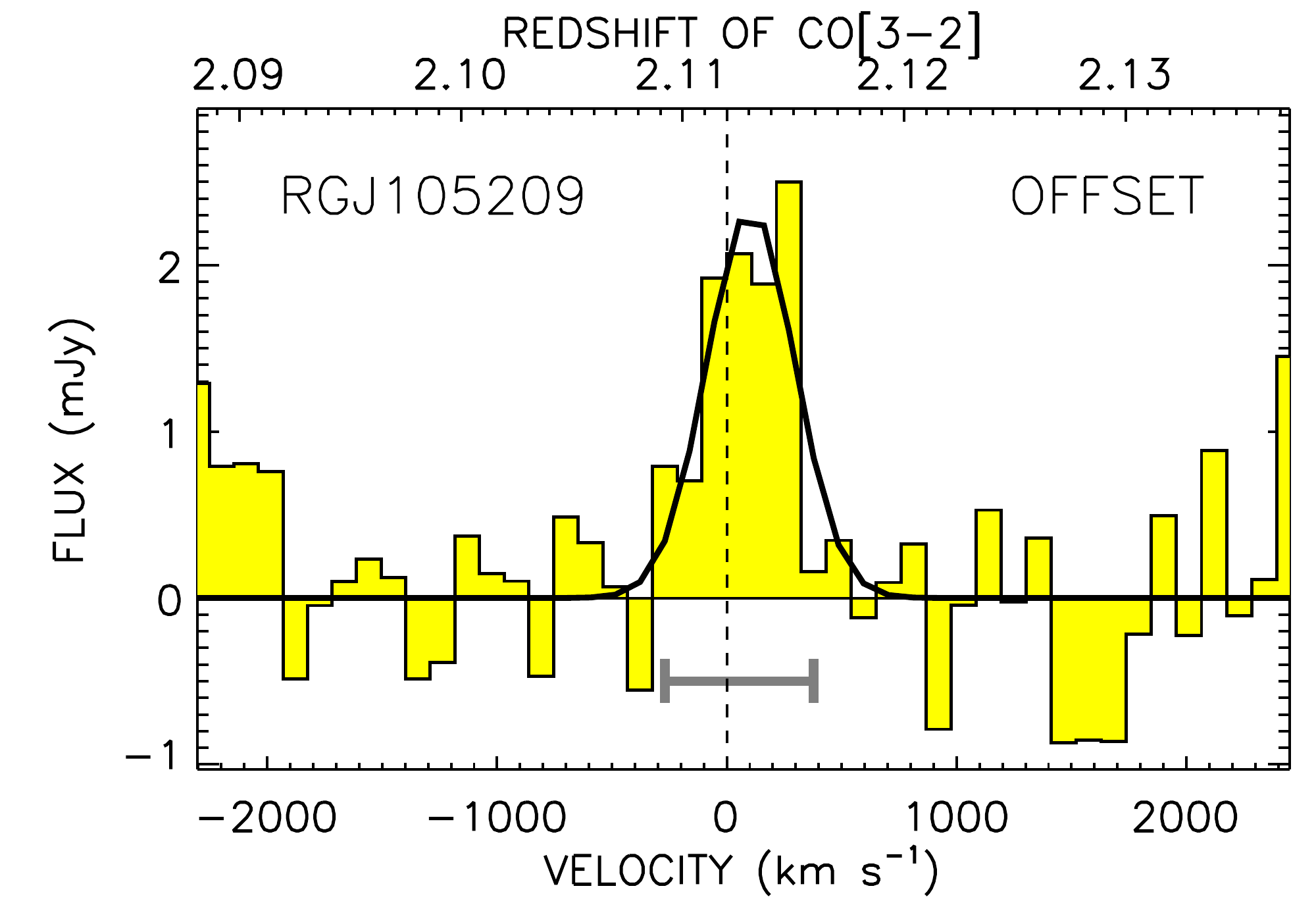}
  \includegraphics[width=0.68\columnwidth]{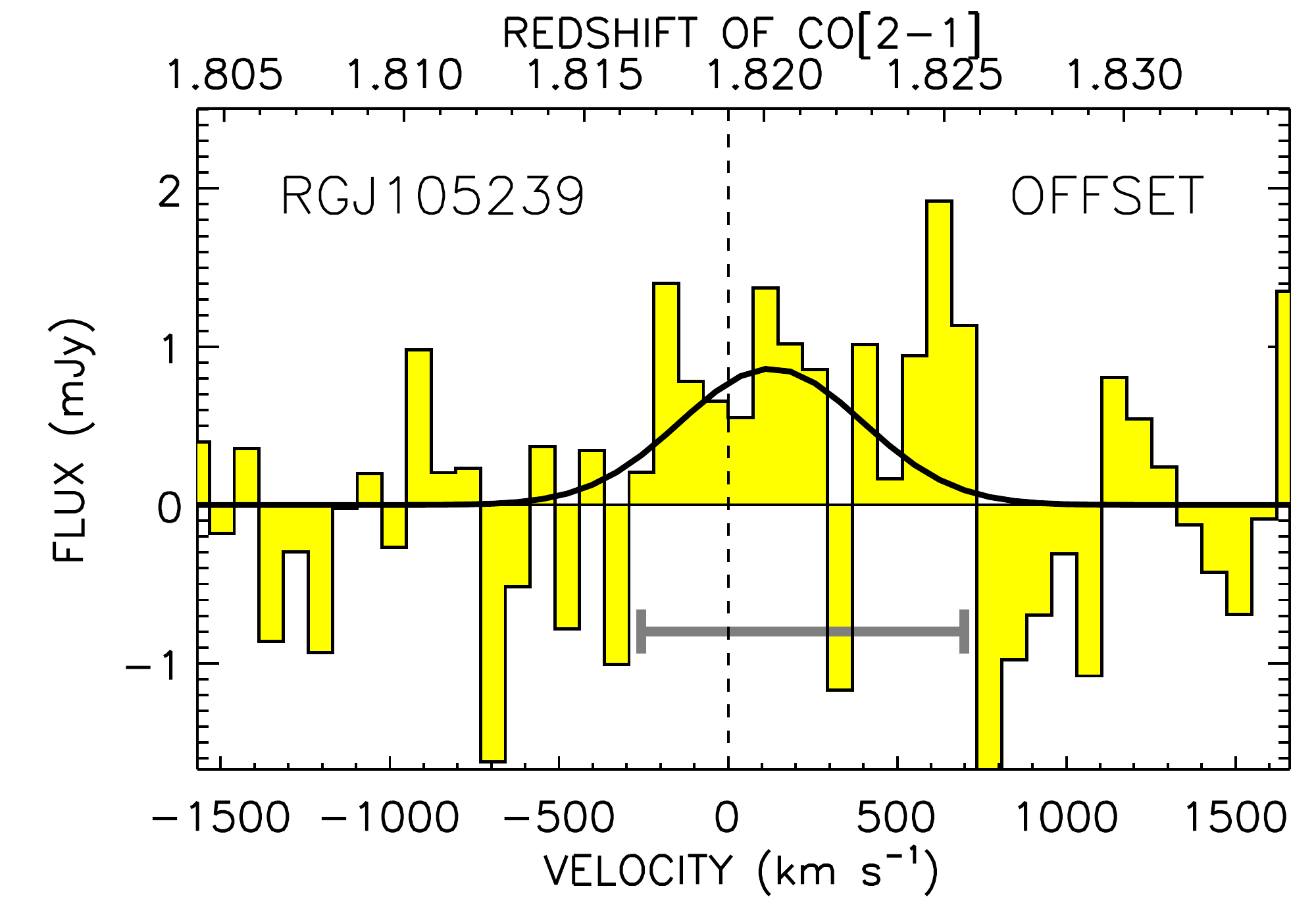}\\
  \includegraphics[width=0.68\columnwidth]{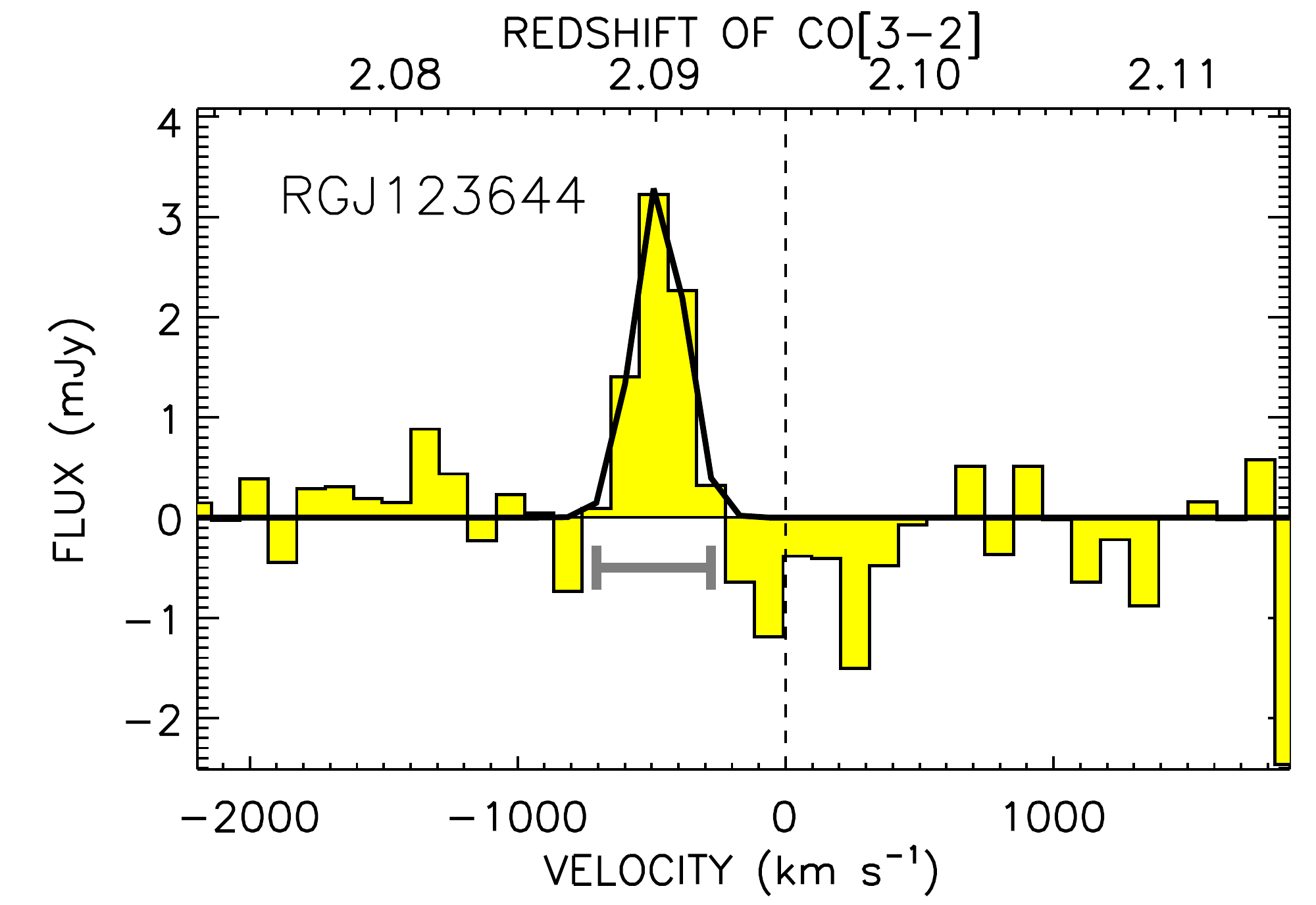}
  \includegraphics[width=0.68\columnwidth]{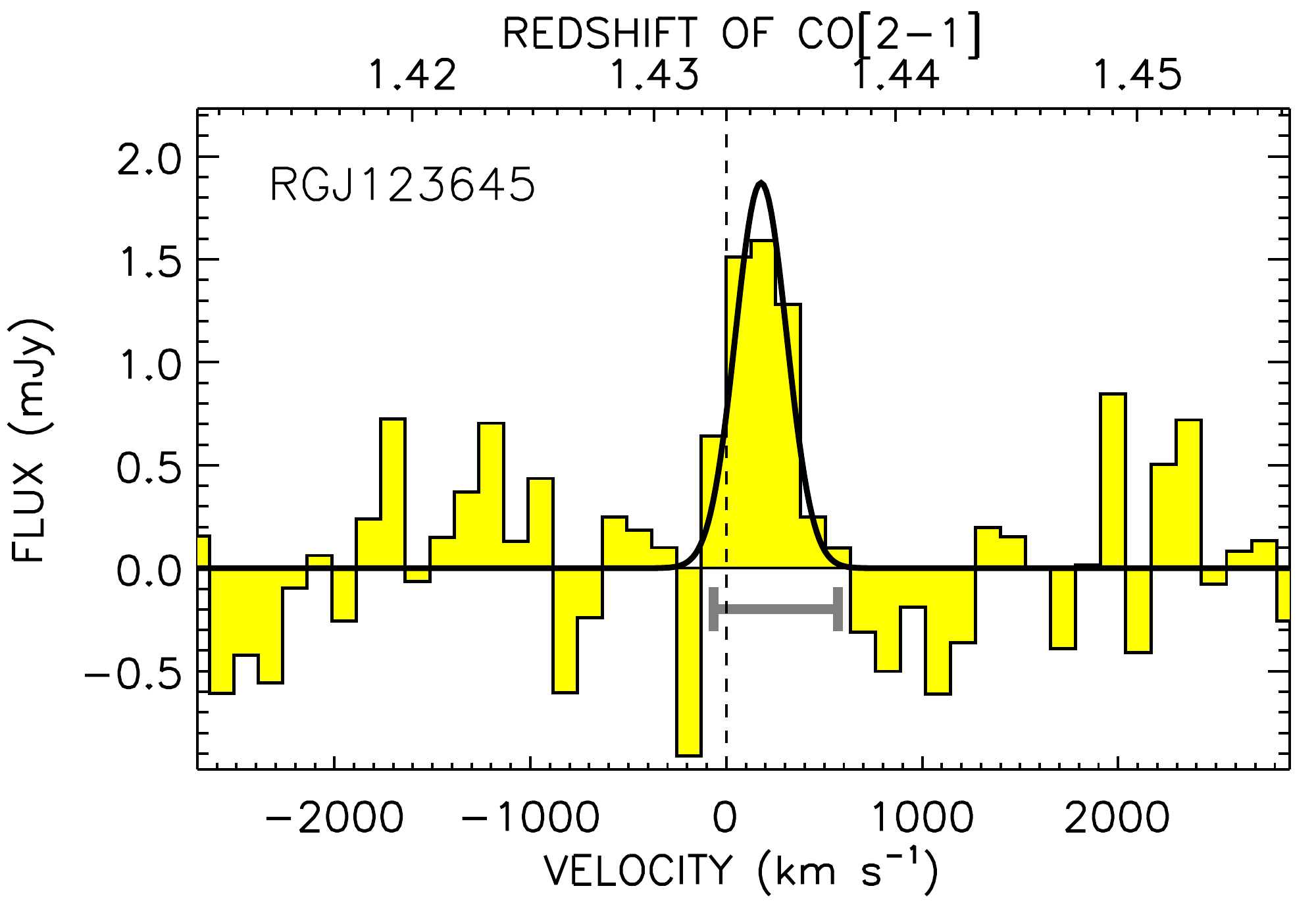}
  \includegraphics[width=0.68\columnwidth]{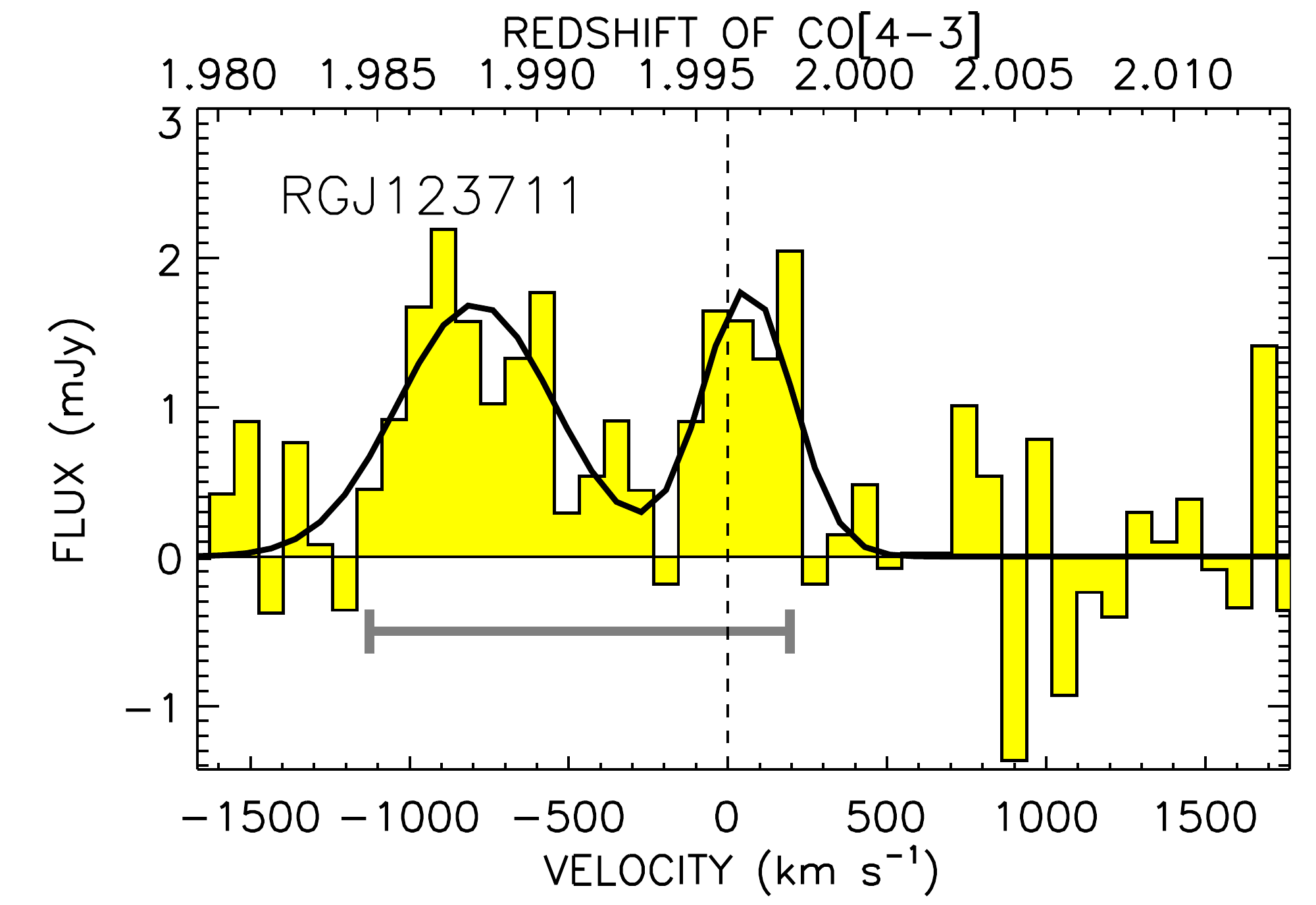}\\
  \includegraphics[width=0.68\columnwidth]{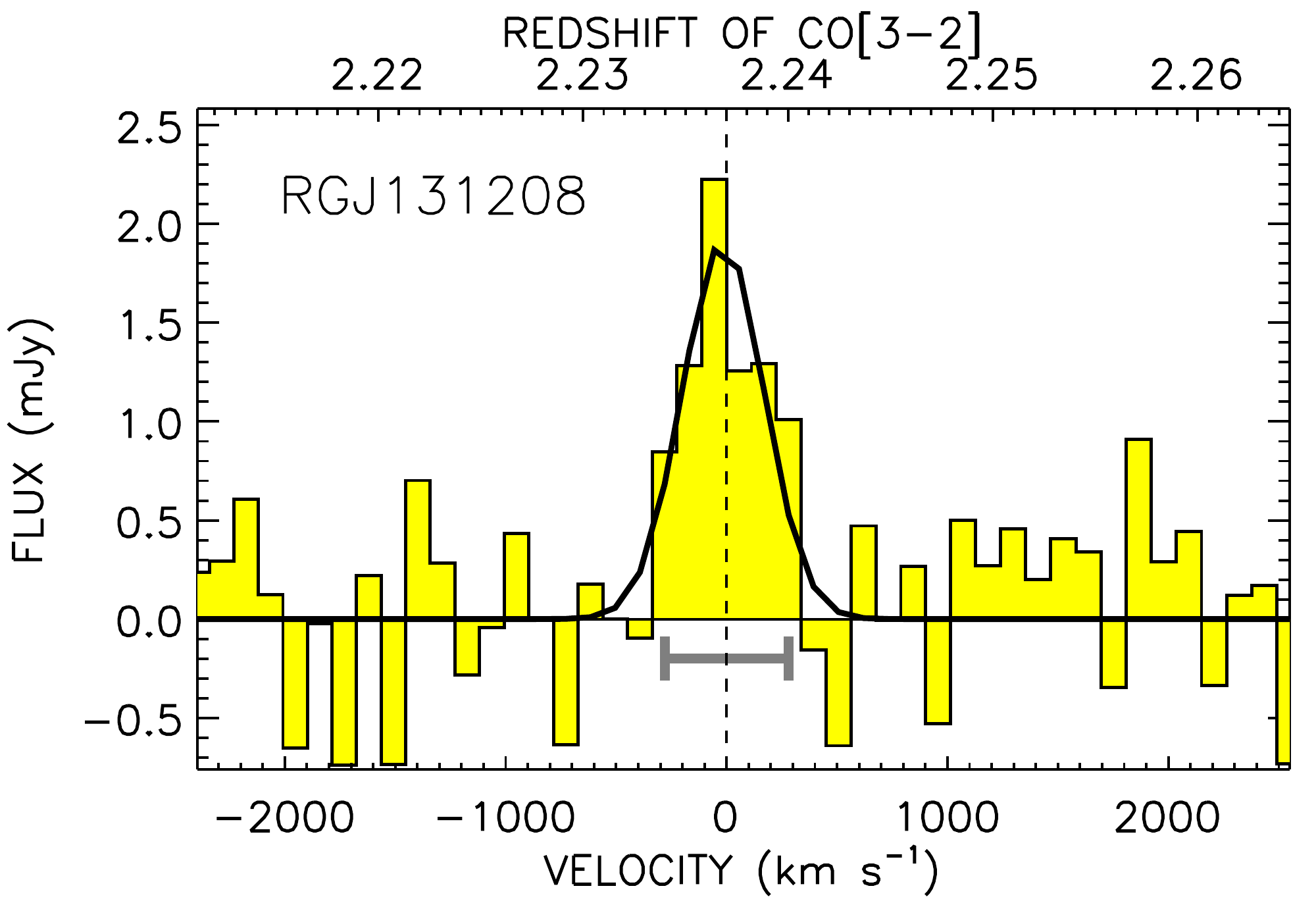}
  \includegraphics[width=0.68\columnwidth]{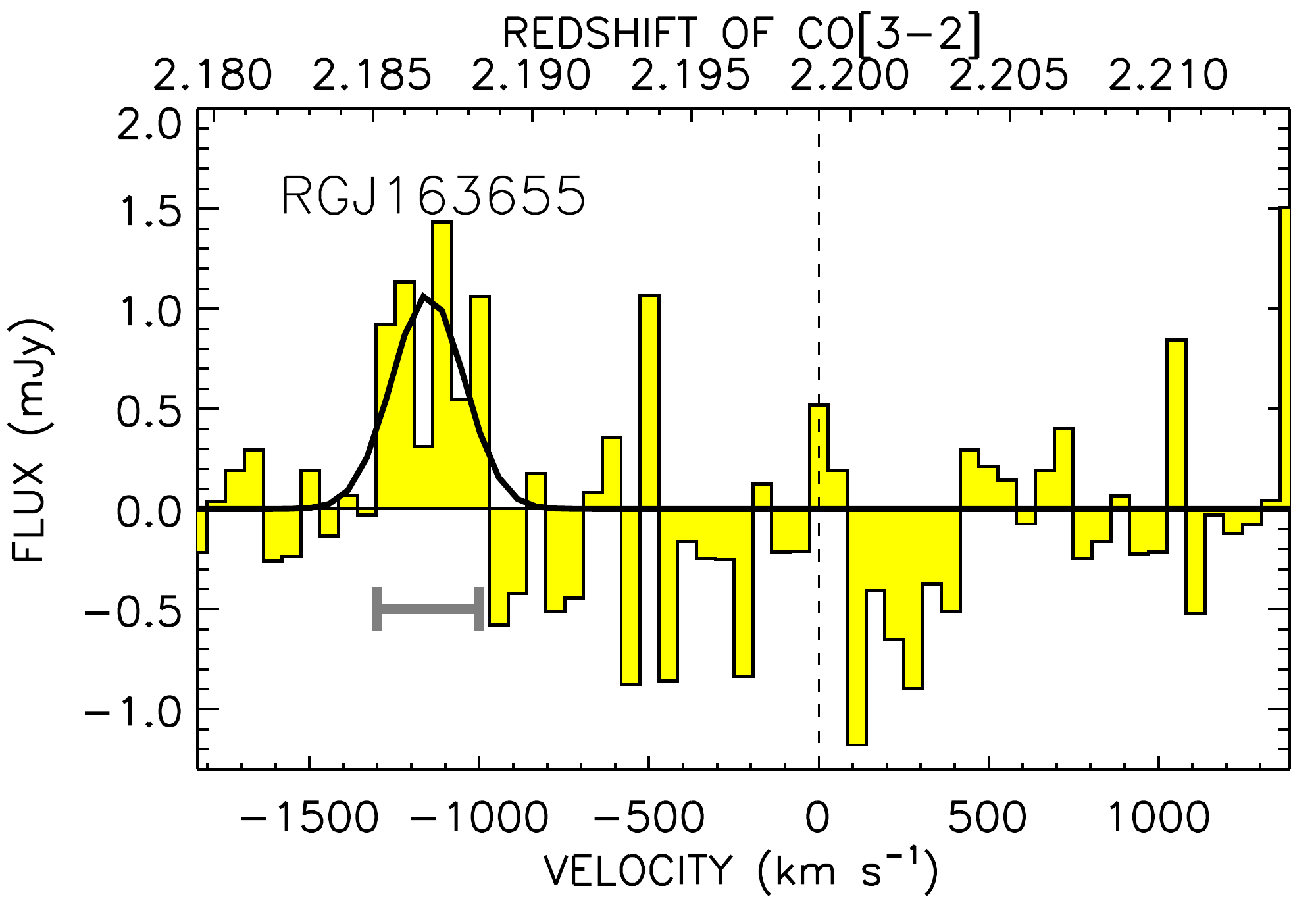}\\
  \includegraphics[width=0.68\columnwidth]{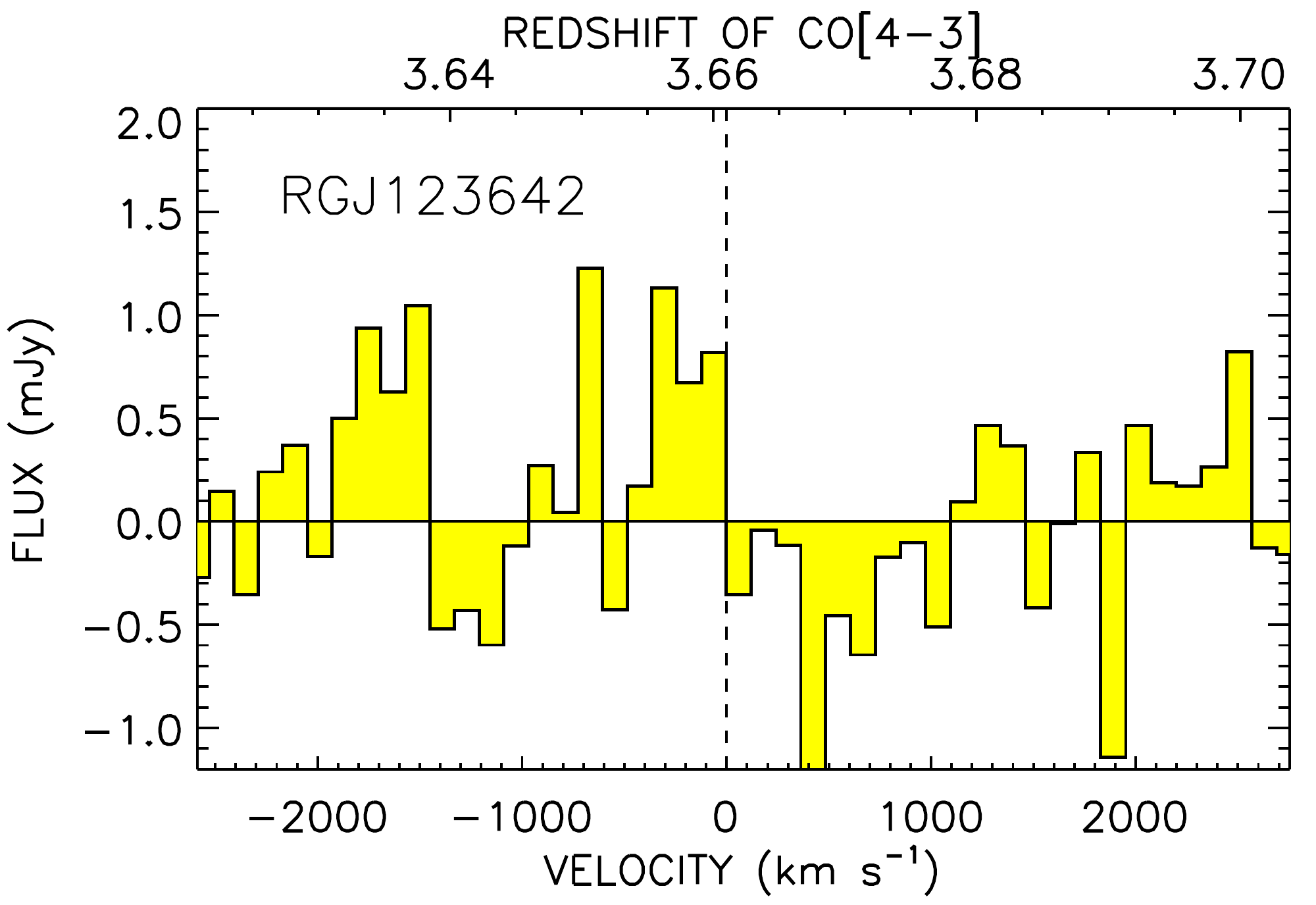}
  \includegraphics[width=0.68\columnwidth]{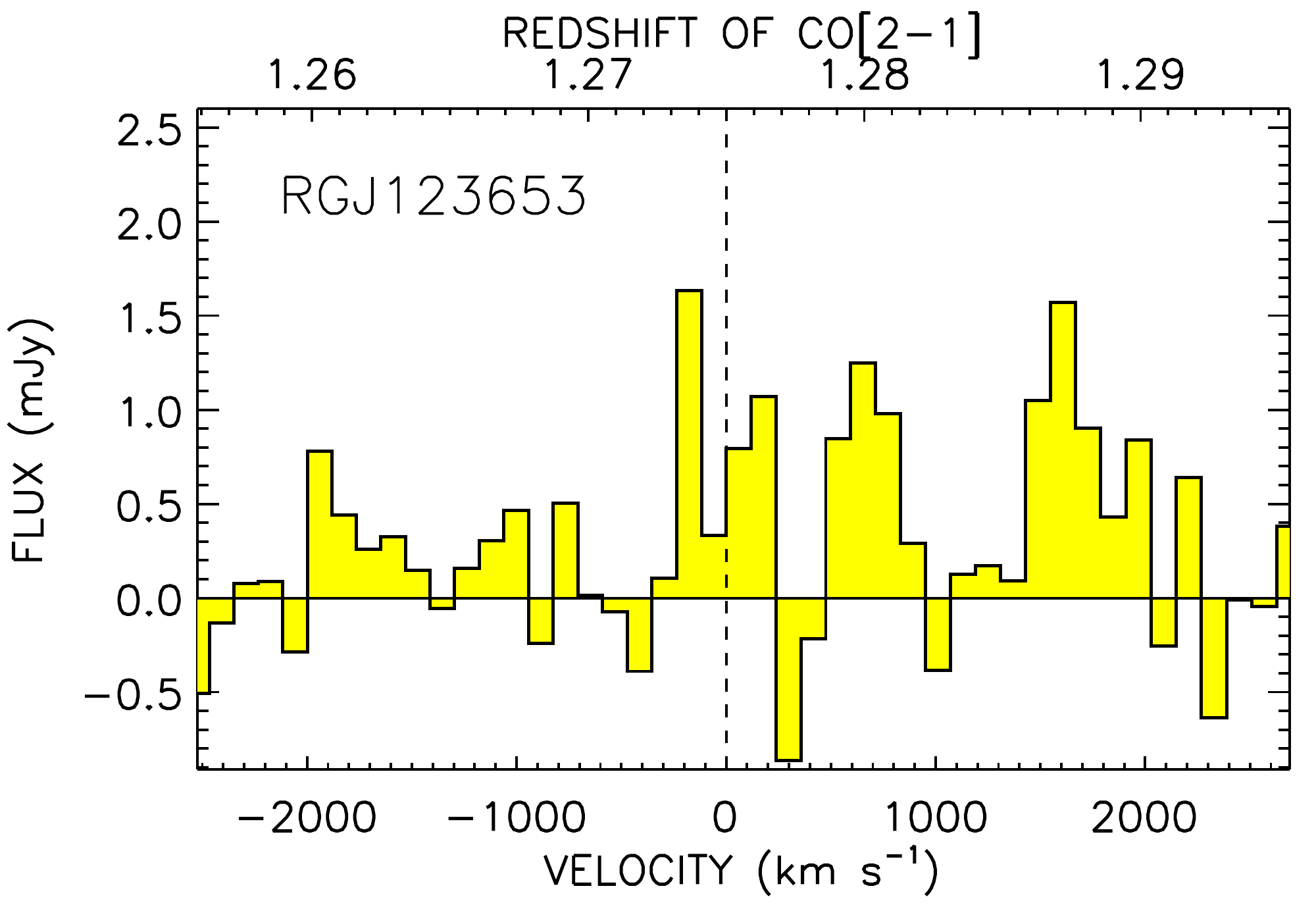}
  \includegraphics[width=0.68\columnwidth]{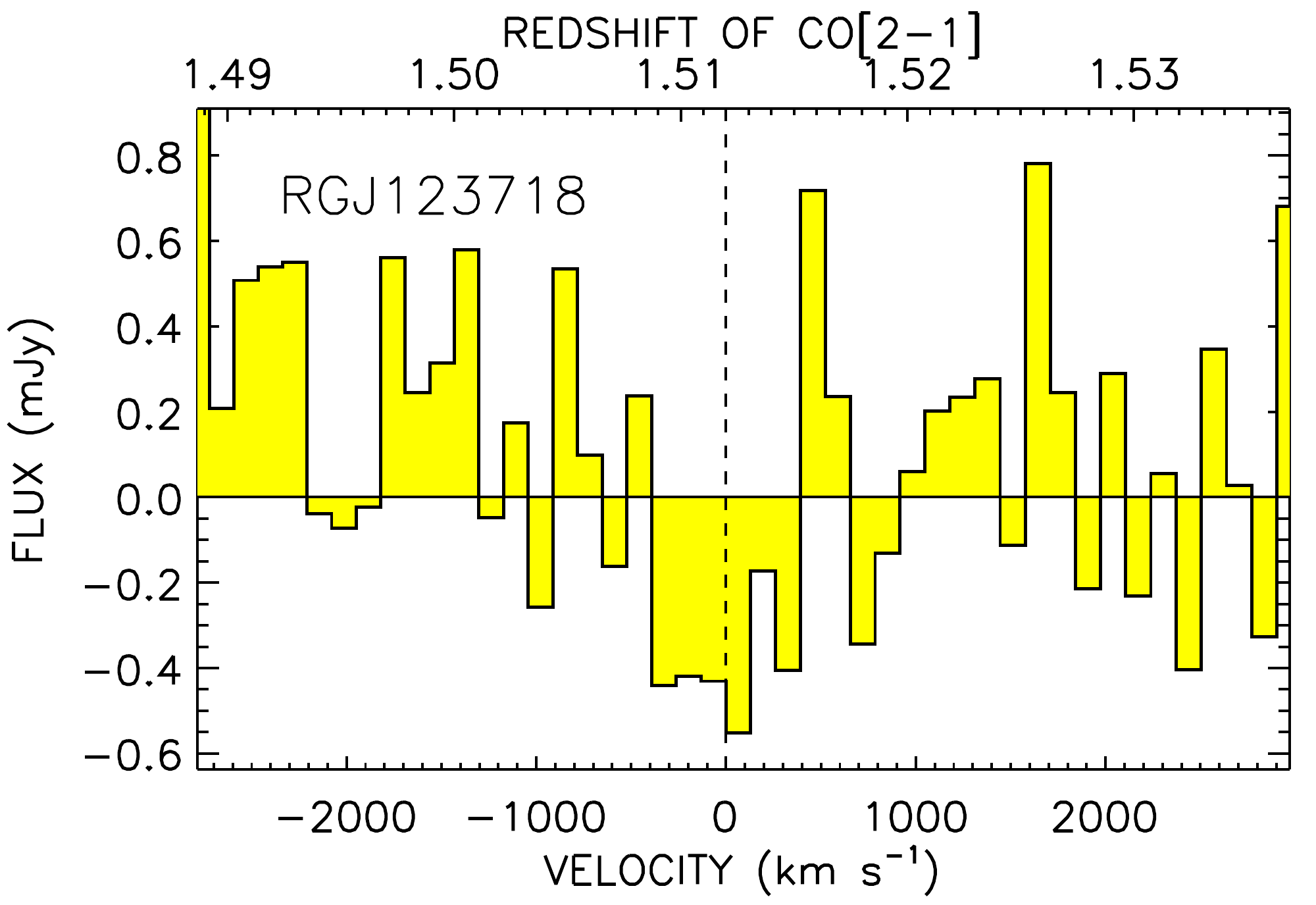}\\
  \includegraphics[width=0.68\columnwidth]{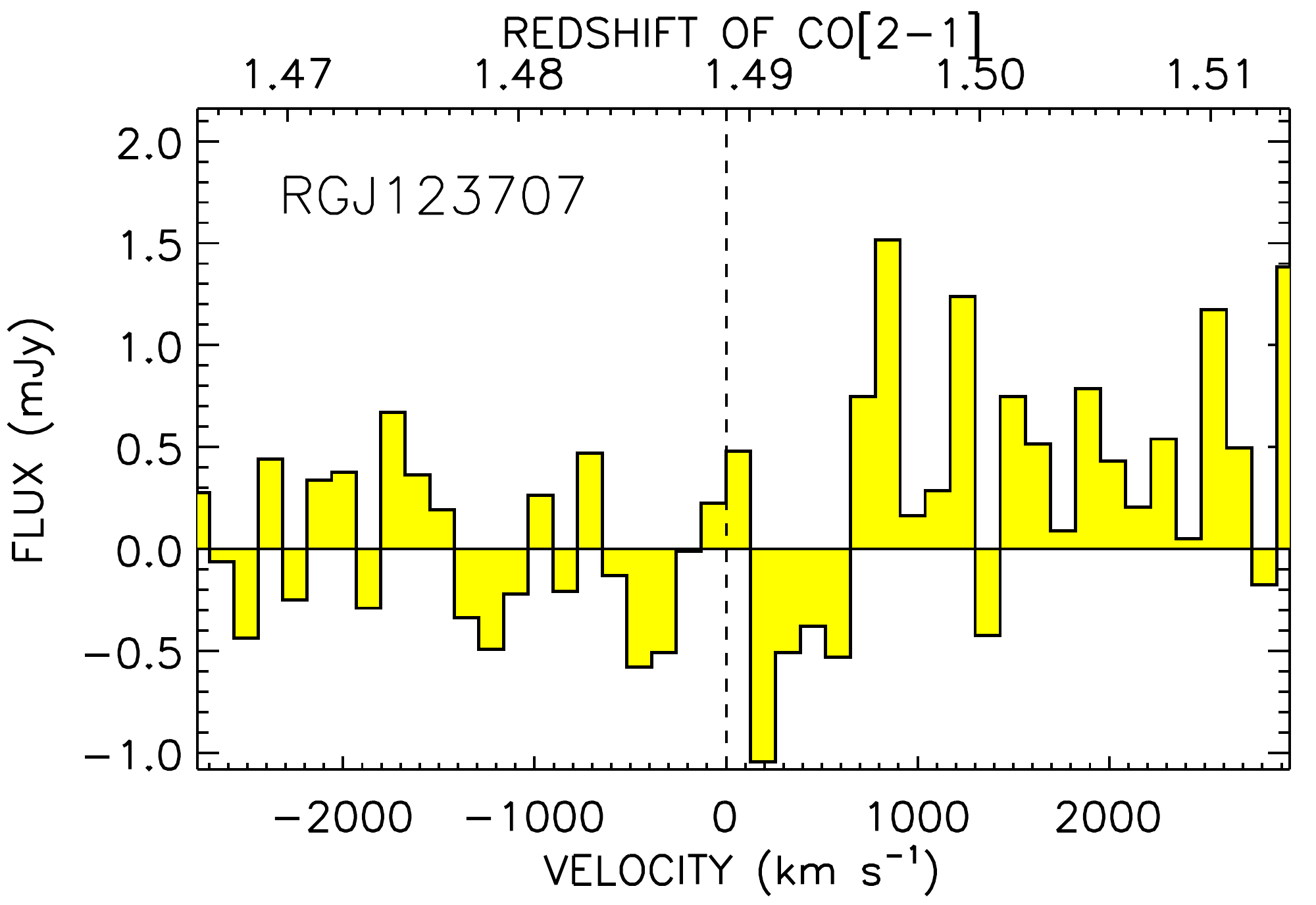}
  \includegraphics[width=0.68\columnwidth]{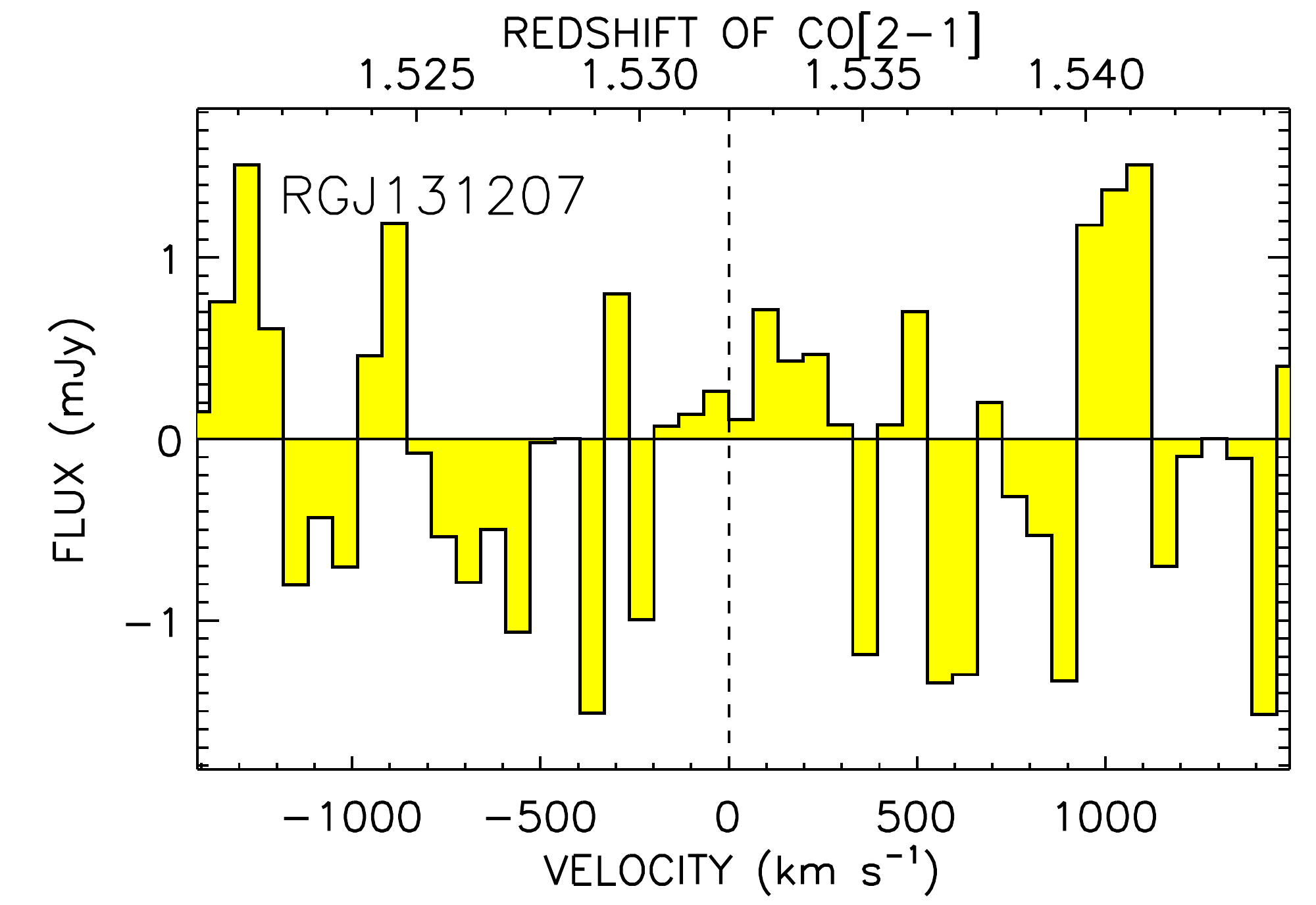}
  \includegraphics[width=0.68\columnwidth]{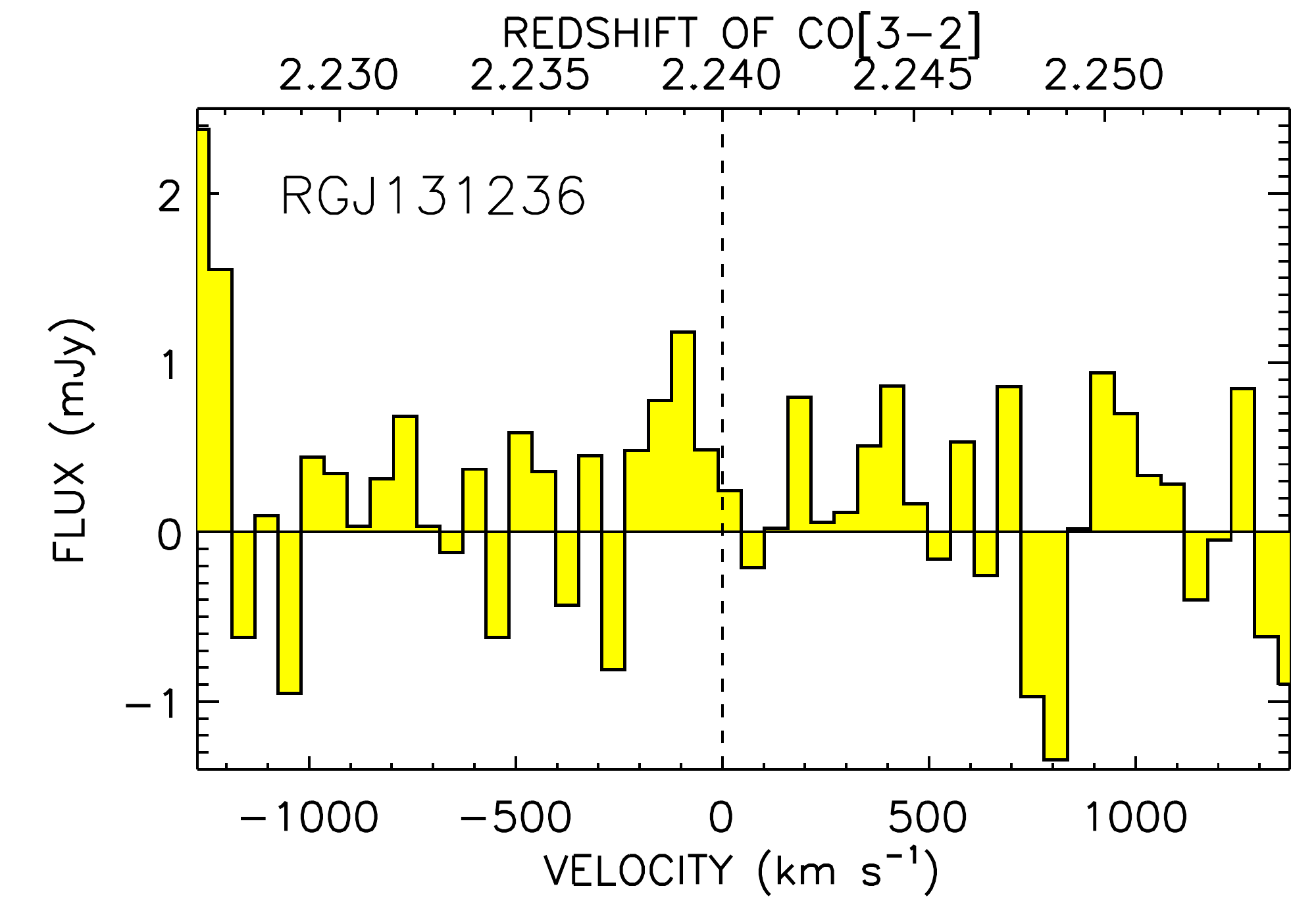}\\
\caption{ The CO spectra for SFRGs.  Spectra are extracted within one
  beamsize centred at the highest S/N peak for detected sources and
  at phase center for non-detections (i.e. the galaxies' radio
  positions).  Vertical dashed lines indicate the galaxies optically
  derived redshifts.  Gray brackets highlight the channel range
  integrated over to produce the maps in Fig~\ref{fig:comaps}.  The
  first three sources labeled ``Offset'' are detected at large offset
  positions $\sim$3-5\arcsec\ from phase center.  At bottom right, we
  show the spectrum for \chan, the SFRG observed in \citet{chapman08a}
  at the redshift 2.240; its revised redshift, 2.224, is at the edge
  of the observed bandwidth.}
\label{fig:cospec}
\end{figure*}

Table~\ref{tab:co} lists the integrated line fluxes and limits for all
SFRGs in our sample.  Out of the twelve sources in our observing
program, seven SFRGs were detected at $\simgt$4$\sigma$.  Including
the literature SFRGs, 10/16 are CO detected, and one `undetected
source' has insufficient data to determine detection (\chan, with an
ambiguous redshift).  Three of our detected sources have significant
positional offsets from their radio positions, thus we have classified
these as `tentative' detections in Table~\ref{tab:co} and excluded
them from calculations in our analysis.  The detection fraction of
SFRGs, 10/16 (63\%, including offset sources), is similar to the
detection fraction of SMGs in \citet{greve05a} and \citet{coppin08a}.

While we do not expect continuum flux from any SFRGs, we test for it
based on FIR detection limits. To test the detectability of blackbody
continuum radiation, we interpolate the best fit FIR SEDs (see section
\ref{ss:dustmass}) to estimate the flux density at the frequency of
PdBI observations ($\nu_{\rm obs}\,=\,80-150\,$GHz).  All SFRGs in the
sample have estimated continuum flux densities $<$0.13\,mJy, which is of order
the RMS noise in our CO spectra, thus not detectable. 

Each remaining SFRG spectrum was fit with a three parameter Gaussian
using a least squares fitting algorithm, where the spectral noise is
simulated as a function of the RMS channel noise \citep[see ][ for
  details]{coppin07a}.  Since no detectable continuum emission is
expected we do not fit line profiles with constant baselines.
Figure~\ref{fig:comaps} shows the velocity-averaged spatial maps of
the mm line observations (with VLA centroids marked with crosses) and
Figure~\ref{fig:cospec} shows the extracted spectra of the CO
detections with overlaid Gaussian fits and the original optical
spectroscopic redshifts.  The 2D maps for CO-detected SFRGs are
integrated over the channel range which produces the highest signal to
noise line profile in the spectrum.  The 2D maps of the undetected
sample are integrated over all velocity channels.  CO detected SFRGs
have CO redshifts which agree with their optical spectroscopic
features.  Every spectrum was also tested against two three-parameter
Gaussian fits (i.e. a double peaked line) in a $\chi$ squared goodness
of fit algorithm.  Only one galaxy's line fits were significantly
improved by fitting to a double peaked Gaussian.  The detection of
\rbbf\ is double peaked with one component centred at z$_{\rm
  \co}$\,=\,1.988 and another at z$_{\rm \co}$\,=\,1.996 which bracket
its optical redshift ($z\,=$\,1.995). The line properties of both
components of \rbbf\ are listed separately in Table~\ref{tab:co}; the
total system is represented by the sum of the two separate gas masses
and sum of dynamical masses (we treat them as separate systems since
the peaks are asymmetrical and more likely indicative of a merger than
rotating disk).  The ratio for double-peaked CO in SFRGs is thus
1/8\,$\approx$\,13\%, comparable to 5/30\,$\approx$\,17\%\ for SMGs.
The weighted mean line width of this sample, excluding `offset'
detections is 320$\pm$80\,\kms (note that it increases to
370$\pm$110\,\kms if the offset sources are detections).

The 2$\sigma$ line intensity limits for the CO-undetected SFRGs are
calculated using the following:
\begin{equation}
I_{\rm CO} < 2\, {\rm RMS}_{\rm ch}\, (\Delta V_{\rm \co}\, dv)^{1/2}
\end{equation}
where ${\rm RMS}_{\rm ch}$ is the channel noise from
Table~\ref{tab:observations}, and $dv$ is the spectral resolution, the
binsize value given in Table~\ref{tab:observations} converted to
\kms\ \citep[as in][]{boselli02a}.  These limits are given in
Table~\ref{tab:co}.  The spatial maps for the undetected SFRGs (in
Figure~\ref{fig:comaps}) are integrated over the entire bandwidth of
observations.  None of the maps are corrected for primary beam
attenuation; given the 40$\times$40\arcsec\ size, the PBA correction
is $<$2 even for all peripheral areas of the map.

The spatial map for \chap\ shows another bright companion CO emitter
$\sim$8\arcsec\ to the northwest of \chap\ not previously identified
in the reduction of Chapman \etal\ 2008.  Extraction of its spectrum
reveals a line centred on the same frequency as the \chap\ \cco\ line,
however there are no radio sources or spectroscopically identified
sources nearby.  If followup CO observations of \chap\ reveal the same
feature in different CO transitions, then high resolution CO will be
needed to precisely measure the companion galaxy's position, and thus
its multi-wavelength properties.

A lack of multiple CO transition data has meant that many past studies
have relied on the assumption of constant brightness temperature to
convert between higher transition line luminosities and L$^\prime_{\rm
  \co(1-0)}$.  However, recent observations of multiple
\co\ transitions in high redshift galaxies has shown that the
transitions are often not in thermal equilibrium
\citep{dannerbauer09a}.  The three $z\sim2.5$ SMGs of \citet{weiss07a}
are consistent with being in thermal equilibrium up to a J=3
transition yet not towards higher-J values, while there are several
other high redshift sources which are much more excited, for example,
FSC\,10214+4724 \citep{scoville95a} and APM\,08279+5255
\citep{weiss07b}
which have nearly constant
brightness temperatures out to J=6) or much less excited (ERO
J16450+4626 turns over at CO(3-2)).  In this paper we adopt the
brightness temperature conversions inferred from the spectral line
energy distributions for three SMGs observed in \citet{weiss07a};
explicitly, we assume the following line flux ratios derived from LVG
models: S$_{\rm CO21}$/S$_{\rm CO10}$\,=\,3.0$\pm$0.1, S$_{\rm
  CO32}$/S$_{\rm CO10}$\,=\,6.8$\pm$0.5 and S$_{\rm CO43}$/S$_{\rm
  CO10}$\,=\,10.0$\pm$0.8.  One SFRG has multiple line data (\rbbf)
and can be analyzed for its excitation.  This work is presented in
section \ref{ss:254excitation}.

More recent work \citep{ivison10c,danielson10a} on \aco\ observations
in SMGs show that the assumption of a constant brightness temperature
can lead to underestimations of gas mass by factors of
$\sim$2$\times$.  They find that the ratio of luminosities between
\cco\ and \aco\ is $r_{3-2/1-0}\,=\,$0.55 and $\sim$0.67 respectively.
Our excitation assumption implies a ratio $r_{3-2/1-0}\,\sim\,$0.78,
slightly higher than these recent studies, but well within the
uncertainties of the $S_{\cco}$ measurements of Ivison \etal\ and
Danielson \etal.  The derived line luminosities for our observations
are listed in Table~\ref{tab:co}, in their observed transition as well
as in our conversion to L$^\prime_{\co(1-0)}$.

\subsection{CO Excitation of \rbbf}\label{ss:254excitation}

\begin{figure*}
\centering
\includegraphics[width=0.89\columnwidth]{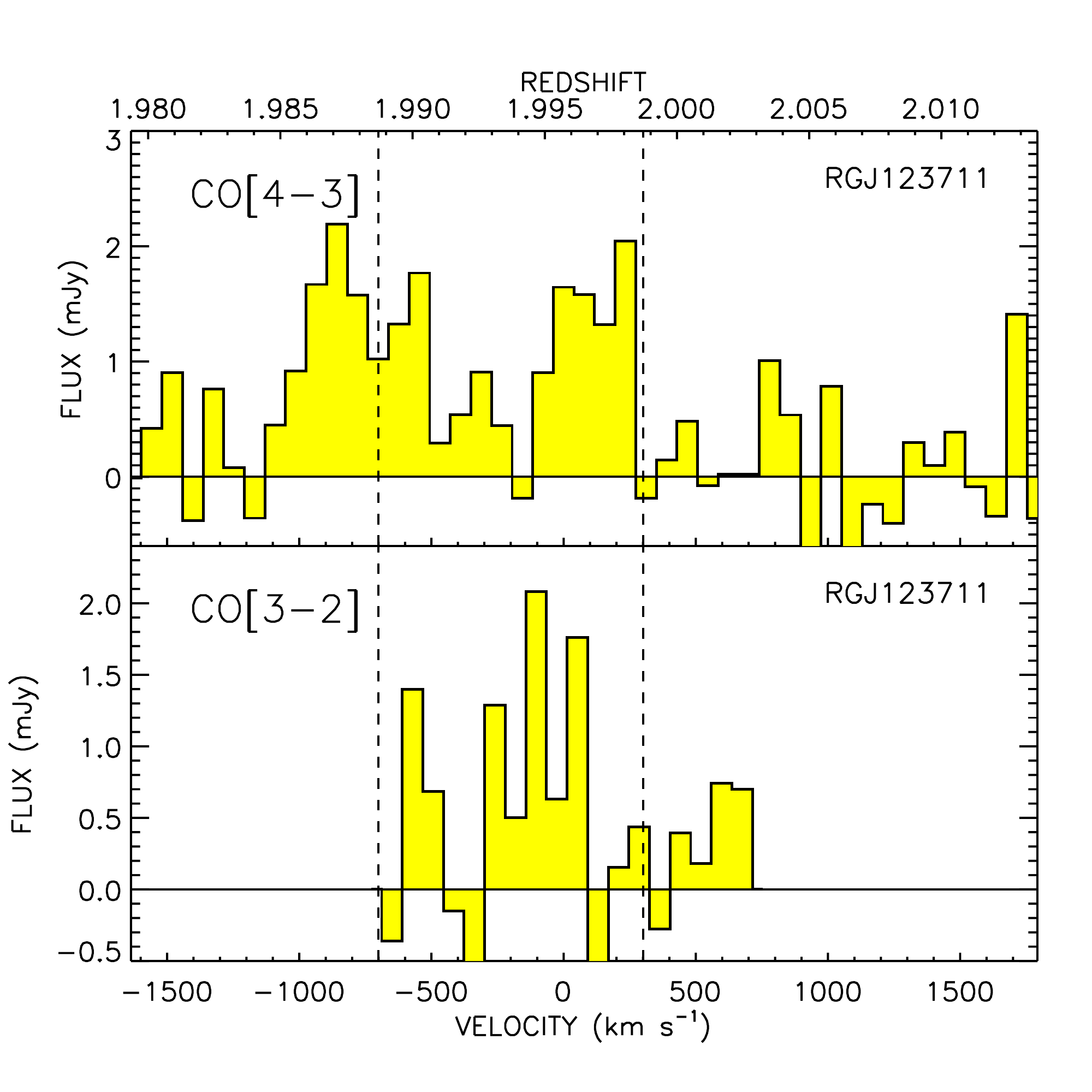}
\includegraphics[width=0.99\columnwidth]{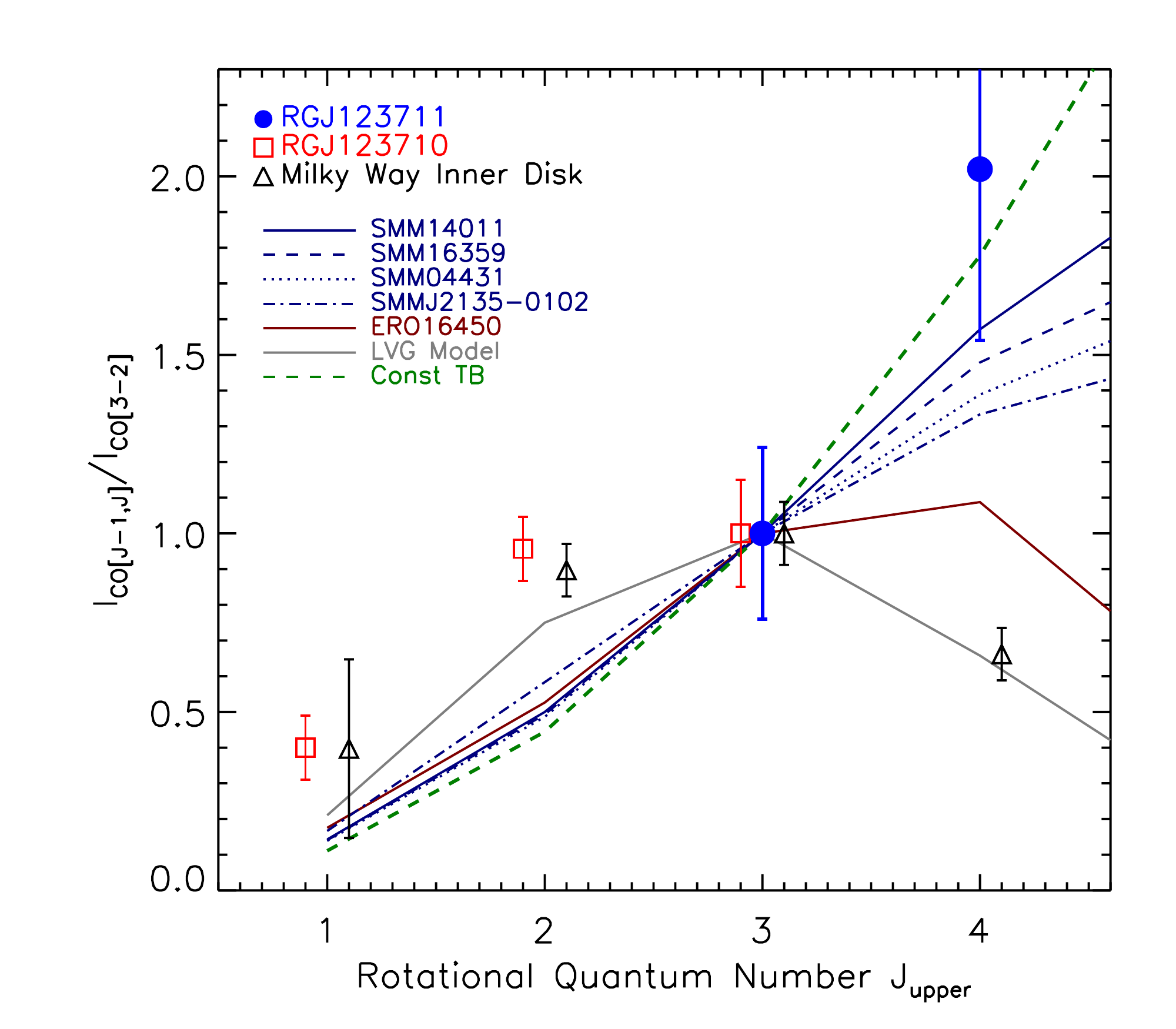}
\caption{ {\it Left:} Both \cco\ and \dco\ observed transitions for
  \rbbf.  Determined from the spectral lines seen in the higher S/N
  CO[4-3] observation, the velocity range of the emitting region which
  overlaps in each data set is marked by two vertical dashed lines at
  -700\,\kms\ and 300\,\kms.  {\it Right:} the CO SLED normalised to
  I$_{CO[3-2]}$ for \rbbf\ overlaid with literature results from other
  high-redshift sources \citep{weiss07a,danielson10a} and the Milky Way
  \citep{fixsen99a}.  One of these sources is \dadb, an SFRG observed
  by \citet{dannerbauer09a}.  We also include LVG model no. 1 from
  Dannerbauer \etal, representing a low-excitation Milky Way type
  source.  By normalising to I$_{CO[3-2]}$, we illustrate that it is
  difficult to draw conclusions on the nature of the CO SLED without
  observing at least three CO transitions, particularly in the region
  where the SLEDs are suspected of turning over near CO[4-3]. }
\label{fig:co254both}
\end{figure*}

The observations of \rbbf\ in \cco\ and \dco\ allow an analysis of the
CO excitation in the system.  Figure~\ref{fig:co254both} shows the
spectrum of both lines and the proposed CO spectral line energy
distribution (SLED) for the source; since the \cco\ observations do
not cover the entire velocity range of the emitting gas, a direct
comparison of the line strengths can only be done in the channels
where data overlap (between -700 and 300\kms) marked by vertical
dashed lines in Figure~\ref{fig:co254both}.  The ratio of fluxes is
then S$_{CO[4-3]}$/S$_{CO[3-2]}$\,=\,1.6$\pm$0.4. We use Monte Carlo
testing to estimate the added uncertainty introduced by not including
the missing portion of \cco\ (from -1150 to
-700\,\kms) and we remeasure this ratio to be
S$_{CO[4-3]}$/S$_{CO[3-2]}$\,=\,2.0$\pm$0.8.  This agrees with
the mean [4--3]/[3--2] ratio observed for SMGs, 1.4$\pm$0.2
\citep{weiss07a} and the expected ratio if constant brightness
temperature is assumed, 1.8, however it appears to be inconsistent
with the ratio for the Milky Way, $\sim$0.7 \citep{fixsen99a}.

We contrast this result with those of \citet{weiss07a},
\citet{dannerbauer09a}, and \citet{danielson10a} in
the right panel of Fig.~\ref{fig:co254both}.  Although we have taken
measurements for $J=$3 and 4 only and other work is restricted to
$J\le$3 transitions, we see that \dada\ and \rbbf, the two SFRGs
included here, seem to exhibit different excitation levels where
\dada\ is more consistent with Milky Way type excitation and \rbbf\ is
consistent with the \citet{weiss07a} SMGs.  The issue of gas
excitation is essential to the physical interpretation of high-$z$
star formers, and more/multiple line transitions are necessary for a
full analysis of the population \citep[e.g. see recent work on low CO
  transitions, like
  \aco\ in][]{riechers06a,hainline06a,ivison10c,carilli10a}.

\subsection{Gas Mass and Dynamical Mass}

To derive \hh\ masses we assume the ULIRG \hh/\co\ gas conversion
factor $X\,=\,M_{\rm \hh}/L^\prime_{\rm \co}\,=\,$ 0.8\,M$_\odot$\,
(K\,km\,s$^{-1}$\,pc$^2$)$^{-1}$ from \citet{downes98a}.  It is worth
noting, however, that the CO to \hh\ conversion factor varies
substantially depending on the type of source being sampled, and it
can range from 0.8 in ULIRGs to 4-5 in normal spiral, late-type
galaxies \citep[see discussion in the Appendix of][]{tacconi08a}.  The
basis of our choice of $X\,=\,$0.8\,\msun\,(K\,\kms\,pc$^{2}$)$^{-1}$
is based on the merger, or disturbed nature of ULIRGs and their high
star formation rates.  Mergers are confined by ram pressure which
pervades the whole system, while quiescent discs are confined
gravitationally leading to gas fragmenting into clouds (per the jeans
mass) and thus much higher ratios of \hh\ mass to unit \co.  We note
that higher gas conversion factors are possible and perhaps likely if
SFRGs were to represent an intermediate stage or `less extreme'
population of galaxies than SMGs.  For this reason, we advise that our
\hh-dependent quantities (like $M_{\hh}$, gas fraction, etc) be taken
with a hint of caution, since the full range of galaxy properties and
conversion factors (from $X_{\rm CO}=\,$0.8--4.5) suggest variations of
$M_{\hh}$ of order 0.75\,dex and variations of gas fractions of order
0.6.  We also include gas mass estimates for $X_{\rm
  CO}=\,$4.5\msun\,(K\,km\,s$^{-1}$\,pc$^2$)$^{-1}$ for contrast in
Table~\ref{tab:derived} but proceed with $X_{\rm CO}$=0.8 for our
analysis.  Using X$_{\rm CO}$=0.8 for ULIRGs, the mean gas mass of the
sample is (2.1$\pm$0.7)$\times$10$^{10}$\,\msun, which is roughly
half the mean \hh\ mass of CO-observed SMGs
\citep[5.1$\times$10$^{10}$\,\msun;][]{greve05a,neri03a}.

Dynamical mass is dependent on the galaxy's inclination
angle; we use the average inclination correction of $\langle sin\,i
\rangle\,=$\,1/2 (in other words $i$\,=\,30$^{o}$, corresponding to a
random distribution in galaxy angles between 0 and 90$^{o}$).  Thus
\begin{equation}
M_{\rm dyn}\,sin^{2}\,i\,= \frac{C \sigma^{2}
r}{G}\,=\,4.215\times10^{4}\,C \Delta V_{\rm \co}^{2} r
\end{equation}
where $C\,=$\,4 \citep[the value adopted for mergers;][]{genzel03a}
and $r$ is given in kpc and $\Delta V_{\rm CO}$ in \kms.  We note
however that the difference between assuming a merger ($C=4$),
spheroid ($C=5$), or rotating disk ($C=3.4$) is on the order of the
uncertainty in M$_{\rm dyn}sin^{2}\, i$ caused by uncertainty in line
width.

Deriving a dynamical mass is dependent on a resolved spatial size
measurement ($r$ in Eq 2), which is unfortunately unavailable with
this low resolution, 3-6\arcsec\ beam size (D-configuration PdBI)
data.  In place of measuring CO sizes explicitly, we assume that the
SFRGs of this sample have gas emitting regions roughly the size of
SMGs \citep{tacconi08a} which have $R_{\rm 1/2}=$2$\pm$1\,kpc.  This
is supported by our analysis of high-resolution MERLIN+VLA radio
imaging discussed in the next section (in
section~\ref{sec:merlinvla}), where SFRGs and SMGs are shown to have
similar radio sizes and SFRG sizes average to 2.3$\pm$0.8\,kpc.
\citet{bothwell10a} have taken high-resolution CO observations of
\rbbf\ and found its size to be consistent with this assumption (its
CO emission extends over 12.6kpc in one direction and $\sim$5kpc in
the other, corresponding to a 3.9kpc effective radius, only 0.2kpc
larger than our measured MERLIN+VLA radio size).  \rbbf\ is perceived
to be a merging system that has two components of roughly equal
spatial extent \simlt2kpc, thus its effective size is about
$\sim$2$\times$ larger than the rest of the galaxies in our sample,
although we still use the MERLIN+VLA sizes as priors for CO size due
to lack of resolved CO data.  For the sources with measured MERLIN+VLA
sizes, we use the $R_{\rm eff}$ as the radius $r$ in Equation 2, and
for sources without MERLIN+VLA data, we assume that $r$ is equal to
the mean of the measured MERLIN+VLA sizes, $r$\,=\,2.3$\pm$0.8\,kpc.

It is important to note that the extent and morphology of the
MERLIN+VLA radio emission can affect the interpretation of the
dynamical mass estimate.  Three objects have large radio/UV offsets or
very extended radio emission: RGJ105209 (whose offset is discussed in
section 2.4), RGJ123707 (extended across a $\sim$7\,kpc$\times$17\,kpc
area), and RGJ123710 (two knots separated by $\sim$8kpc).  If the CO
gas were to trace the MERLIN+VLA morphology perfectly, as is our a
priori assumption, then the radius of each knot of radio emission
would be used to estimate dynamical masses and they would be summed.
Depending on the distribution of these knots, the true dynamical mass
would be calculated by considering the size of each knot and the
distance separating them, however, within the uncertainty of our
measurements and assumptions (e.g. inclination angle, line-width
uncertainty and uncertainty on the value of $C$) we consider our
approach of circularising the MERLIN+VLA sizes and inferring a radius
as accurate.  We caution that dynamical masses are potentially
slightly underestimated, due to the size and distribution of the CO
gas relative to radio emission.  For example, radio sizes are
potentially inconsistent with CO sizes, as with the B-configuration
data of \dada\ and \dadb\ presented in \citet{daddi10a}, which suggest
larger dynamical mass estimates by a factor proportional to the
radius.

When applied to the whole SFRG sample, we derive a median SFRG
dynamical mass of M$_{\rm dyn} =
\,$(7.2$^{+6.7}_{-3.4}$)$\times$10$^{10}$\,\msun$/\langle sin^{2}\,i
\rangle\, \approx \,$2.9$\times$10$^{11}$\,\msun.  The same value of
$C$ is used to compare with mean dynamical mass of SMGs, which is
M$_{\rm dyn}(SMG) =
$(1.5$^{+1.4}_{-0.4}$)$\times$10$^{11}$\,\msun$/\langle
sin^{2}\,i\rangle\,\approx\,$6.0$\times$10$^{11}$\,\msun.

\begin{figure*}
  \centering
  \includegraphics[width=0.66\columnwidth]{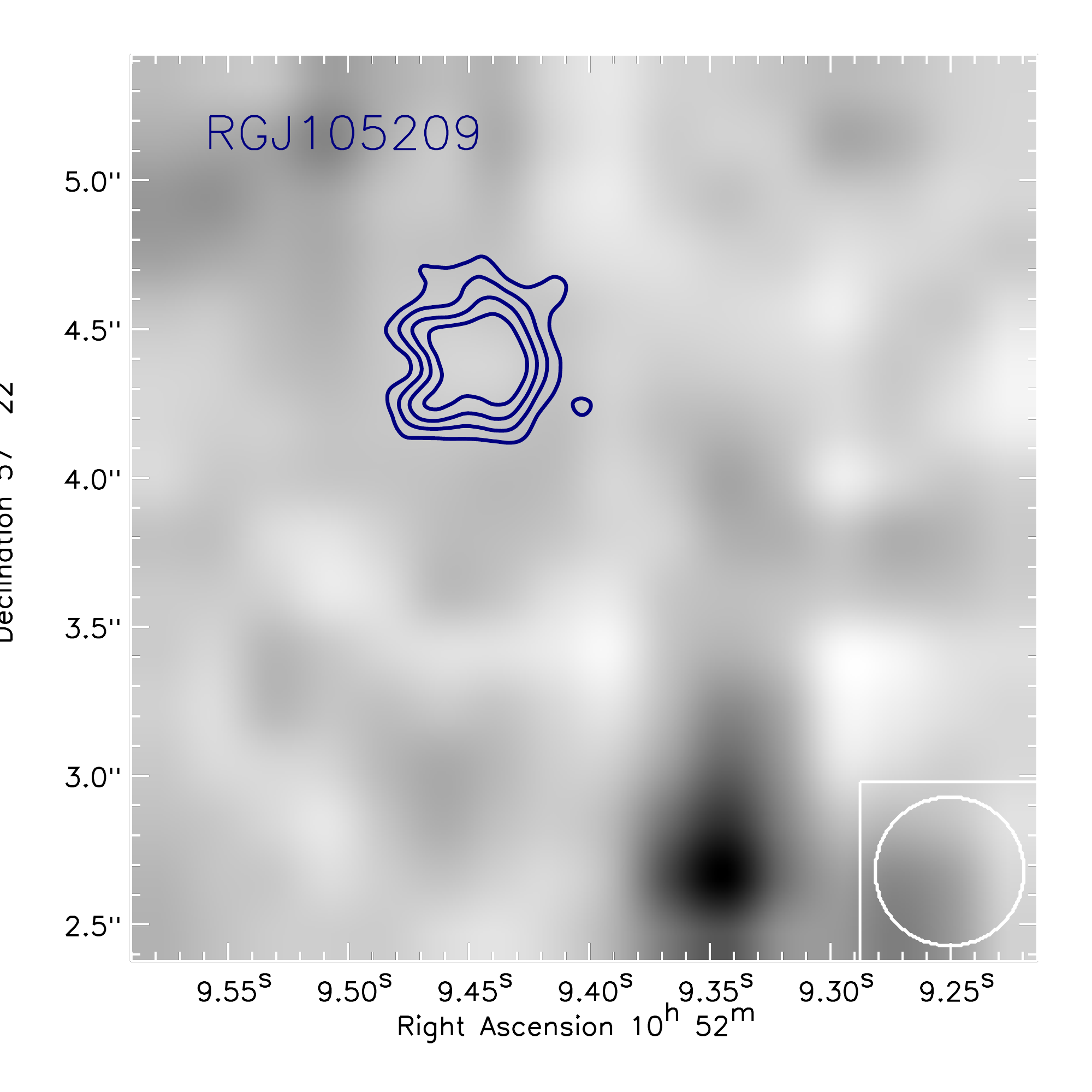}
  \includegraphics[width=0.66\columnwidth]{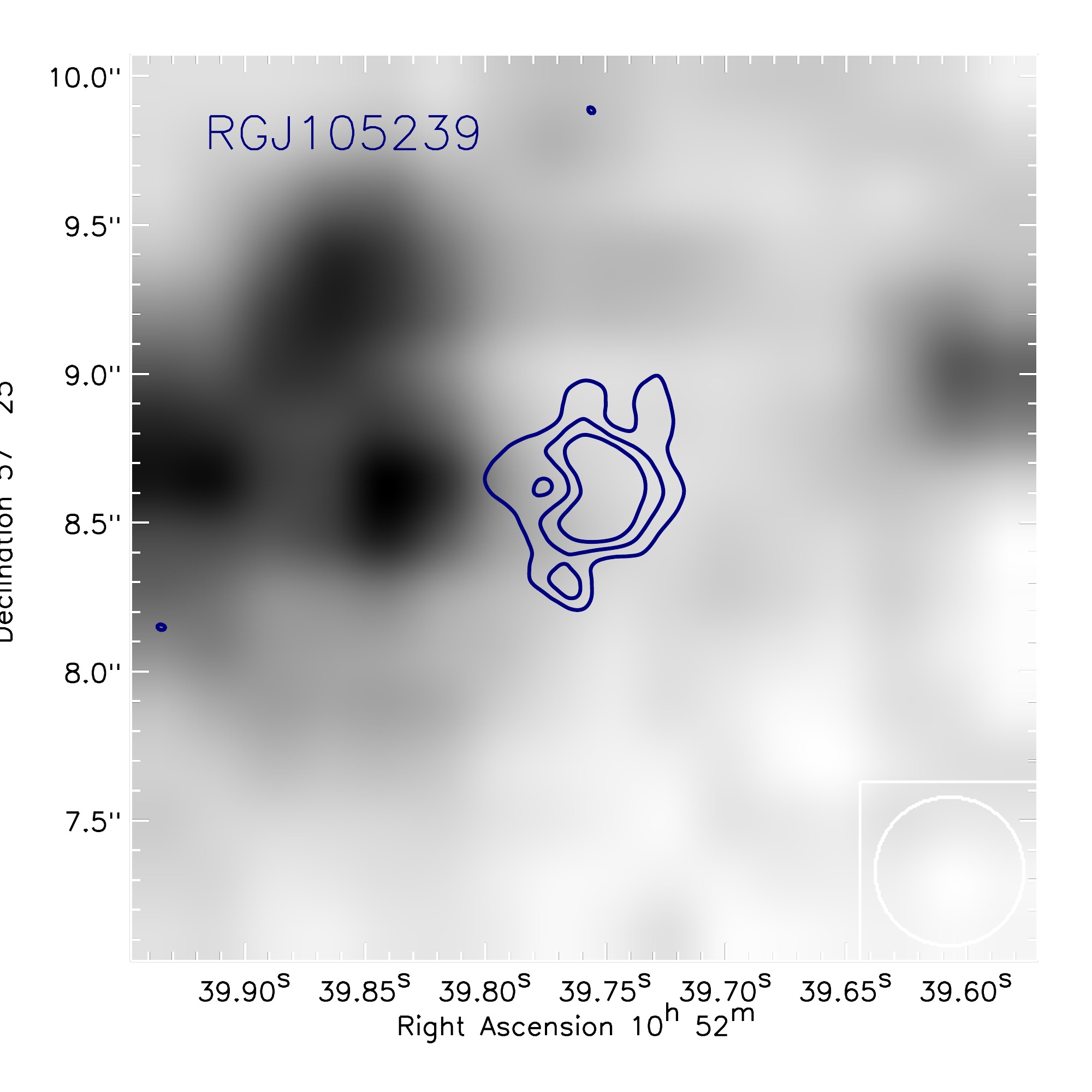}
  \includegraphics[width=0.66\columnwidth]{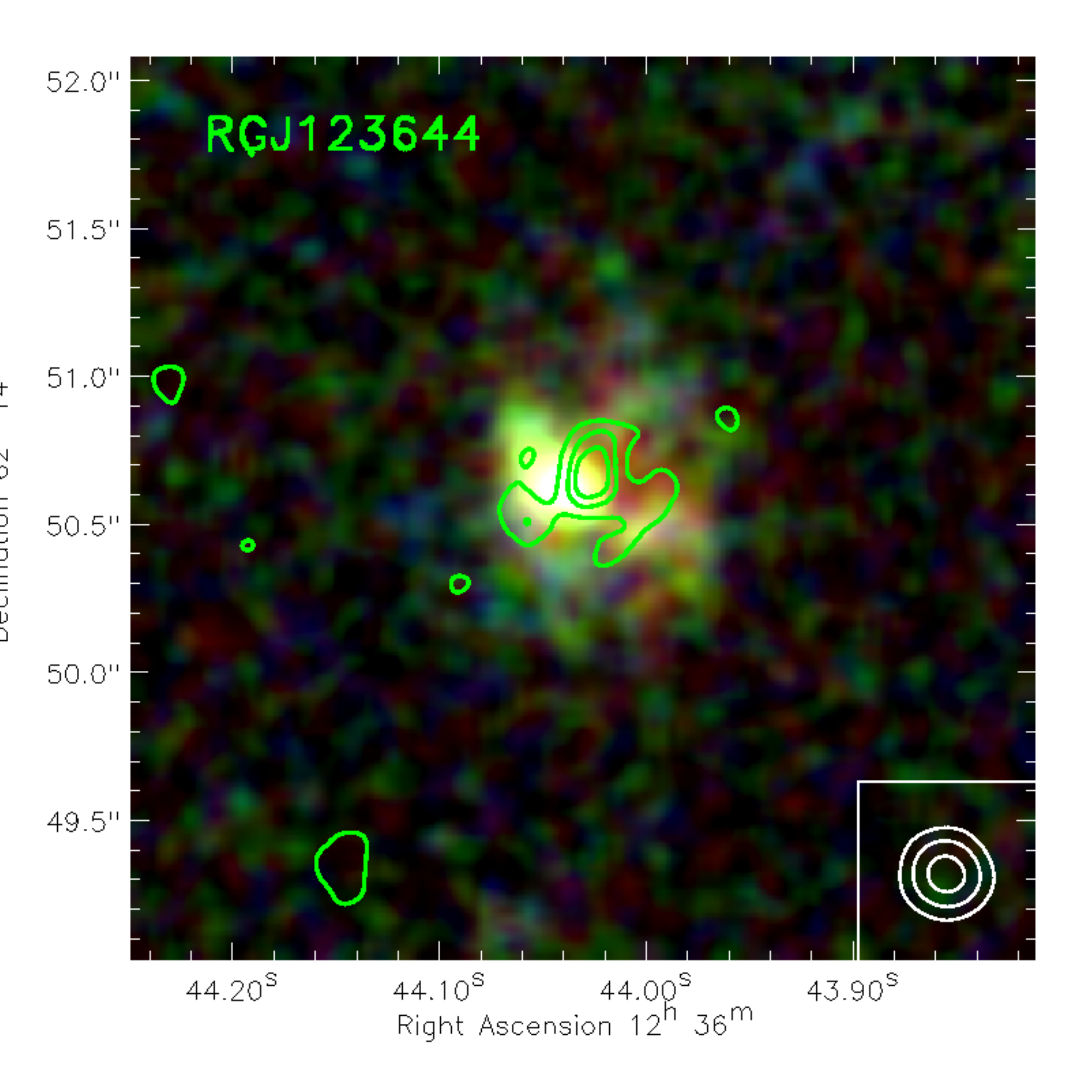}\\
  \includegraphics[width=0.66\columnwidth]{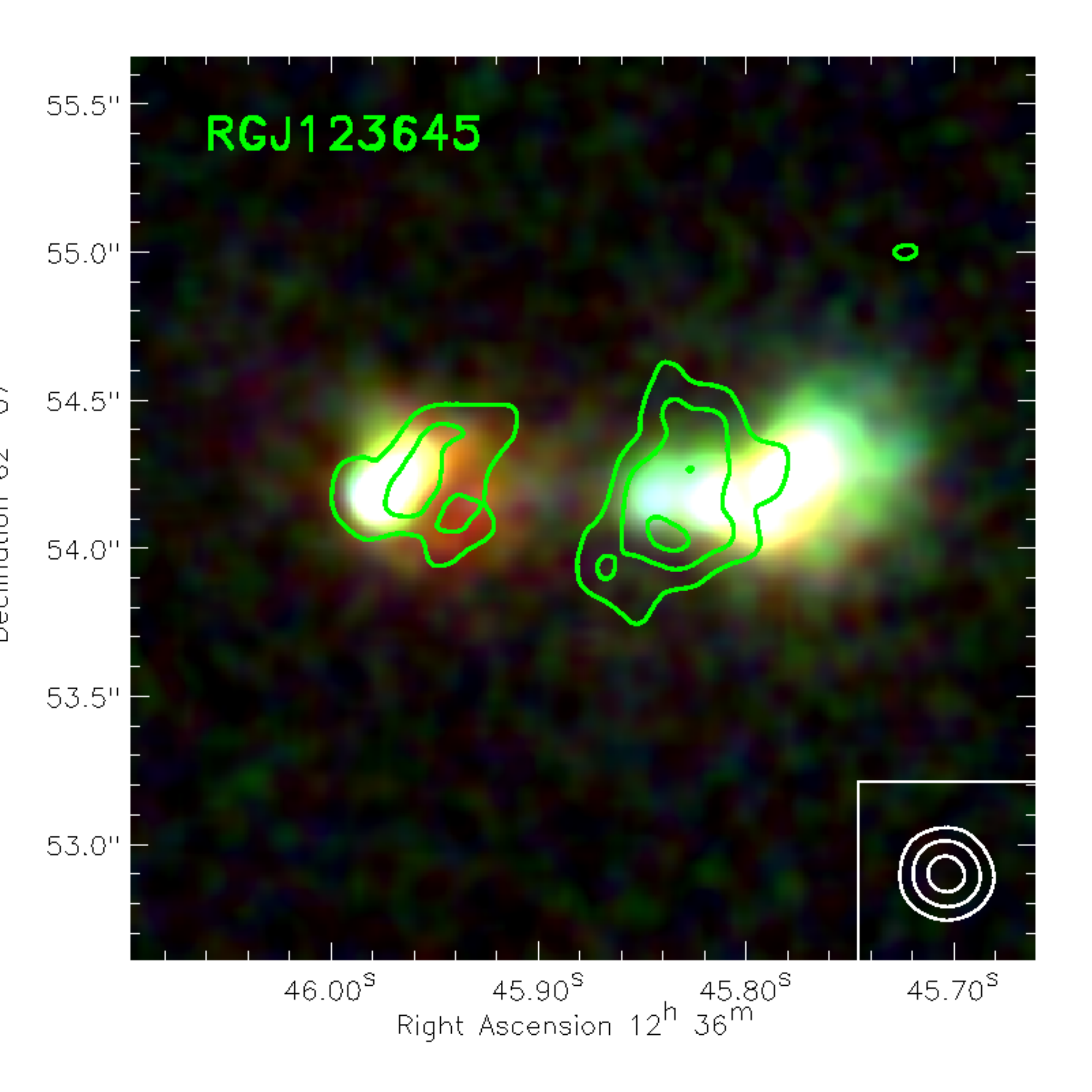}
  \includegraphics[width=0.66\columnwidth]{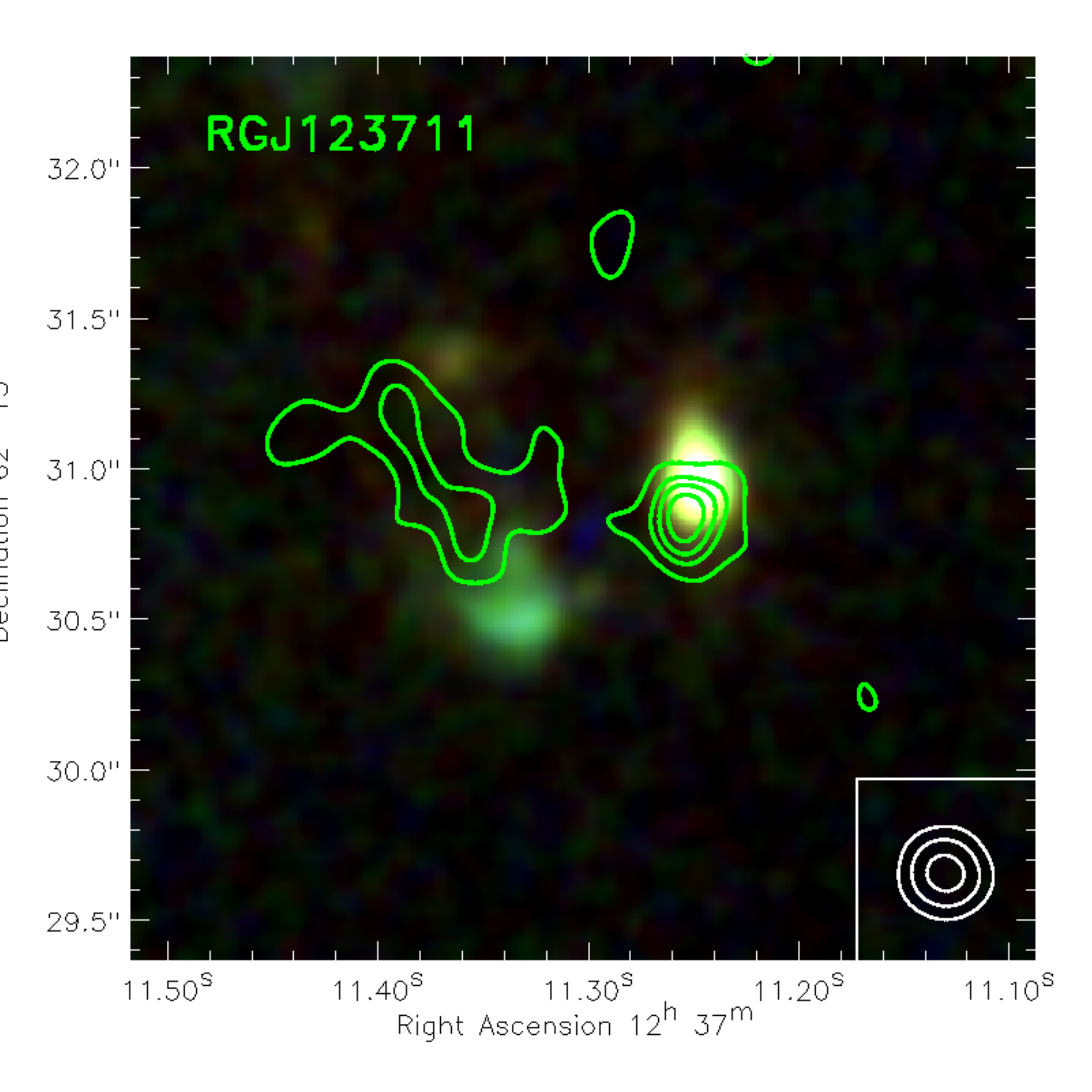}
  \includegraphics[width=0.66\columnwidth]{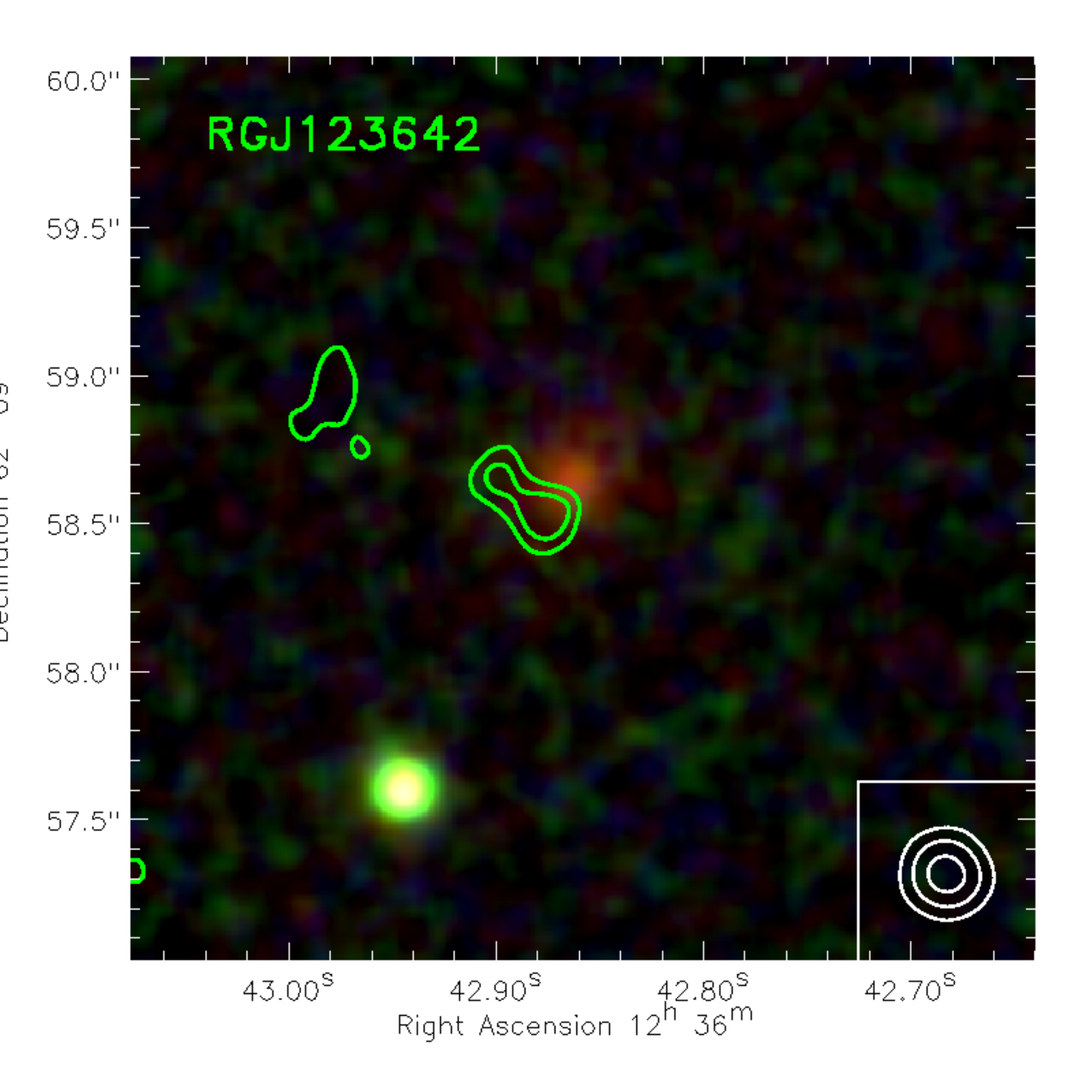}\\
  \includegraphics[width=0.66\columnwidth]{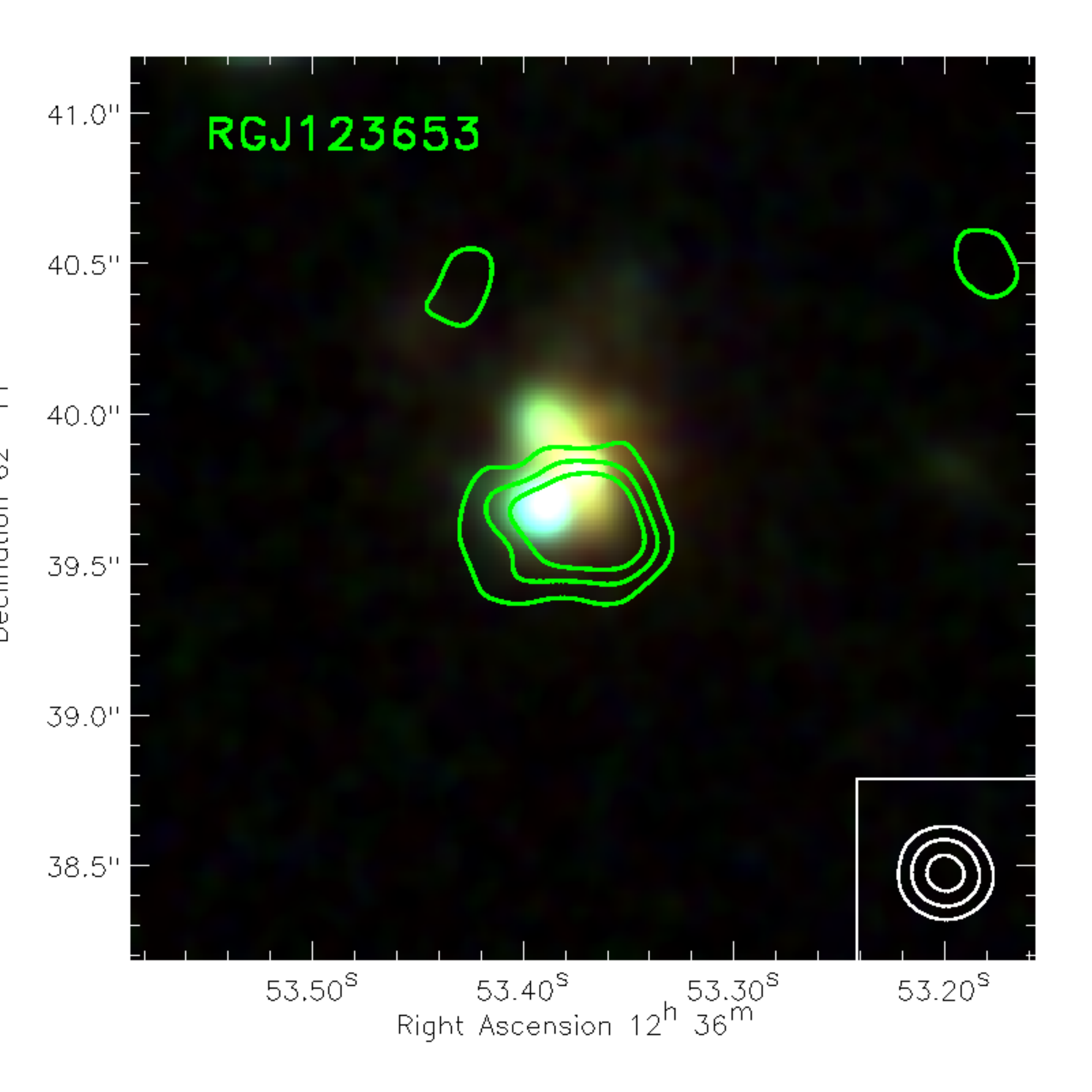}
  \includegraphics[width=0.66\columnwidth]{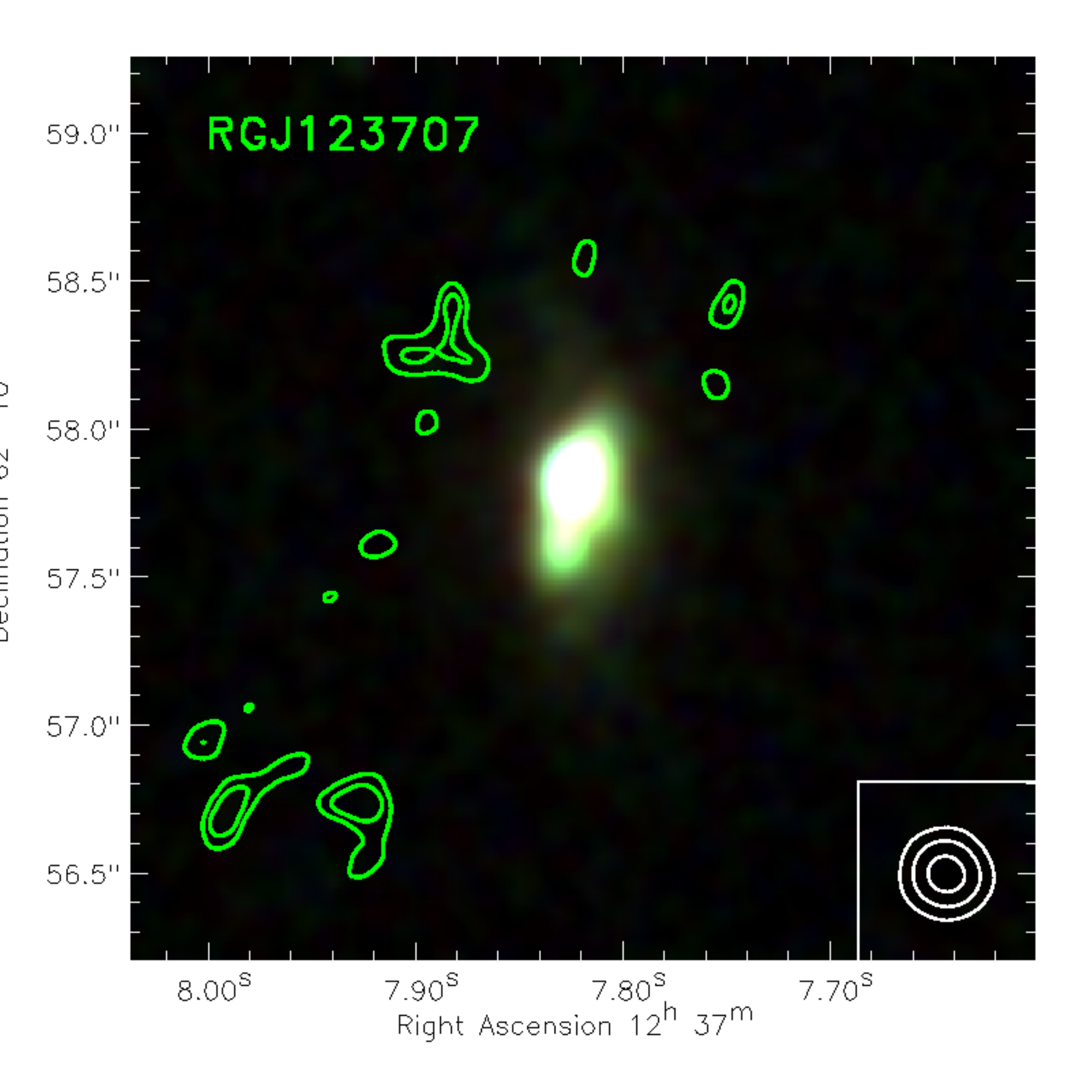}
  \includegraphics[width=0.66\columnwidth]{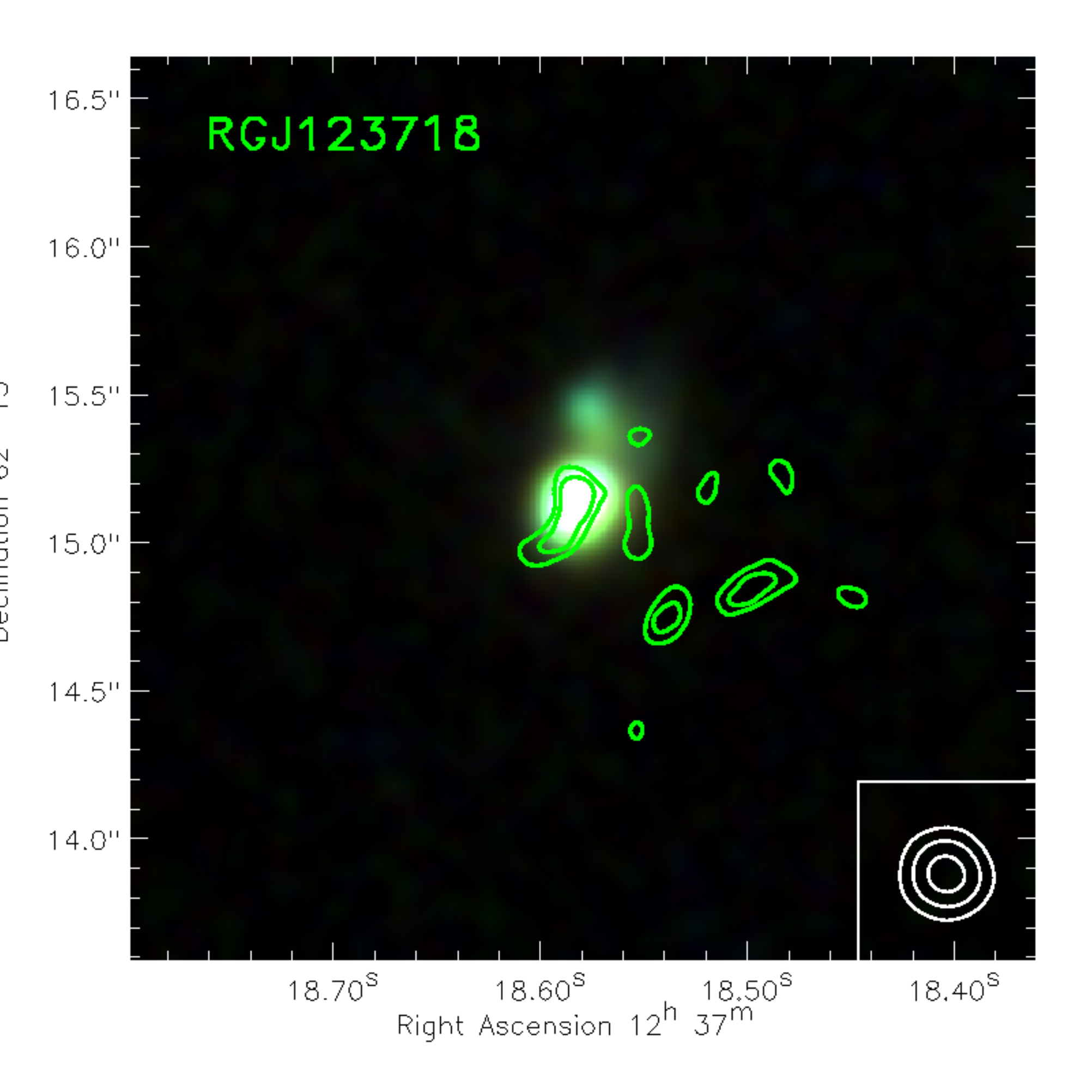}\\
  \includegraphics[width=0.66\columnwidth]{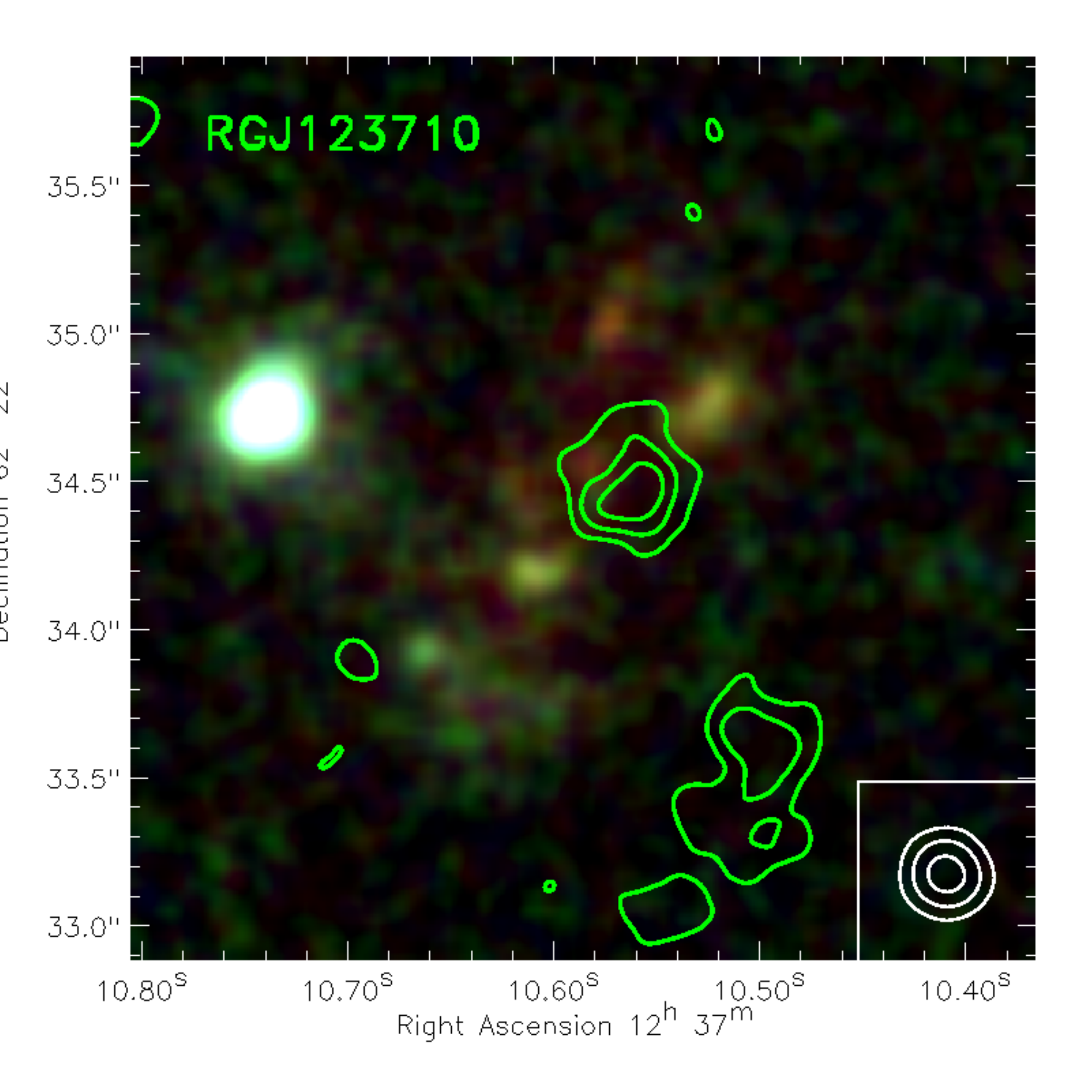}
  \includegraphics[width=0.66\columnwidth]{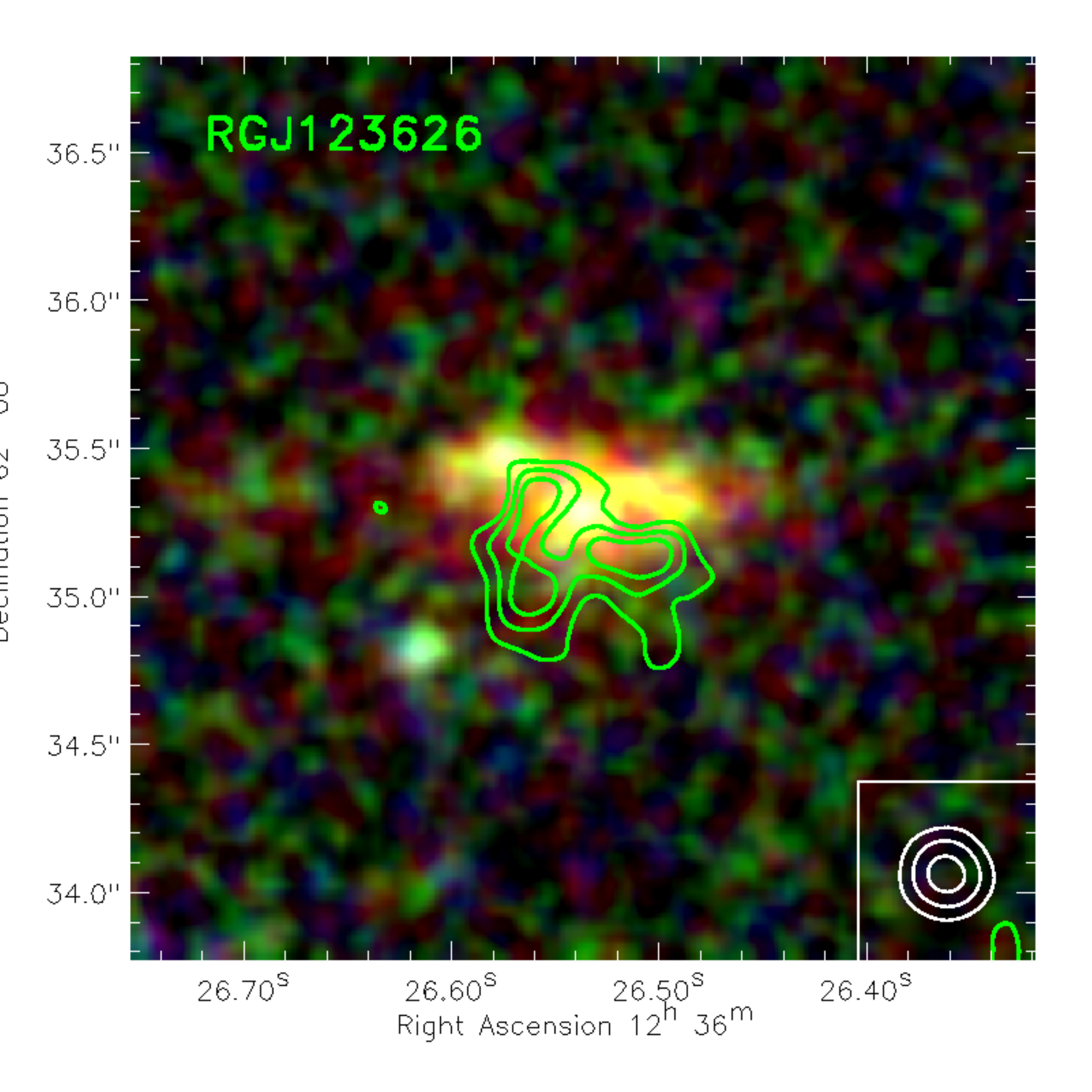}\\
  \caption{ Contours of MERLIN+VLA high-resolution 0.4-0.5\arcsec\ radio
    emission overlayed on 3\arcsec\,$\times$\,3\arcsec\ optical imaging
    for the CO-observed SFRGs.  The optical imaging of GOODS-N sources
    is from $HST$-ACS $B$, $V$, and $z$ bands (presented in tricolor).  The Lockman Hole
    sources have ACS $i$-band (black and white cutouts).  The MERLIN+VLA beam shape is shown in the
    lower right of each panel in white (outer contour is the FWHM).
    The contour levels plotted integer multiples of $\sigma$ starting
    at 3$\sigma$ (except for \rbbf\ and \rdbf, where only 3, 5, and
    7$\sigma$ contours are shown for clarity).  Structure in radio
    emission is seen in these galaxies on 1-8\,kpc scales. }
  \label{fig:merlin}
\end{figure*}

\subsection{MERLIN+VLA Radio Morphology}\label{sec:merlinvla}

We use the high-resolution MERLIN+VLA radio maps to map the extension
of starburst emission and assess the contribution of AGN by
considering their radio morphology, which is shown as contour overlays
on optical imaging in Fig.~\ref{fig:merlin}.  With resolutions of
0.4\arcsec\ and 0.5\arcsec\ per beam (for GOODS-N and the Lockman Hole
respectively), the smallest resolvable structure at z$\sim$1.5 would
be 4-5\,kpc across.  The typical size of an AGN emission region in a
\uJy\ radio galaxy at $z>1$ \citep[see][]{casey09a} at radio
wavelengths is much less than 1kpc, implying that an AGN dominated
source would be completely unresolved in MERLIN+VLA radio maps
\citep[e.g. as seen in the high stellar mass, giant elliptical systems
  of ][]{casey09a}.  Fig.~\ref{fig:merlin} shows that each of these
galaxies has extended emission regions on large scales irregularly
spread across large regions of the galaxy, suggestive of spatially
distributed star formation, unlikely to be generated by AGN.  About
10-60\%\ of the total flux from the source is estimated to be resolved
out by the high-resolution imaging, which is dependent on the
galaxies' extended or compact morphologies.  We measure an effective
radius of the star forming area by isolating the regions where
MERLIN+VLA radio emission is significant to $>$3$\sigma$, and we then
take the square root of the surface area over pi (effectively
circularised), R$_{\rm eff}$, which is given in Table
\ref{tab:derived}).  While the morphologies are irregular (and are not
in fact circular), we find that the effective radii average to
2.3$\pm$0.8\,kpc, which agrees with the size measurements of SMGs in
\citet{biggs08a} and \citet{chapman04b}.  We use the agreement of
MERLIN+VLA SFRG and SMG sizes to partly justify our assumption of
similar CO sizes between the populations.

Since as much as 60\%\ of the radio flux can be resolved out in the
high-resolution maps, we caution that this effective radius might be
an underestimate, and that it could increase by factors of
1.1-1.5$\times$ if all of the flux is accounted for; however, this
adjustment is not made to our effective radii measurements since it is
highly uncertain and relies on assumptions of the distribution of star
formation activity in the outskirts of each source.  Similarly, this
underestimate of effective radius would propagate to the derivation of
dynamical masses.

\begin{table*}
\begin{center}
\caption{Other Derived Properties of CO-observed SFRGs}
\label{tab:derived}
\begin{tabular}{c@{ }c@{ }c@{ }c@{ }c@{ }c@{ }c@{ }c@{ }c@{ }c@{ }c@{ }c@{ }c@{ }c@{ }c@{ }c}
\hline\hline
NAME & {\it z} & L$_{\rm FIR}$ & SFR$_{\rm radio}$ & SFR$_{\rm UV}$ & M$_{\rm \hh}$[U] & M$_{\rm \hh}$[Sp] & M$_{\rm dyn}$sin$^2i$ & SFE  & R$_{\rm eff}$ & $\Sigma_{\rm SFR}$& M$_\star$ & Class$^{\rm X}$ & Class$^{\rm IR}$ & Class$^{\rm CO}$ \\
        &  & {\scriptsize (10$^{12}$\lsun)}       & {\scriptsize (\sfr)}  & {\scriptsize (\sfr)}    & (\msun) & (\msun) & (\msun) & {\scriptsize (\lsun/\msun)}    & (kpc) & ($\dagger$) & (\msun)  &  &  &         \\
\hline
\tett... & 1.362 & 3.3$^{+1.0}_{-0.8}$   & 570$^{+170}_{-120}$     & 9$^{+5}_{-3}$     & 1.2$\times$10$^{10}$    & 6.5$\times$10$^{10}$     & 1.2$\times$10$^{11}$ & 280     & ... & ... & 2.3$\times$10$^{11}$   & SB  & SB  & SB  \\
\rfbf... & 2.113 & 4.4$^{+2.2}_{-1.5}$   & 750$^{+380}_{-250}$     & 20$^{+30}_{-10}$  & 3.0$\times$10$^{10}$    & 1.7$\times$10$^{11}$     & 7.4$\times$10$^{10}$ & 150     & 2.2 & 49  & 1.0$\times$10$^{11}$  & SB  & SB  & SB  \\
\tptt... & 1.820 & 2.2$^{+1.8}_{-1.0}$   & 380$^{+300}_{-170}$     & 90$^{+30}_{-20}$  & 2.6$\times$10$^{10}$    & 1.5$\times$10$^{11}$     & 9.6$\times$10$^{10}$ & 80      & 2.3 & 23  & 1.2$\times$10$^{11}$  & SB  & SB  & SB  \\
\dada... & 1.465 & 1.9$^{+1.5}_{-0.5}$   & 320$^{+250}_{-140}$     & 7$^{+5}_{-3}$     & 1.8$\times$10$^{10}$    & 1.0$\times$10$^{11}$     & 5.8$\times$10$^{10}$ & 110     & 2.8 & 13  & 2.8$\times$10$^{10}$   & SB  & SB  & SB  \\
\rhbf... & 3.661 & 1.1$^{+1.7}_{-0.7}$   & 1900$^{+3000}_{-1100}$  & 30$^{+20}_{-10}$  & $<$1.0$\times$10$^{10}$ & $<$5.9$\times$10$^{10}$  & ...                  & $>$110  & 1.4 & 310 & 4.1$\times$10$^{10}$ & SB+ & SB  & SB \\
\rabf... & 2.090 & 5.0$^{+3.5}_{-2.1}$   & 850$^{+600}_{-350}$     & 30$^{+30}_{-20}$  & 2.0$\times$10$^{10}$    & 1.1$\times$10$^{11}$     & 2.1$\times$10$^{10}$ & 250     & 2.7 & 37  & 1.1$\times$10$^{11}$  & SB+ & SB+ & SB \\
\rgbf... & 1.434 & 3.8$^{+1.4}_{-1.0}$   & 660$^{+250}_{-180}$     & 30$^{+8}_{-6}$    & 1.9$\times$10$^{10}$    & 1.1$\times$10$^{11}$     & 5.2$\times$10$^{10}$ & 200     & 3.0 & 23  & 4.9$\times$10$^{10}$   & SB  & SB  & SB \\
\rdbf... & 1.275 & 2.9$^{+0.9}_{-0.7}$   & 500$^{+160}_{-120}$     & 40$^{+10}_{-10}$  & $<$5.0$\times$10$^{9}$  & $<$2.8$\times$10$^{10}$  & ...                  & $>$580  & 2.8 & 20  & 3.6$\times$10$^{10}$   & SB  & AGN & SB+ \\
\rcbf... & 1.489 & 1.2$^{+1.6}_{-0.7}$   & 210$^{+280}_{-120}$     & 50$^{+20}_{-10}$  & $<$7.1$\times$10$^{9}$  & $<$4.0$\times$10$^{10}$  & ...                  & $>$170   & 1.3 & 39  & 2.3$\times$10$^{10}$   & SB+ & SB  & SB  \\
\dadb... & 1.522 & 2.1$^{+1.8}_{-1.0}$   & 350$^{+310}_{-170}$     & 10$^{+10}_{-5}$   & 2.7$\times$10$^{10}$    & 1.5$\times$10$^{11}$     & 2.0$\times$10$^{10}$ & 80      & 1.9 & 30  & 2.4$\times$10$^{10}$   & SB  & SB  & SB  \\
\rbbf... & 1.996 & 8.8$^{+6.1}_{-3.6}$   & 1500$^{+1000}_{-600}$   & 30$^{+10}_{-9}$   & 3.0$\times$10$^{10}$    & 1.7$\times$10$^{11}$     & 1.1$\times$10$^{11}$ & 290     & 3.7 & 34  & 1.2$\times$10$^{11}$   & SB+ & AGN & SB \\
\ribf... & 1.512 & 0.8$^{+1.5}_{-0.5}$  & 140$^{+250}_{-90}$      & 40$^{+10}_{-10}$  & $<$5.0$\times$10$^{9}$  & $<$2.8$\times$10$^{10}$  & ...                  & $>$160   & 1.1 & 36  & 1.6$\times$10$^{11}$  & SB  & AGN & SB  \\
\tott... & 1.532 & 2.5$^{+0.6}_{-0.5}$   & 420$^{+100}_{-80}$      & 9$^{+7}_{-4}$     & $<$7.8$\times$10$^{9}$  & $<$4.4$\times$10$^{10}$  & ...                  & $>$320  & ... & ... & 6.0$\times$10$^{9}$   & SB  & SB  & SB \\
\rebf... & 2.237 & 5.6$^{+1.9}_{-1.4}$   & 960$^{+330}_{-250}$     & 20$^{+9}_{-6}$    & 2.6$\times$10$^{10}$    & 1.5$\times$10$^{10}$     & 7.5$\times$10$^{10}$ & 220     & ... & ... & 1.8$\times$10$^{10}$  & SB  & AGN & SB  \\
\chan... & 2.224 & 6.6$^{+3.3}_{-2.2}$   & 1100$^{+600}_{-400}$    & 40$^{+20}_{-20}$  & $<$5.0$\times$10$^{9}$  & $<$2.8$\times$10$^{10}$  & ...                  & $>$1300 & ... & ... & 2.4$\times$10$^{10}$  & AGN & SB  & AGN \\
\chap... & 2.187 & 6.8$^{+2.1}_{-1.6}$   & 1200$^{+400}_{-300}$    & 70$^{+40}_{-30}$  & 8.0$\times$10$^{9}$     & 4.5$\times$10$^{10}$     & 2.5$\times$10$^{10}$ & 850     & ... & ... & 4.1$\times$10$^{10}$  & SB  & SB+ & AGN \\
\hline\hline
\end{tabular}
\end{center}
{\small {\bf Table Notes.}  Derived gas properties of SFRGs.  L$_{\rm
    FIR}$ is derived from radio luminosity, and the associated SFR is
  thus labelled $SFR_{\rm radio}$.  Conversion to gas mass assumes the
  typical ULIRG conversion factor
  $X=$\,0.8\,\msun\,(K\,\kms\,pc$^2$)$^{-1}$ \citep{downes98a} for
  $M_{\hh}$ [U], the default assumption of gas mass for this sample,
  although we note that higher conversion factors
  ($\sim$4.5\,\msun\,(K\,\kms\,pc$^2$)$^{-1}$) might be appropriate
  for a subset of these galaxies which exhibit normal spiral galaxy
  characteristics \citep[see][]{daddi10a}, and these gas masses are
  given in $M_{\hh}$ [Sp].  The total dynamical mass is calculated by
  assuming a size of CO emission, which we do not directly measure,
  but assume is equal to the effective radius of radio emission, given
  as $R_{\rm eff}$, and when radio is not available we assume a radius
  of $\sim$2.3\,kpc, the mean MERLIN+VLA radius, consistent with SMG
  sizes.  The star formation efficiency (SFE) is $SFR_{\rm radio}$
  divided by the ULIRG molecular gas mass, M$_{\rm \hh}$.  SFR$_{UV}$
  has not been corrected for extinction.  The star formation rate
  density, $\Sigma_{\rm SFR}$, has units of
  \msun\,yr$^{-1}$\,kpc$^{-2}$ ($\dagger$).  Since these galaxies were
  selected in the radio, we provide a measure of their starburst and
  AGN content by classifying them into starburst (SB), AGN, and
  starburst/AGN mix (SB+) in five separate AGN selection criteria.
  Class$^{\rm X}$ marks any galaxy with X-Ray luminosity
  $>$5$\times$10$^{43}$\,erg\,s$^{-2}$ as ``AGN'',
  $<$1$\times$10$^{43}$\,erg\,s$^{-2}$ as ``SB'' and intermediate
  luminosity galaxies as ``SB+''.  X-ray data is taken from
  \citet{alexander03a}, \citet{brunner08a}, \citet{ueda08a}, and
  \citet{mushotzky00a}.  Class$^{\rm IR}$ marks galaxies with
  significant 8\um\ flux excess (relative to the stellar population
  fits) as ``AGN'', galaxies with marginally significant
  $<$2$\sigma$\ 8\um\ excess as ``SB+'' and galaxies with no
  8\um\ excess as ``SB'' (see Figure~\ref{fig:midir}).  Class$^{\rm
    CO}$ marks galaxies with unusually high FIR-to-CO ratios as
  ``AGN'' ($log(L_{FIR}/L_{CO}^\prime)\,>\,$2.7), ``SB+''
  (2.5$\,<\,log(L_{FIR}/L_{CO}^\prime)\,<\,$2.7), and at low FIR-to-CO
  ratios, ``SB'' ($log(L_{FIR}/L_{CO}^\prime)\,>\,$2.5).  A fourth
  class may be defined as a measure of the star formation rate
  density, $\Sigma_{\rm SFR}$; all of the galaxies for which we have
  $\Sigma_{\rm SFR}$ measurements satisfy the ``SB'' criteria however,
  with star formation densities $<$200\,\msun\,yr$^{-2}$.  A fifth
  class is defined by their rest-UV/optical spectroscopic properties,
  which, by method of their selection, is uniformly ``SB.''

}
\end{table*}

\subsection{Star Formation Rates, Densities and Efficiencies}\label{ss:sfrsfe}

We estimate SFRs from VLA radio luminosities, using the
radio/FIR correlation for star forming galaxies
\citep[e.g.][]{helou85a,condon92a,sanders96a}:
\begin{equation}
L_{\rm FIR}\,=\,(3.583\times10^{-48})\,D_{L}^{2}\,S_{1.4}\,(1\,+\,z)^{(1\,-\,\alpha)}
\end{equation}
where $D_L$ is luminosity distance in cm, $L_{\rm FIR}$ is evaluated
from 8-1000\um\ and is in \lsun, $S_{1.4}$ is the radio flux density
at 1.4\,GHz in units of \uJy.  A Salpeter initial mass function is
assumed.  The factor of (1\,+\,z)$^{(1 - \alpha)}$ accounts for
bandwidth compression and the radio K-correction to the rest-frame
luminosity; $\alpha$ is the synchrotron slope index, here taken to be
0.75 \citep{yun01a}.  Since this is an empirical relation defined by
the mean ratio $q$ between FIR and radio luminosities, there are a few
different versions in use in the literature.  This relation uses the
bolometric ratio between FIR flux and radio flux of $q_{\rm
  IR}\,=\,$2.46 \citep[e.g.][]{ivison10a,ivison10b,casey10a}.  The
relation is traditionally defined by $q$\,=\,2.35 \citep{sanders96a}
and was used by \citet{chapman05a} to calculate FIR luminosities for
the redshift-identified SMGs (only differing by factors of
20\%\ between different assumed $q$ ratios).  In most cases, the FIR
luminosities derived from the radio are consistent with the submm
detection limits and any flux density measurements at shorter
wavelengths in the FIR (e.g. at 70\um\ or at 350\um).

The FIR luminosity is then converted into a star formation rate (SFR)
using the following relation:
\begin{equation}
SFR (M_{\odot}\,yr^{-1})\,=\,1.7\times10^{-10}\,L_{\rm FIR}\, ($\lsun$)
\end{equation}
from \citet{kennicutt98a}.  The inner quartile (25-75 percent) of star
formation rates is 400-1600\Mpy\ with median 700\Mpy.  Both derived
quantities, L$_{\rm FIR}$ and SFR, are given in
Table~\ref{tab:derived}.  We also derive UV-inferred SFRs using
$v$-band and $i$-band optical photometry.  Since all SFRGs are quite
optically faint ($i>22$) and FIR luminous, they are potentially
subject to significant (yet uncertain) extinction factors.  While not
taken into account for the UV-inferred SFRs themselves, the large
extinction factors may be deduced from the large disparities between
SFR$_{\rm radio}$ and SFR$_{\rm UV}$ given in table~\ref{tab:derived}
with a typical ratio of SFR$_{\rm radio}$/SFR$_{\rm UV}$\,=\,30 (we do
not consider SFR$_{\rm radio}$/SFR$_{\rm UV}$ to have AGN
contamination based on the previous section).

Additional evidence that extinction caused by dust is significant
comes in the comparison between rest-UV morphologies and their MERLIN+VLA
radio morphologies as seen in Fig.~\ref{fig:merlin}.  Since the
brightest and/or bluest components of the rest-UV emission and radio
emission do not often coincide, it indicates that the dustiest, most
FIR-luminous regions in the galaxy could be highly obscured.

Total star formation rate densities, $\Sigma_{\rm SFR}$, are estimated
by dividing these SFRs (from VLA fluxes) by the surface areas of
MERLIN+VLA radio emission regions.  We measure an effective star
formation surface area by isolating the regions where MERLIN+VLA radio
emission is significant to $>$3$\sigma$ (the values for effective
radius of radio emission and SFR density, $\Sigma_{\rm SFR}$, are
given in Table~\ref{tab:derived}).  The median total SFR density is
30$^{+40}_{-20}$\,\Mpy\,kpc$^{-2}$.  Note that this is potentially an
overestimation of the SFR density since the radio flux contained
within the $>$3$\sigma$ MERLIN+VLA region constitutes a fraction of
the total VLA radio flux (it is inferred that this fraction is within
40-90\%).  For this reason we also compute alternate star formation
rate densities using the total integrated MERLIN+VLA flux within the
$>$3$\sigma$ emission area.  We convert the flux to a radio
luminosity, then FIR luminosity, then SFR, and divide it by the area.
The median value for this alternate SFR density is
10$^{+20}_{-10}$\,\Mpy\,kpc$^{-2}$ which is on average, three times
less than the total $\Sigma_{\rm SFR}$.

Comparing either SFR density measurements to their theoretical
maximum$-$the maximum gas density divided by the local dynamical time
\citep[see equation 5 of][ we use
  t$_{dyn}$=4$\times$10$^{7}$yr]{elmegreen99a}$-$we can determine if
the implied SFR density exceeds the theoretical prediction.  While
local ULIRGs with $\Sigma_{\rm SFR}\approx$200\,\Mpy\,kpc$^{-2}$ are
forming stars at their theoretical maximum \citep[e.g.][]{tacconi06a},
only one of the SFRGs exceeds this limit, which is a dust opacity
Eddington limit for SFR density \citep[see][]{thompson05a}.  This
source is \rhbf, the highest redshift source in our sample having a
low S/N in the radio, meaning the measured effective radius is
unusually small.  Even by assuming unresolved radio profiles, only
\rhbf\ exceeds the maximal starburst density, further highlighting
that even starburst emission can dominate unresolved radio emission.
In contrast, it would be unlikely for extended radio emission (this
faint at these redshifts) to be driven by emission from AGN.

The star formation efficiency (SFE) can be calculated by dividing the
FIR luminosity by the \hh\ gas mass.  This calculation is contingent
on the gas reservoir being in the same region as the starburst, which
is an assumption which needs to be investigated in more detail through
future high-resolution multiple-$J$ CO maps and resolved FIR emission
maps.  The mean SFE for SFRGs is 280$\pm$260\,\lsun/\msun.  Put
another way, the median depletion timescale (defined as M$_{\rm
  \hh}$/SFR, assuming 100\%\ efficiency) for SFRGs is $\sim$34\,Myr
with inner quartile 20\,Myr - 55\,Myr.  It should be noted here that a
different gas conversion factor
(e.g. 4.5\,\msun\,(K\,km\,s$^{-1}$\,pc$^2$)$^{-1}$) would increase the
depletion timescale and decrease the star formation efficiency by 5.6
times.

\subsection{Dust Temperature and Dust Mass}\label{ss:dustmass}

We fit FIR spectral energy distributions (SEDs) to the FIR flux
density limits at 70\um, 350\um, 850\um, and 1200\um\ (see
Table~\ref{tab:observations} for details). Our fitting method follows
the methodology of \citet{chapman04a}, \citet{chapman05a}, and
\citet{casey09b}: a single dust temperature modified black body fit.
We fix the dust emissivity $\beta$ to 1.5 and use the FIR/radio
correlation to infer a FIR luminosity from the radio flux.  Since we
are limited by a lack of data in the FIR for most of our sample (many
only have limits), our derived dust temperatures are not well
constrained.  However if the uncertainty in our assumption of the
radio/FIR correlation and dust emissivity are taken as givens, then we
deduce temperature uncertainties on the order of 20-30K.  This is
calculated by assuming $L_{\rm FIR}$ scales with $L_{\rm radio}$; a
higher $L_{\rm radio}$ implies a high $L_{\rm FIR}$, and a FIR SED
with a high $L_{\rm FIR}$ can then be constrained by the FIR flux
density measurements, even in the case of upper limits.  Despite the
large uncertainties, our measurements (whose statistical mean is
~65\,K) are consistent with a selection of warmer dust galaxies, even
though there may be outliers.


We also estimate dust mass for the SFRG sample by using the following
relation
\begin{equation}
M_{\rm dust} (M_\odot) \simeq \frac{S_{\rm obs}D^2_L}{\kappa_{\nu} B(\nu_{\rm obs},T_d)}
\end{equation}
where $\kappa_{\nu}$ is the dust mass absorption coefficient as a
function of rest wavelength:
$\kappa_{850}$\,=\,0.15\,m$^2$\,kg$^{-1}$,
$\kappa_{250}$\,=\,0.29$\pm$0.03\,m$^2$\,kg$^{-1}$ \citep{wiebe09a}
and $\kappa_{70}$\,=\,1.2\,m$^2$\,kg$^{-1}$ \citep{weingartner01a}.
Again, the lack of FIR data means that our dust mass calculations are
not well constrained; our SFRGs have a mean dust mass upper limit (at
2$\sigma$) of $\langle M_{\rm dust} \rangle<$2$\times$10$^{9}$\,\msun,
which is not a well constrained measurement.
A quick comparison to the mean gas mass (2$\times$10$^{10}$\,\msun)
implies mean dust-to-gas ratios of $<$1/10 (well within expectation).

\subsection{Stellar Mass}\label{ss:stellarpops}

\begin{figure*}
\centering
\includegraphics[width=1.99\columnwidth]{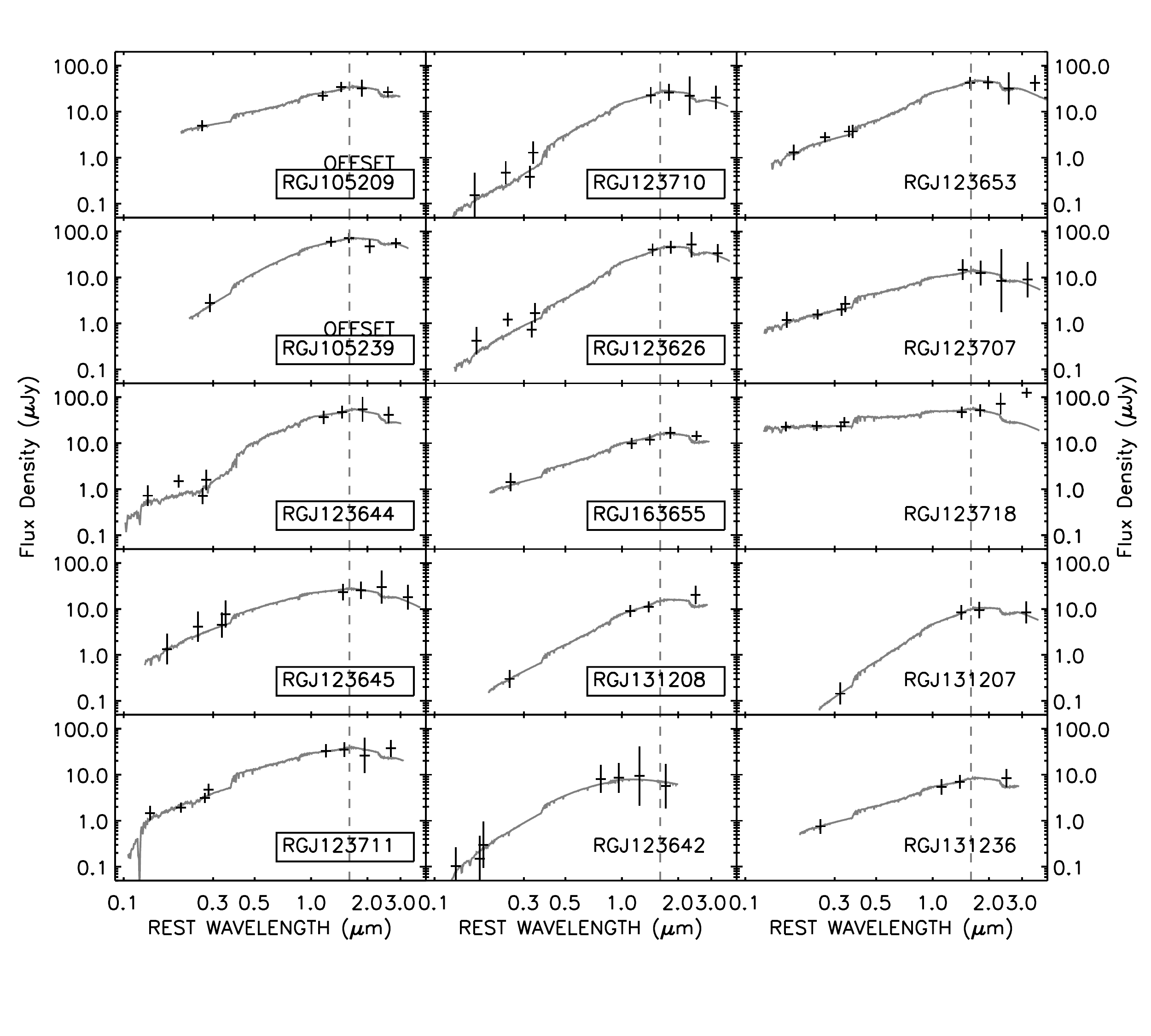}
\caption{ Stellar population models are fit to SFRG optical and
  near-IR photometric points.  We estimate stellar mass using the
  rest-frame H-band magnitudes which are inferred from the fits; the
  method is described in section~\ref{ss:stellarpops}. The SFRGs which
  are detected in CO have their names enclosed in boxes.}
\label{fig:midir}
\end{figure*}

To estimate the galaxies' stellar masses, we combine the photometric
points in the optical ($HST$ ACS $B$, $V$, $i$, and $z$ for GOODS-N,
$HST$ ACS $i$ for Lockman Hole, and Subaru $i$ for SSA13) and the
mid-IR ($Spitzer$-IRAC 3.6\um, 4.5\um, 5.8\um\ and 8.0\um).  For
$z=$1-3, these photometric points cover rest-frame 1.6\um\ where
stellar emission peaks.  We use the {\sc hyperz} photometric redshift
code \citep{bolzonella00a} to fit this photometry to several stellar
population SEDs \citep{bruzual03a}.  Using the best fit stellar
population SED, we estimate the rest-frame H-band magnitude.  While
rest-frame K-band is often used to measure stellar mass for
high-redshift galaxies \citep[using the method outlined by
][]{borys05a}, recent work has shown that stellar masses derived from
K-band are overestimated due to increased contribution from AGN power
law and decreased contribution from star light at longer wavelengths,
\citep[see][]{hainline09a}.  The mean light to mass ratio, $L_{H}$/M,
can range from 5-10\,\lsun/\msun, and using the \citet{bruzual03a}
models, is set at 5.6\,\lsun/\msun for SMGs as in Hainline \etal.  The
mean rest-frame H-band absolute magnitude is -25.4$\pm$0.8 with an
inferred average stellar mass of 7$\times$10$^{10}$\,\msun.  These
masses are consistent with the H-band derived SMG stellar masses
(H-band absolute magnitude averages -25.8$\pm$0.2 and masses average
2$\times$10$^{11}$) found by \citet{hainline09a} and a factor of
$\sim$2 lower than the K-band SMG masses derived by \citet{borys05a}.

Figure~\ref{fig:midir} illustrates the best stellar population models
with respect to the galaxies' photometric data.  If an excess flux
density is detected above the stellar model SED at observed 8\um\ then
an AGN might be contributing significantly to the near infrared
luminosities (see section \ref{ss:agn} for an analysis of AGN
contamination).  While there are potential flux excess in seven of the
15 sources illustrated, the excess is only significant $>$2$\sigma$ in
\rdbf\ and \ribf.  The former is discussed in detail in
\citet{casey09b} who conclude an insignificant AGN contribution based
on combined evidence from X-ray (very low flux, consistent with
starbursts), extended radio emission (inconsistent with AGN radio
emission), and no clear powerlaw dominating the near to mid-IR data.
The latter has extended radio emission, no X-ray detection and is the
faintest radio galaxy of our sample ($\sim$15\uJy), and therefore, it
is unlikely that it is dominated by a very powerful AGN.  There is no
clear relation between a system's stellar mass and its resulting
detection in CO.

The galaxies' formation timescales are given by $\tau_{form} \propto
M_\star$/SFR; the median $\tau_{\rm form}$ of the sample is 10\,Myr.
While this quantity could be overestimated if M$_\star$ is
overestimated, the formation timescales likely represent a lower limit
on the time it took to build up the stellar population since the star
formation rates are hypothetically near their peak during the ULIRG
phase ($\simgt$200\,\Mpy) in comparison to most galaxies at the same
epoch (1-10\,\Mpy). The hypothesis of SFRGs being near the peak is
based on the fact that their SFRs are not sustainable beyond
$\sim$100\,Myr, but it is unlikely that these systems are that young;
it is more probable that SFRGs evolved more slowly, building up
stellar mass gradually until a point, when a trigger led to an extreme
star-bursting phase.

\section{Discussion}\label{s:discussion}

Here we explore the relationship between the CO observations in SFRGs
and their other multi-wavelength properties, drawing on comparisons
with other galaxy populations.  CO observations provide a unique and
independent probe of the galaxies' star formation properties, by
constraining the molecular gas reservoirs which fuel the star
formation.  In Table~\ref{tab:summary} we have summarised many of the
measured physical properties of SFRGs relative to SMGs for quick reference.
Important to note, however, is that the SMGs of Table~\ref{tab:summary} are 
those with CO observations, limited to the mostly bright SMG subsample.
After comparing with other populations, we discuss the implications
that these observations have for the study of all high-z ULIRGs and
how improved targeted observations, in both molecular interstellar
medium lines and FIR continuum, will enable thorough, unhindered
analysis of ULIRG evolution and extreme star formation at z$\sim$2.

The original motivation for segregating submm-faint and submm-bright
populations for CO observations is the premise that they exhibit
similar extreme starburst qualities, thus molecular gas qualities, but
differ in dust distribution whereby slightly warmer dust systems are
undetectable at 850\um\ (the usual SMG-selection band).  Surveying SFRGs
in CO provides confirmation, through detection of vast gas reservoirs,
that SMGs are not the only significant population of ULIRGs at
high-$z$.  The basic physical premise of the comparison is that the
dust distribution in SFRGs would need to be clumpier or more compact
to be heated to slightly higher dust temperatures.  Since SFRG radio
morphologies on average seem to be similarly extended and irregular,
SFRGs are suggested to be less homogeneously diffuse than SMGs
\citep{menendez-delmestre09a,hainline09a}, with more concentrated
clumps, yet spread over the same large area.

\begin{table}
\begin{center}
\caption{SFRG Properties Summarised Relative to SMGs}
\label{tab:summary}
\begin{tabular}{l@{ }cc}
\hline\hline
Property & SFRGs & SMGs$^a$ \\
\hline
$\langle z\rangle$                        & 1.8$\pm$0.7                    & 2.4$\pm$0.6 \\
L$_{\rm FIR}$ (\lsun)                     & (4$\pm$3)$\times$10$^{12}$     & (8$\pm$4)$\times$10$^{12}$ \\
SFR (\Mpy)                                & 700$\pm$500                    & 1400$\pm$500 \\
$\langle$ R$_{\rm eff} \rangle$ (kpc) $^b$ & 2.3$\pm$0.8                   & 2.7$\pm$0.4 \\
$\Sigma_{\rm SFR}$ (\Mpy\,kpc$^{-2}$) $^b$ & 30$^{+40}_{-20}$              & 60$^{+140}_{-40}$ \\
T$_{\rm dust}$ (K) $^c$                   & 66$\pm$15                      & 41$\pm$5 \\
M$_{\rm dust}$ (\msun) $^c$               & $<$2$\times$10$^9$             & 9$\times$10$^8$ \\
H-band Mag                                & -25.4$\pm$0.8                  & -25.8$\pm$0.2 \\
M$_{\rm \star}$ (\msun)                   & (7$\pm$6)$\times$10$^{10}$     & (2$\pm$1)$\times$10$^{11}$ \\
$\langle \tau_{form} \rangle$ (Myr)       & 10$^{+15}_{-6}$                 & 14$^{+8}_{-4}$ \\
M$_{\rm \hh}$ (\msun)                     & (2.1$\pm$0.7)$\times$10$^{10}$  & (5.1$\pm$2.8)$\times$10$^{10}$ \\
$\Delta V_{\rm \co}$ (\kms)               & 320$\pm$80                     & 530$\pm$150 \\
SFE (\lsun/\msun)                         & 280$\pm$260                    & 450$\pm$170 \\
M$_{\rm dyn}\,sin^{2}\,i$ (\msun) $^d$    & (7.2$^{+6.8}_{-3.4}$)$\times$10$^{10}$ & (1.5$^{+1.4}_{-0.4}$)$\times$10$^{11}$ \\
$\langle \tau_{\rm depl} \rangle$ (Myr)   & 34$\pm$24                      & 40$^{+50}_{-30}$ \\
f$_{gas}=\langle M_{gas}/M_{dyn} \rangle$ $^e$ & 0.07$^{+0.11}_{-0.02}$    & 0.09$^{+0.09}_{-0.07}$ \\
f$_{stars}=\langle M_{\star}/M_{dyn} \rangle$ $^e$ & $\sim$0.2             & $\sim$0.3 \\
$S_{\rm CO(4-3)}/S_{\rm CO(3-2)}$ $^f$    & 2.0$\pm$0.8                    & 1.4$\pm$0.2 \\
$f_{\rm AGN}$ $^g$                        & 0.3$\pm$0.1                    & 0.4$\pm$0.2 \\
\hline\hline
\end{tabular}
\end{center}
{\small

$^a$ All aggregate properties of SMGs are measured with respect to the
CO-observed subset from \citet{neri03a}, \citet{greve05a}, and
\citet{tacconi06a}.

$^b$ Effective radius and SFR density are measured for MERLIN+VLA imaged SFRGs only.
The effective size of SMGs is from \citet{biggs08a} and the SFR
density is the SMG SFR divided by the mean R$_{\rm eff}$.

$^c$ T$_{\rm dust}$ and M$_{\rm dust}$ fits for SFRGs are described in
section \ref{ss:dustmass}; while both measurements are highly
uncertain due to poor FIR flux density constraints, their calculation
is useful for comparison with SMGs from \citet{chapman05a} and
\citet{kovacs06a}.

$^d$ Dynamical mass assumes an effective radius of 2\,kpc for SFRGs
and SMGs without radio size measurements; for SMGs, this size is
supported by measurements from \citet{tacconi08a}, and for SFRGs the
size has been shown to be similar for one SFRG \citep{bothwell10a} and
is similar to their MERLIN+VLA radio sizes (R$_{\rm eff}$).

$^e$ The gas and stars fraction represent the fraction of total mass
which is in gas (or in stars), internal to each galaxy.  Note: the gas
fraction does not include the 40\%\ correction for helium.

$^f$ The ratio of CO line fluxes represents the source excitation.
The value for SFRGs is based on the single measurement of
\rbbf\ described in section \ref{ss:254excitation}, and the SMG
measurement is taken from the three SMGs measured by \citet{weiss07a}.

$^g$ The AGN fraction (within a population) is estimated in the SFRG
sample as described in section~\ref{ss:sfes}, while the AGN fraction
of CO-observed SMGs is taken from their rest-UV spectral
classification (note none of the SFRGs have AGN spectral signatures).
The entire SMG population is estimated to have an AGN fraction of
$\sim$0.25 \citep[see][]{alexander05a}.

}
\end{table}

\subsection{Multi-wavelength Properties of SFRGs and AGN fraction}\label{ss:agn}

Table~\ref{tab:derived} provides a summary of SFRGs' physical
properties derived from multi-wavelength data.  It includes
AGN/starburst classifications for each individual SFRG, which help
shed light on the complex nature of the population, and the potential
for AGN contamination.  ``Class$^{\rm X}$'' classifies SFRGs based on
X-ray luminosity, ``Class$^{\rm IR}$'' classifies according to
8\um\ flux excess in the near-IR, and ``Class$^{\rm CO}$'' classifies
according to FIR-to-CO luminosity ratio (where an unusually high ratio
can be accounted for by an AGN contributing significantly to radio
luminosity, thus overestimated FIR luminosity).  A fourth class and
fifth could also be defined, as rest-UV/optical spectroscopic features
and as a function of the star formation rate density as traced by
MERLIN+VLA morphologies, however we have noted that all of the
samples' morphologies are consistent with starbursts in both criteria.
While we leave the reader the flexibility to interpret the AGN content
of the SFRG sample as (s)he sees fit, we infer a total SFRG AGN
fraction of 0.3$\pm$0.1 to first order (included in
Table~\ref{tab:summary}), the mean AGN dominated fraction from each of
the three Table~\ref{tab:derived} classifications.  This agrees with
earlier measurements and estimates for SFRGs given in \citet{casey09a}
and \citet{casey09b}.  Also, the possibility exists that weak beaming
of low-luminosity radio jets could overestimate radio luminosity by
factors of a few for some SFRGs \citep[please see][for the detailed
  discussion of radio beaming in more compact SFRGs]{casey09a}.
Again, we caution AGN contamination has the potential to strongly bias
the CO observations/interpretation of individual objects.

Figure~\ref{fig:merlin} shows that SFRGs have faint optical
luminosities, suggestive of heavy reddening or obscuration by dust.
This idea is supported further by the large discrepancy between star
formation rates derived from radio luminosity versus rest-UV flux
densities.  Even after correction for dust extinction using the UV
slope and dust extinction models of \citet{calzetti94a}, the UV
derived SFR can be a factor of 10-100 times lower than the
radio-inferred SFR demonstrating that SFRGs are far dustier than
normal dust extinction laws predict (which also holds for SMGs).
Their radio morphologies are primarily extended and irregular, thus
attributable to star formation and not compact AGN as is also the case
for SMGs \citep{biggs08a,chapman04b}.  We also note the effective
MERLIN+VLA radii between SFRGs that have been CO detected ($R_{\rm
  eff}\,=\,$2.7$\pm$0.6\,kpc) and those that have not ($R_{\rm
  eff}\,=\,$1.7$\pm$0.8\,kpc). While the difference is insignificant,
it could be explained by observing that the sources with lower
effective radii have lower S/N in the radio, implying lower FIR
luminosities.  If we assume the inferred radio/FIR and the ULIRG
$L^\prime_{\rm CO}$/$L_{\rm FIR}$ relations, we predict integrated CO
fluxes of 0.1-0.3\,Jy\,\kms\ for CO-undetected sources based on radio
flux density (requiring particularly sensitive observations with our
setup).

Figure~\ref{fig:midir} showed that most SFRGs have only minor AGN
contribution in the near-IR; most emission is dominated by stars.  The
stellar masses of SFRGs average to 7$\times$10$^{10}$\,\msun,
consistent with SMG masses derived by \citet{hainline09a} using rest
H-band luminosity but a factor of $\sim$2$\times$ lower than the SMG
masses of \citet{borys05a} from rest K-band luminosity.

\subsection{Comparing to SMGs}\label{ss:brightSMGs}

The comparison of molecular gas properties between SFRGs and SMGs is
the main focus of this work, however, a strong selection bias is
folded into this comparison.  The first SMGs to be observed in CO
\citep{neri03a,greve05a,tacconi06a} were among the brightest SMGs with
spectroscopic redshifts ($L_{\rm FIR}\simgt$10$^{12.5}$\,\lsun) since
brighter $L_{\rm FIR}$ systems had a higher likelihood of being CO
detected.  We note that the luminosity distribution of SMGs plotted in
Fig.~\ref{fig:radiodist} includes many lower luminosity SMGs which
have recently been observed in CO, in parallel to our SFRG observing
programs, but have not yet been analyzed or published (Bothwell \etal, in
preparation).  Most of the 'bright-SMGs' (those with published CO
spectra) have AGN signatures in their rest-UV spectra, a property
which would exclude them from SFRG selection if they were submm-faint.
This means that similarly bright radio sources which are submm-faint,
comparable to the 'bright-SMGs,' were excluded from our sample due to
AGN contamination.

Besides the luminosity bias introduced by weeding out AGN as revealed
by rest-UV/optical spectroscopy, a further bias exists due to
spectroscopic incompleteness of the SFRG population.  Since their
radio emission was more likely thought to be dominated by AGN, SFRGs
were not followed up in rest-UV/rest-optical spectroscopy nearly as
thoroughly or completely as SMGs. This likely means that the absolute
brightest SMGs were CO observed while a large sample of bright SFRGs
could have been excluded from CO observations due to a lack of
reliable redshift information or potentially strong AGN contamination.
Furthermore, the original SFRG selection of \citet{chapman04a} had the
added criteria of a faint optical magnitude $i>$\,23, which made
redshift measurement from rest-UV spectra more difficult.  There has
been some anecdotal indication, however, that warmer dust systems
might not exist at the highest luminosities with the same frequency as
colder dust systems, as revealed by 250\um\ selected HyLIRG
populations \citep{casey10a}.

We note that the mis-identification of optical counterparts to SMGs
has potentially led to a lower CO-detection rate for that population
than other CO-observed galaxy populations; some of the radio galaxy
SMG-counterparts might not be starbursts and might have intrinsically
low FIR luminosities, thus CO luminosities.  We note that SFRGs
(although not faced with the issue of matching FIR positions to a
correct radio counterpart) could also suffer from the selection of
non-starburst radio galaxies.

In the sections below, we frequently discuss how SFRGs relate to SMGs:
both the CO-observed 'bright' subsample of SMGs and extrapolations
based on preliminary analysis on the fainter, more numerous sample of
SMGs (Bothwell, private communication).  The overall luminosity bias
existing in these distinct samples must be kept in mind when
population comparisons are made.

\subsection{Star Formation Efficiencies}\label{ss:sfes}

\begin{figure*}
  \centering
  \includegraphics[width=0.99\columnwidth]{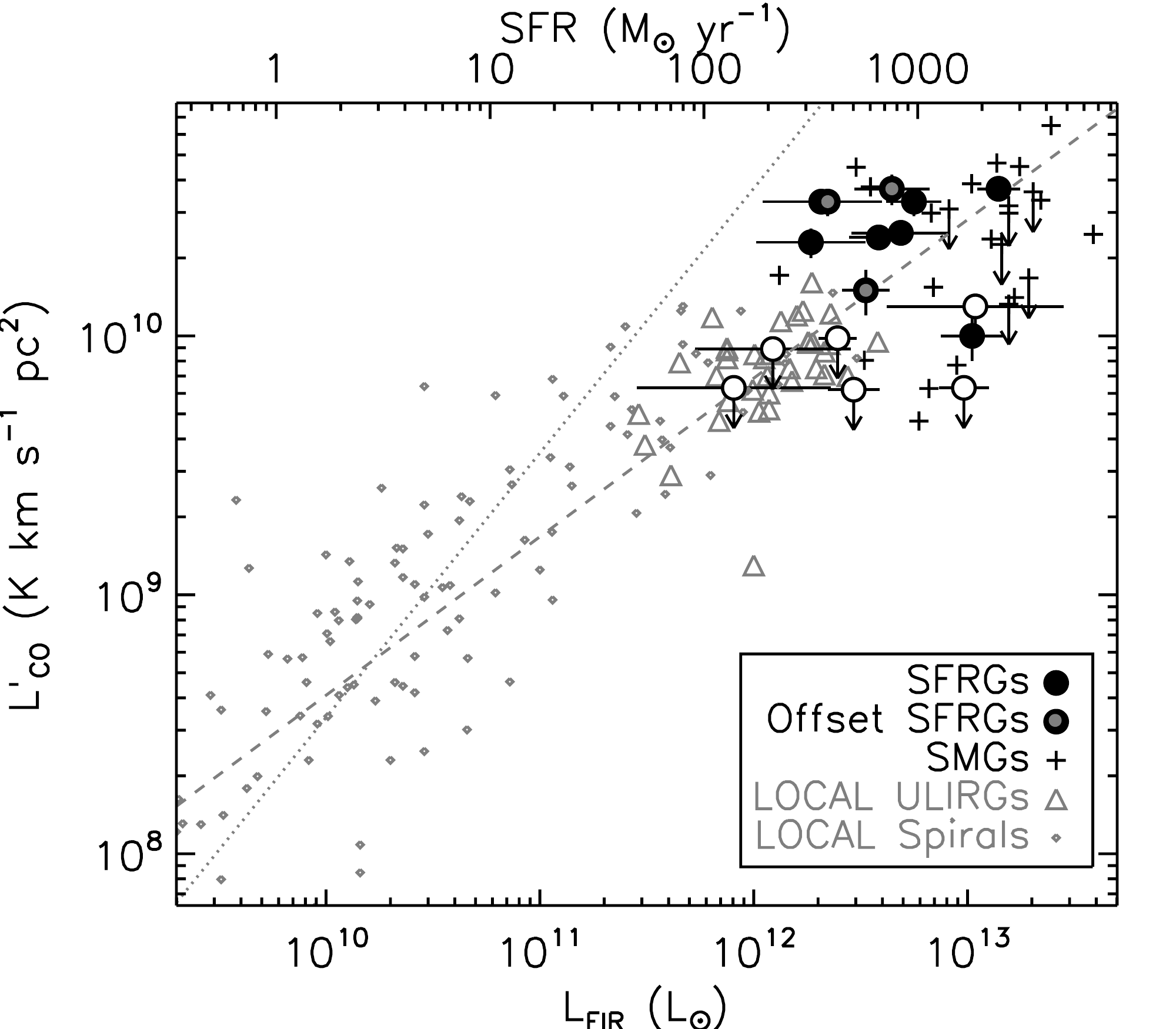}
  \includegraphics[width=0.99\columnwidth]{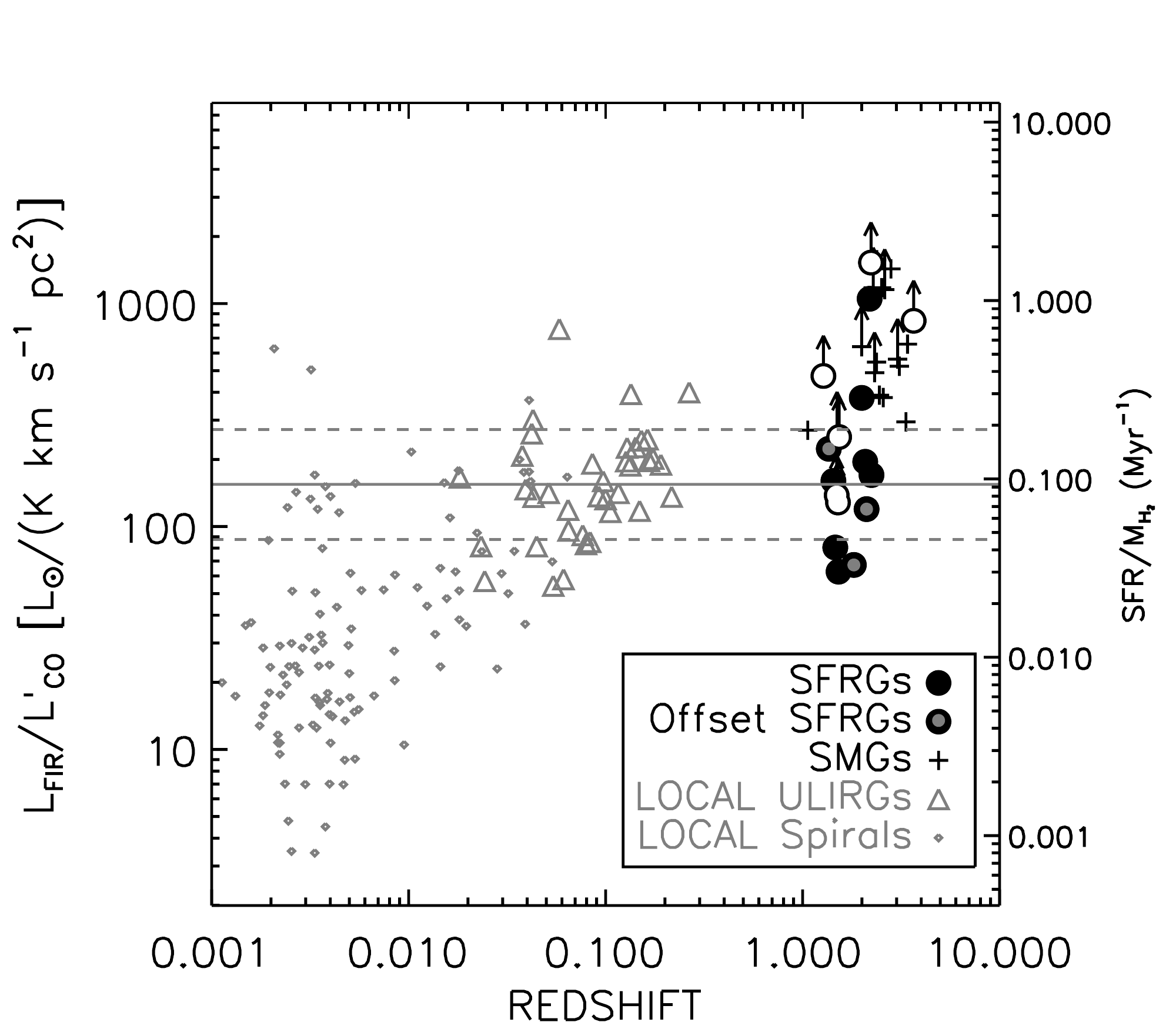}
  \caption{{\it Left:} FIR Luminosity against CO Luminosity,
    L$^\prime_{CO[1-0]}$.  SFRGs ({\it black circles}) lie in the same
    luminosity space as local ULIRGs \citep[{\it gray
        triangles};][]{solomon97a}, while SMGs
    \citep[$crosses$;][]{greve05a} have higher luminosities.  SFRGs
    detected in CO are solid while SFRGs without CO detection are open
    circles. The tentative identifications or ``offset'' CO sources
    are circles with gray centres.  We also overplot data of spiral
    galaxies for perspective \citep[{\it gray
        diamonds};][]{solomon88a,gao04a}, and include the best-fit
    observed relations between L$_{\rm FIR}$ and L$^\prime_{\rm CO}$
    for local spirals (dotted line) and local ULIRGs/SMGs (dashed
    line).  {\it Right:} the star formation efficiency, given by
    L$_{\rm FIR}$/L$_{\rm CO}$, is plotted with redshift.  SFRGs
    appear to share the same range of SFEs as SMGs despite their
    fainter luminosities.  The mean SFE and 1-$\sigma$ bounds of local
    ULIRGs are illustrated by the horizontal solid and dashed lines.
  }
  \label{fig:lfirlco}
\end{figure*}

The relationship between star formation rate and molecular gas mass is
paramount to a galaxy's evolutionary interpretation.  This is measured
by comparing the FIR luminosity with CO line luminosity, as we show in
Figure~\ref{fig:lfirlco}.  In this context, SFRGs appear to have
similarly high CO luminosities as SMGs, and most SFRGs lie slightly
above the `ULIRG' star formation efficiency powerlaw relation,
$L^\prime_{\rm CO}\,\propto\,L_{\rm FIR}$$^{0.61}$, which is followed
by both local ULIRGs and SMGs.  However, SFRGs are inconsistent with
the star formation efficiency relation which describes local spiral
galaxies \citep[$L^\prime_{\rm CO}\,\propto\,L_{\rm
    FIR}$$^{0.93}$;][]{solomon88a,gao04a}.  The SMGs shown on
Fig~\ref{fig:lfirlco} are those of \citet{neri03a}, \citet{greve05a}
and \citet{tacconi06a}.  As mentioned in section~\ref{ss:brightSMGs},
SFRGs are a factor of $\sim$2 less luminous in radio (thus L$_{\rm
  FIR}$) than CO-observed SMGs.  Due to their relative high
luminosities with respect to local ULIRGs, the SMGs have been
described as the {\it scaled-up} high-redshift analogues to local
ULIRGs \citep{tacconi06a}.  In contrast, SFRGs probe a luminosity
regime \simlt10$^{12.5}$\,\lsun\ closer to the locus of local ULIRGs
at $\sim$10$^{12}$\,\lsun.  While SFRGs might seem to be better
analogues of local ULIRGs than SMGs in luminosity space, we note that
the stellar and gas properties of the two populations are quite
distinct: local ULIRGs being more compact with lower stellar masses
than SFRGs and SMGs \citep{dasyra06a}, and they have lower CO
luminosities by a factor of $\sim$2-3.

The star formation efficiencies (SFEs) of SMGs and SFRGs span the
range 70-1000\,\lsun /\msun.  The median SFE of the CO-detected SFRG
sample is 280$\pm$260\,\lsun\,\msun$^{-1}$ which is statistically
consistent with the mean SFE for SMGs,
450$\pm$170\,\lsun\,\msun$^{-1}$, although both values incorporate the
large uncertainties of the gas conversion factor and FIR-derived SFR.
\citet{chapman08a} highlights that the two pilot program
\cco\ detections of \rbbf\ and \chap\ have exceptionally high SFEs and
hypothesised that SFRGs, with further observation, might show
similarly high SFEs compared to SMGs.  \citet{daddi08a} analyzed the
CO content of two $BzK$ selected galaxies (\dadb\ and \dada, also
selected as SFRGs and included in our analysis) and claimed that they
had relatively low, Milky Way/``normal spiral'' efficiencies,
emphasising the difference between them and the high-efficiency
ULIRGs.  Our large sample of SFRGs, including both the Chapman et
al. and Daddi et al. subsamples reveals a much wider spread in star
formation efficiencies, suggestive of a wide range in gas states and a
possible range of galaxy states, although more SFRGs are consistent
with the less-efficient Daddi et al. sample.

The SFRGs not detected in CO and those far below the ULIRG $L_{\rm
  FIR}$/$L_{\rm CO}^\prime$ relation would appear to be very efficient
star formers (less gas to fuel their high SFRs), however this assumes
that AGN contamination is minimal.  AGN contamination is more likely
than super-efficient star formation and happens when AGN boosts the
radio-inferred FIR luminosity, thus inferred star formation rate; as
the empirical relation between $L_{\rm FIR}$ and $L_{\rm CO}^\prime$
suggests, AGN contaminated sources would be fainter in CO gas than
predicted.  We then can classify SFRGs in terms of CO luminosity to
FIR luminosity (i.e. the ratio of $L_{\rm CO}^\prime$/$L_{\rm FIR}$),
where low ratios are designated 'AGN' in 'CLASS$_{CO}$' in
Table~\ref{tab:derived}.  While the scatter of local ULIRGs around the
$L_{\rm CO}^\prime$/$L_{\rm FIR}$ relation is minimal, $\sim$0.3\,dex,
both SFRG and SMG populations have more significant scatter below the
relation; this is consistent with many SFRGs and SMGs having powerful
AGN which boosts radio (thus FIR) luminosity.

The scatter of SFRGs above the ULIRG $L_{\rm CO}^\prime$/$L_{\rm FIR}$
relation is suggestive of low star formation efficiencies as presented
by \citet{daddi08a} and \citet{daddi10a}.  We caution that sources
like \dadb\ and \dada\ which have these low star formation
efficiencies, it might be more appropriate to assume that the gas
properties are more similar to spiral/disk galaxies than ULIRGs,
particularly when converting to molecular gas mass using the
CO/\hh\ conversion factor.  This factor differs greatly between ULIRGs
(0.8 $M_\odot$/(K\,\kms\,pc$^{2}$)) and Milky Way type disk galaxies
(4.5 $M_\odot$/(K\,\kms\,pc$^{2}$)).  Since the inferred gas masses
differ so greatly given these different assumptions, we give $M_{\hh}$
for both ULIRG and spiral/disk galaxy assumptions in
Table~\ref{tab:derived} but proceed with our interpretation using the
ULIRG conversion factor.

\subsection{Line Widths: Implications for Merger Stage}

\begin{figure}
  \centering
  \includegraphics[width=0.90\columnwidth]{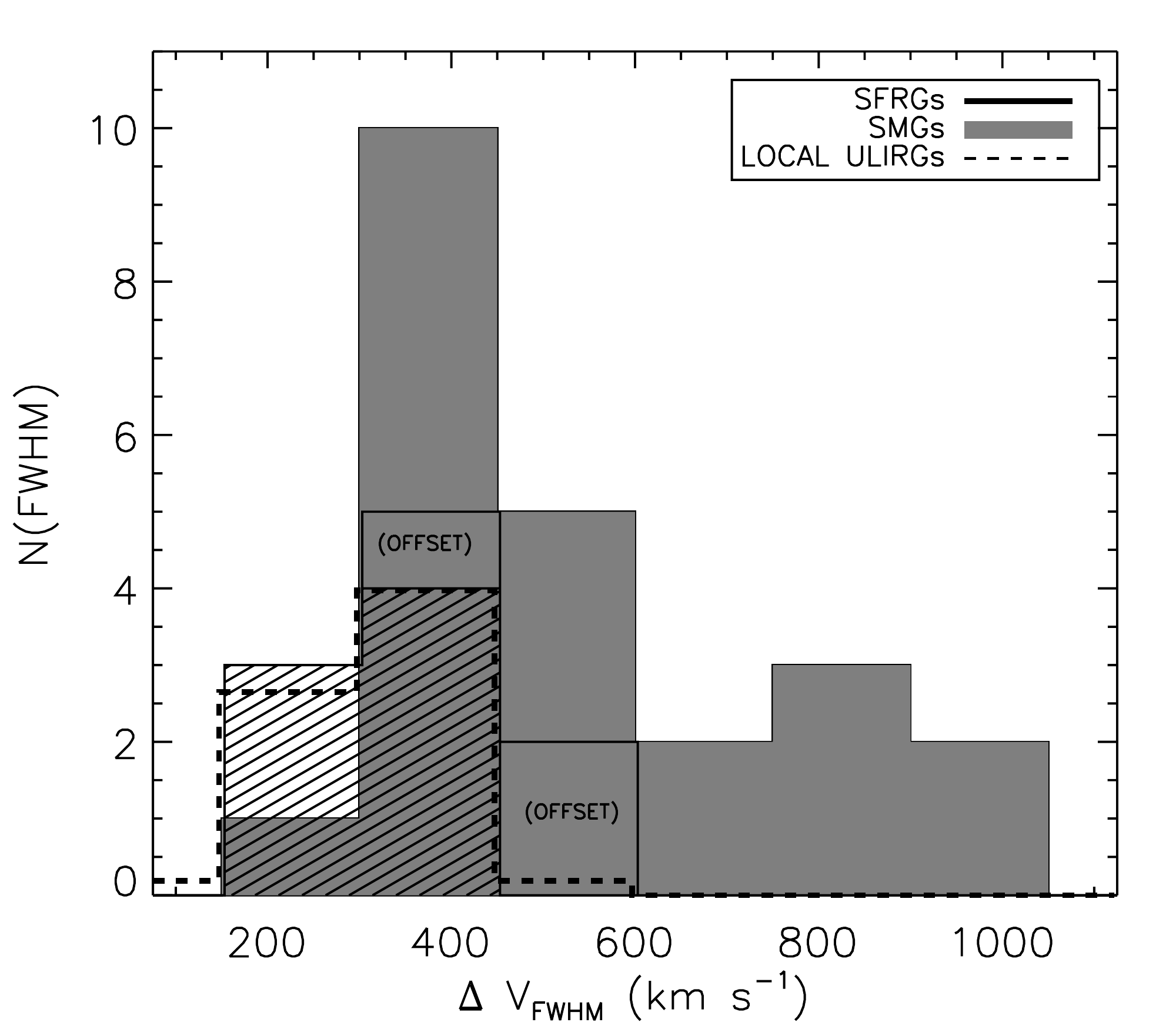}
  \caption{The distribution in CO line widths for the SFRG sample
    compared to local ULIRGs \citep{solomon97a} and SMGs.  The
    offset/tentative SFRGs (open histogram) are added on top of the
    the remaining SFRG sample (hashed area).  The distribution for
    local ULIRGs has been re-normalised with respect to the total
    number of SFRGs for a more clear comparison. The SMG distribution
    has been corrected for overestimation in line widths which is
    caused by fitting a single Gaussian to a double-peaked CO line,
    and the SMG distribution includes the samples of \citet{neri03a},
    \citet{greve05a}, and \citet{tacconi06a}, as well as some yet
    unpublished SMG observations (Bothwell \etal, in preparation). }
  \label{fig:nfwhm}
\end{figure}

The distribution in CO line widths provides important insight into the
galaxies' dynamics.  Figure~\ref{fig:nfwhm} shows the $\Delta V_{\rm \co}$ full
width at half maximum (FWHM) distributions for our SFRGs, SMGs (both
'bright' SMGs and the fainter sample observed in CO only recently;
Smail, private communication) and local ULIRGs \citep{solomon97a}.
None of the line widths presented here have made any inclination angle
assumptions. The line widths of SMGs were adjusted to correct for the
prior exclusion of double-peaked Gaussians \citep[see the details of
  this correction in][]{coppin08a}; this has reduced the mean SMG line
width from 600\,\kms\ to 530\,\kms.  Despite the adjustment, SMGs seem
quite distinct from SFRGs and local ULIRGs by having a high-FWHM tail
in its distribution.  This high-FWHM tail is seen only in bright
subsample of SMGs originally surveyed in CO gas.  While it could be
attributed to selection bias, in that wide CO features are only
detectable in the brightest objects where S/N is much higher, we
highlight that there was significant improvement in receiver
sensitivity between observations of these bright and wide SMG CO lines
and the fainter SFRG observations, so the data have comparable S/N.
This raises the possibility that only the brightest subsample of
high-$z$ ULIRGs ($L_{\rm FIR}\simgt$10$^{13}$\,\lsun) have wide CO
line widths ($\simgt$500\,\kms). 

Broad dispersion-dominated CO lines (and irregular double peaked
profiles) in the highest luminosity systems is suggestive of early
stage major mergers where two gas rich disks are infalling.  Local
ULIRGs in contrast have a much narrower line width distribution and
are a factor of $\sim$5-10 fainter in $L_{\rm FIR}$.  For this reason,
local ULIRGs are often said to be in a late starburst phase, at a
coalesced point during a merger \citep[when progenitors have
  coalesced into a single system; for a review see][]{sanders96a}.
SFRGs and more modest-luminosity SMGs are difficult to place in this
evolutionary sequence, but their populations are not likely to be
exclusively dominated by either beginning or ending merger sub-stages.

The lower dynamical masses of local ULIRGs (typical sizes
$R\,=$\,1\,kpc and $\Delta V\,=$\,300\,\kms) could be due to
downsizing$-$where extreme starbursts today are less massive than
those at high-$z$.  Both SFRGs and SMGs are consistent with this
picture since they seem to be a factor of $\sim$2 larger
\citep[$R_{1/2}\,\simgt$2\,\kpc, see measurements of CO size
  in][]{tacconi08a,daddi10a,bothwell10a}.  The subsample of SMGs which
have very broad features may be in a particular stage of merger where
their line profile becomes broadened, perhaps in observation of two
distinct components with very small physical separation or gas.  This
would appear to increase our dynamical mass estimates of these
systems. However, the radial `size' and merger correction factor $C$
would both need to be reasessed (both of which have not been measured
on a case by case basis for these sources) before physical
interpretation of higher dynamical mass estimates were made.

Although more observations need to be taken to conclude, an alternate
explanation for the narrower line widths observed in modest-luminosity
sources, including SFRGs, is that the population consists of fewer
major mergers than the very bright systems.  Observational evidence
indicates that anywhere from 50-90\%\ of local ULIRGs have undergone
recent mergers \citep[e.g.]{lawrence89a,melnick90a,clements96a}, but
that a sizable fraction might be triggered by other mechanisms.
Recent work from \citet{genzel08a} suggests that high star formation
rates in secularly evolving disk galaxies may be caused by rapid
rotation or smooth accretion of material from its surroundings
(e.g. minor merging or tidal accretion).  In addition, theoretical
work indicates that the high star formation rates and IR luminosities
in ULIRGs could often be generated by minor mergers and turbulent disk
processes \citep[e.g.][]{monaco04a,dib06a}.  ULIRGs driven by secular
processes would exhibit narrow CO line widths, consistent with the
SFRGs presented in this paper.  High spatial resolution gas
observations are needed to determine the true nature of their dynamics
however.

Recent observations and simulation work have reiterated the idea that
$>$10$^{13}$ ULIRGs are more often in early-stage major mergers.
\citet{tacconi08a} and \citet{engel10a} show that most SMGs at
z$\sim$2 exhibit disturbed gas morphologies rather than smoothly
rotating disks, and simulations and semi-analytic SMG models tell us
that major mergers likely initiate most ultraluminous phases of
high-$z$ star formation seen in SMGs
\citep{narayanan09a,swinbank08a,baugh05a}.  However, recent work from
\citet{dave10a} indicate that ULIRGs may also be driven by continual
bombardment by very low mass fragments onto a
$\sim$10$^{11}$\,\msun\ galaxy.  Cold streams feeding continual gas
buildup \citep[e.g.][]{dekel09a} has also been raised as a possible
origin.  The tail of large CO line widths in SMGs provides a crucial
piece of observational evidence that some SMGs are much more highly
disturbed and represent a different phase than SFRGs and local ULIRGs.

We note that the mean gas fraction, defined as gas mass to dynamical
mass ratio, of SFRGs is $\langle$M$_{\rm gas}$/M$_{\rm
  dyn}\rangle$=\,$f_{\rm gas}$=\,0.07$^{+0.11}_{-0.02}$, consistent
with the same ratio for SMGs, which have $\langle$M$_{\rm
  gas}$/M$_{\rm dyn}\rangle\,\sim\,$0.09$^{+0.09}_{-0.07}$ (after
correction for a 30$^{o}$ inclination angle). While this comparison is
between SFRGs which have lower luminosities than CO-observed SMGs, the
same molecular gas fraction suggests that the two populations are
likely in similar evolutionary stages.  We note, however, that if we
assume a CO/\hh\ gas conversion factor consistent with spirals instead
of ULIRGs, the gas mass fraction increases substantially to
$\sim$0.60.  While overall, SFRGs and SMG properties are more
consistent with ULIRGs, it is possible that a few outliers, for
example \dadb\ and \dada\ described in \citep{daddi08a} and a few of
the SMGs exhibiting unusually low SFEs, are much more gas rich than
the ULIRGs which comprise the rest of the populations.


\subsection{Comparison with Other Populations}

An evolutionary transition stage between ULIRG and quasar has been
explored extensively by work on Dust Obscured Galaxies
\citep[DOGs;][]{dey08a,pope08b}, a population of galaxies with warm
dust temperatures and more modest SFRs than the brightest SMGs,
consistent with $\sim$10$^{12}$\,\lsun\ ULIRGs.  Ten of fourteen SFRGs
with 24\um\ flux measurements satisfy DOG selection.  While DOG
selection is quite broad and selects nearly all SFRGs and many SMGs
(requiring 24\um\ flux densities $>$100\,\uJy\ and red optical to IR
colors), a subset of DOGs with spectroscopic redshifts have detailed
near-IR to FIR photometric constraints \citep{bussmann09a} which show
that AGN may contribute significantly to their bolometric
luminosities, and as a result, most have warm dust temperatures
($T_d$\,\simgt\,45\,K).  Due to their high AGN fraction
\citep[dependent on luminosity and only $\ll$0.5 in the faintest,
  S$_{24}\,<\,$0.5\,mJy subset, e.g.][]{pope08b} and heavy dust
obscuration, many DOGs are believed to lie at the transition phase
between SMGs (or star-forming ULIRG) and luminous quasar
\citep{pope08b} and overlap with the SFRG population.

While SFRGs might have warm dust temperatures like some DOGs, we find
that most are dominated by star formation and not AGN.  This is
largely a function of the aggressive selection of SFRGs, meant to weed
out strong AGN by their spectral indicators in the rest-UV/optical and
the observation of minimal 8\um\ flux excess and of extended radio
emission.  High star formation rates and a low AGN fraction (with
respect to a higher AGN fraction in DOGs) are strong evidence that
SFRGs are at a similar ULIRG phase to SMGs despite their warm dust.

\begin{figure}
  \centering
  \includegraphics[width=0.90\columnwidth]{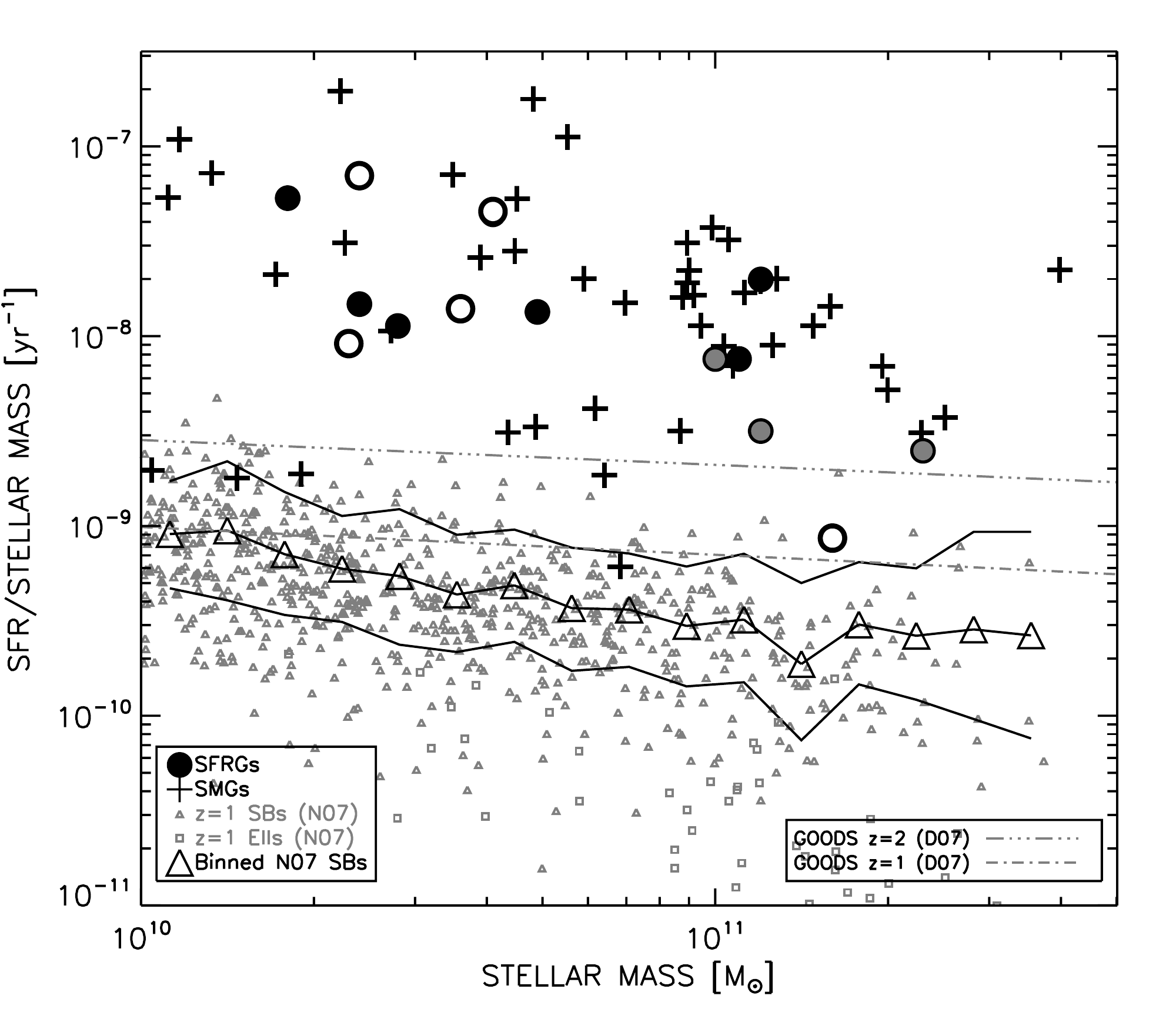}
  \caption{ The star formation rate per unit stellar mass against
    stellar mass.  We compare SFRGs ({\it large circles}) to SMGs
    ({\it crosses}) and z$\sim$1 starburst galaxies \citep[gray
      triangles and squares;][]{noeske07a}.  We also overplot the
    derived redshift dependent relations (at $z$=1 and $z$=2,
    dot-dashed lines) found for GOODS galaxies in \citet{daddi07a}.
    SFRGs not detected in CO are open circles while detected SFRGs are
    filled.  Like SMGs, SFRGs have very large star formation rates per
    stellar mass compared with ``blue sequence'' galaxies, although
    their mean stellar mass ($\sim$7$\times$10$^{10}$\,\msun) is less
    than the mean SMG stellar mass
    ($\sim$2$\times$10$^{11}$\,\msun). }
  \label{fig:noeske}
\end{figure}

The molecular gas fractions of z$\sim$2 normal starburst galaxies have
been estimated at $f_{\rm gas}\sim$0.4-0.5 \citep[assuming a
  $L^\prime_{\rm CO}/L_{\rm FIR}$ prior;][]{erb06a}.  The gas fraction
in SMGs (and a few BX active galaxies) has been measured to be $f_{\rm
  gas}\sim$0.3-0.5 \citep[these values do not take inclination into
  account, which is a factor of 1/4][]{tacconi08a}.  We measure an
internal gas fraction of our SFRG sample of $\sim$0.07, which is
consistent with SMGs, both the Tacconi et al. estimate ($f_{\rm
  gas}\sim$0.1 if corrected for inclination) and our reassessment of
the same data ($f_{\rm gas}\,=\,$0.09$^{+0.08}_{-0.06}$).  All $f_{\rm
  gas}$ measurements for high-$z$ ULIRGs ($f_{\rm gas}\sim$0.1) are
lower than the inferred gas fractions for normal z$\sim$2 galaxies
($f_{\rm gas}\sim$0.5) from \citet{erb06a}.  This could indicate that the more
modest-luminosity galaxies, consisting of gas-rich disks, have low
dynamical masses but proportionately high gas mass, meaning they would
be good progenitor candidates for ULIRG systems, if set on collision
courses with other gas-rich disks.  During the ULIRG starburst phase,
the gas mass would start to decrease with rapid star formation.  In
contrast, we recognize that by using a higher $X_{\rm \co}$ conversion
factor more consistent with quiescent disks, then the measured gas
fractions of these ULIRGs would increase by $\sim$6$\times$, making
their gas fractions consistent with those of the estimate for normal
$z\sim$2 galaxies.  Future observations of more modest-luminosity
galaxies in CO gas \citep[e.g. like the recent work of][]{tacconi10a}
are needed to to truly understand the sequencing and gas properties of
ULIRGs and their progenitors.

Figure~\ref{fig:noeske} highlights the unusually high star formation
rates per unit mass of SFRGs and SMGs above normal starbursts
\citep[AEGIS samples;][]{noeske07a}, for a wide range of stellar
masses.  The relation between stellar mass and specific star formation
rate has been shown to evolve with redshift, e.g. \citet{daddi07a};
however, SMGs and SFRGs still lie at higher star formation rates than
galaxies of equal mass at the same redshift
\citep[e.g.][]{da-Cunha10a}.  This enhanced SFR fraction per unit mass
reveals that SFRG and SMG star formation processes seem fundamentally
different than SF processes in more modest luminosity galaxies,
despite the overall range of stellar mass spanning almost two orders
of magnitude.

The $BzK$ active galaxy selection \citep{daddi04a} is meant to select
moderate star forming ($\sim$100-200\,\Mpy), massive
($\sim$10$^{11}$\,\msun) galaxies at high redshift$-$systems which are
typically below the ULIRG star formation rate threshold
($\simgt$200\,\Mpy).  \citet{daddi10a} detect several active $BzK$
galaxies in \bco\ with surprisingly high gas masses given their star
formation rates, indicating that they exhibit gas properties more
consistent with normal Milky Way type galaxies, but at 'scaled-up'
luminosities, stellar masses and gas masses.  It is important to note
however that the active $BzK$ galaxies observed in CO were all
radio-detected, in other words, most of them have SFRs above the ULIRG
cutoff ($>$200\,\Mpy) and might otherwise be characterized as SMGs or
SFRGs.  All SFRGs in this paper satisfy the active $BzK$ selection
criterion, indicating that $BzK$ selection might probe both massive
star-bursting galaxies and massive, extreme, dusty starbursts.
Similarly, it appears as if SFRG selection might select both high-$z$
ULIRG merging systems {\it and} extreme gas-rich disk galaxies.

\subsection{Volume Density}\label{ss:volumedensity}

It is probable that the luminosity bias described in
section~\ref{ss:brightSMGs} has significant effect on how we can interpret
the CO observations of either population.

Spectroscopic incompleteness in the SFRG population (as discussed in
section~\ref{ss:brightSMGs}) makes volume density difficult to
calculate.  We estimate that about 35\,\%\ of a complete sample of
\uJy\ radio galaxies ($\sim$0.7\,arcmin$^{-2}$) with
S$_{1.4}>$20\,\uJy\ have not been observed spectroscopically
\citep[see][]{chapman03b}.  The majority of these are submillimetre
faint (since most spectroscopic observations of \uJy\ radio galaxies
has been of SMGs).  Of the submm-faint galaxies which were
spectroscopically observed, 65\,\%\ have confirmed redshifts, roughly
half of which are low luminosity AGN and the other half are star
forming galaxies.  This means that by completing spectroscopic
observations of known submm-faint \uJy\ radio galaxies in well
surveyed fields, the source density of star-forming, submm-faint
ULIRGs with redshifts would likely increase by $\sim$30\,\%,
increasing the source density of all high-$z$ ULIRGs by $\simgt$15\%,
which could also increase the ULIRG contribution (including SMGs) to
the cosmic star formation rate density at its peak
\citep[e.g. see][]{bouwens09a,goto10a}.

Future observations from {\sc SCUBA2} and the {\it Herschel Space
  Observatory} in the 50-500\,\um\ range will dramatically improve the
census of warmer dust ULIRGs (and all dusty starburst galaxies) at
z$\sim$2 through continued discovery, thus allowing more thorough
follow up of the submm-faint (and 50-500\um\ bright) radio sample.
With improved interferometric millimetre line observations at low and
high frequencies from the Atacama Large Millimeter Array (ALMA), we
will be able to target these high redshift sources in multiple CO
transitions, thus removing an observational bias towards certain
J-transitions of CO and enabling more accurate calculation of gas
masses. Without such strong temperature biases in gas and dust
observations, a more complete interpretation of high redshift
ultraluminous galaxies will finally be possible.

\section{Conclusions}\label{s:conclusions}

We have presented CO molecular gas observations of a sample of
submillimetre-faint, star-forming radio galaxies (SFRGs).  Due to
their non-detection at submillimetre wavelengths and lack of dominant
AGN, these ultraluminous, \uJy\ radio galaxies are thought to be
dominated by star formation but have warmer dust temperatures than
SMGs.  Out of 16 CO-observed SFRGs (12 from this paper and 4 from the
literature), 10 are detected with a mean CO luminosity of
$L^\prime_{\rm
  CO[1-0]}\,\sim\,$2.6$\times$10$^{10}$\,K\,km\,s$^{-1}$\,pc$^2$,
which is slightly less luminous than the CO-observed SMG sample,
despite being $\sim$2$\times$ less luminous in the radio.  We
attribute the luminosity difference to a selection bias but suggest
that physical driving mechanisms might differ between the very bright
($>$10$^{13}$\,\lsun) and moderately bright ($\sim$10$^{12}$\,\lsun)
populations.

High-resolution radio imaging from MERLIN+VLA shows that the radio
emission in the SFRG sample is resolved and extended with mean
effective radii $\sim$2\,kpc, suggesting that the SFRG radio
luminosities are dominated by star formation rather than AGN.  The
MERLIN+VLA sizes constraints are consistent with similarly analyzed SMG
MERLIN+VLA sizes.  While we note that AGN do not dominate our sample, it
is possible that several of our sources have non-negligible AGN due in
part to their selection as radio galaxies.  Due to limited FIR data,
we use the FIR/radio correlation to derive $L_{FIR}$ and then compute
extinction-free star formation rates from the FIR.

The star formation efficiencies (SFEs) of SFRGs are comparable within
large uncertainties to those of SMGs and local ULIRGs, even though a
few sources appear to have very high SFEs \citep[like those in
][]{chapman08a} or very low SFEs \citep[like those in ][]{daddi08a}.
Those with perceived very high SFEs are more likely AGN dominated than
super efficient; their FIR luminosities as calculated from the radio
are probably overestimated.

SFRGs have narrower CO line widths than the bright subsample of SMGs
at the same redshifts ($\Delta V_{\rm SFRG}\,\sim\,$320\,\kms\ and
$\Delta V_{\rm SMG}\,\sim\,$530\,\kms), suggesting that SFRGs might
have less disturbed dynamical environments.  The line width
distribution is potentially suggestive of different evolutionary
stages or processes between SFRGs and SMGs; however, the observed
difference with SMGs could be due to a S/N or luminosity bias.  The
former would mean that intrinsically broad lines would have
underestimated FWHMs due to low S/N.  The latter is due to the more
thorough spectroscopic sampling of the SMG population.  SMGs have
higher spectroscopic completeness and also include many objects with
AGN signatures in the rest-UV/optical.  Any SFRGs which have similar
AGN signatures were culled from the sample, thus eliminating some of
the potentially brightest SFRGs (most of the bright SMGs which have
been surveyed in CO contain optical AGN).  While less luminous SMGs
(at the same luminosities of SFRGs) exist, few have been observed in
CO, thus it is difficult to rule out that CO luminosity might relate
to the distribution in CO with line width.

Despite selection biases, we have explored the possible physical
scenarios triggering warm-dust ULIRGs in contrast to the well studied
cold-dust ULIRGs.  SFRGs appear to bridge the gap between the
properties of $>$10$^{13}$\,\lsun SMGs and $\sim$10$^{12}$\,\lsun.
Quantitatively, their extended radio emissions suggest sizes
consistent with SMGs, implying much larger dynamical masses than local
ULIRGs. We show that SFRGs have the same AGN fraction as SMGs and are
therefore unlikely to represent a `post-SMG' AGN turn-on phase.
Luminous SMGs have been characterised as an early infall stage during
a major merger, and local ULIRGs are often described at late-type
major mergers.  Here, we suggest that SFRGs (and less luminous SMGs)
span the range of states during peak merger interaction.

\section*{Acknowledgments}
We thank the anonymous referee for detailed, helpful comments which
helped improve this paper greatly.  This work is based on observations
carried out with the IRAM Plateau de Bure Interferometer.  IRAM is
supported by INSU/CNRS (France), MPG (Germany) and IGN (Spain).  We
acknowledge the use of GILDAS software ({\tt
  http://www.iram.fr/IRAMFR/GILDAS}).  This work is also based, in
part, on observations by the University of Manchester at Jodrell Bank
Observatory on behalf of STFC, and the VLA of the National Radio
Astronomy Observatory, a facility of the National Science Foundation
operated under cooperative agreement by Associated Universities, Inc.
CMC thanks the Gates Cambridge Trust for support, IRS thanks STFC for
support, and KC is supported by an STFC Postdoctoral Fellowship.
Support for this work was provided by NASA through Hubble Fellowship
grant HST-HF-51268.01-A awarded by the Space Telescope Science
Institute, which is operated by the Association of Universities for
Research in Astronomy, Inc., for NASA, under contract NAS 5-26555.

\label{lastpage}
\end{document}